\documentclass[longauth]{aa}  

\usepackage{graphicx}
\usepackage[dvipsnames]{xcolor}
\usepackage{lineno}
\usepackage{caption}
\usepackage{subcaption}
\usepackage{arydshln}
\usepackage{txfonts}
\usepackage{orcidlink}
\usepackage{verbatim}
\usepackage{multirow}
\usepackage{multicol}
\usepackage{float}
\usepackage{upgreek}
\usepackage{placeins}
\usepackage{hyperref}
\hypersetup{colorlinks=true, linkcolor=blue, urlcolor=blue, filecolor=blue, citecolor=blue}
	
\usepackage{soul}
\let\ACMmaketitle=\maketitle
\renewcommand{\maketitle}{\begingroup\let\footnote=\thanks \ACMmaketitle\endgroup}

\newcommand{\mjup}[0]{M_\mathrm{Jup}}
\newcommand{\rjup}[0]{R_\mathrm{Jup}}
\newcommand{\rearth}[0]{R_\oplus}
\newcommand{\Searth}[0]{S_\oplus}

\begin{document} 
   \title{Three Hot Jupiters transiting K-dwarfs with a significant heavy element mass}

    \author{
    Y.G.C. Frensch\inst{1}\orcidlink{0000-0003-4009-0330},
    F. Bouchy\inst{1}\orcidlink{0000-0002-7613-393X},
    G. Lo Curto\inst{2}\orcidlink{0000-0002-1158-9354},
    S. Ulmer-Moll\inst{3}\orcidlink{0000-0003-2417-7006},
    S.G. Sousa\inst{4}\orcidlink{0000-0001-9047-2965},
    N.C. Santos\inst{4,5}\orcidlink{0000-0003-4422-2919},
    K.G. Stassun\inst{6}\orcidlink{0000-0002-3481-9052},
    C.N. Watkins\inst{7}\orcidlink{0000-0001-8621-6731},
    H. Chakraborty\inst{1}\orcidlink{0000-0002-5177-1898},
    K. Barkaoui\inst{8,9,10}\orcidlink{0000-0003-1464-9276},
    M. Battley\inst{1,11}\orcidlink{0000-0002-1357-9774},
    W. Ceva\inst{1}\orcidlink{0009-0009-7691-6027},
    K.A. Collins\inst{7}\orcidlink{0000-0001-6588-9574},
    T. Daylan\inst{12}\orcidlink{0000-0002-6939-9211}, 
    P. Evans\inst{13}\orcidlink{0000-0002-5674-2404},
    J.P. Faria\inst{1}\orcidlink{0000-0002-6728-244X},
    C. Farret Jentink\inst{1}\orcidlink{0000-0001-9688-9294},
    E. Fontanet\inst{1}\orcidlink{0000-0002-0215-4551},
    E. Fridén\inst{1}\orcidlink{0009-0002-2548-4948},
    G. Furesz\inst{14}\orcidlink{0000-0001-8467-9767},
    M. Gillon\inst{8},
    N. Grieves\inst{1}\orcidlink{0000-0001-8105-0373},
    C. Hellier\inst{15}\orcidlink{0000-0002-3439-1439},
    E. Jehin\inst{16},
    J.M. Jenkins\inst{17}\orcidlink{0000-0002-4715-9460},
    L.K.W. Kwok\inst{1}\orcidlink{0000-0003-4493-510X},
    D.W. Latham\inst{7}\orcidlink{0000-0001-9911-7388},
    B. Lavie\inst{1}\orcidlink{0000-0001-8884-9276},
    N. Law\inst{18},
    A.W. Mann\inst{18}\orcidlink{0000-0003-3654-1602},
    F. Murgas\inst{10,19}\orcidlink{0000-0001-9087-1245},
    E. Palle\inst{10,19}\orcidlink{0000-0003-0987-1593},
    L. Parc\inst{1}\orcidlink{0000-0002-7382-1913},
    F. Pepe\inst{1}\orcidlink{0000-0002-9815-773X},
    A. Popowicz\inst{20}\orcidlink{0000-0003-3184-5228},
    F.J. Pozuelos\inst{21}\orcidlink{0000-0003-1572-7707},
    D.J. Radford\inst{22}\orcidlink{0000-0002-3940-2360},
    H.M. Relles\inst{7}\orcidlink{0009-0009-5132-9520},
    A. Revol\inst{1},
    G. Ricker\inst{14}\orcidlink{0000-0003-2058-6662},
    S. Seager\inst{14,9,23}\orcidlink{0000-0002-6892-6948},
    M. Shinde\inst{1},
    M. Steiner\inst{1}\orcidlink{0000-0003-3036-3585},
    I.A. Strakhov\inst{24}\orcidlink{0000-0003-0647-6133},
    T.-G. Tan\inst{25}\orcidlink{0000-0001-5603-6895},
    S. Tavella\inst{1}\orcidlink{0009-0005-0003-2429},
    M. Timmermans\inst{8},
    B. Tofflemire\inst{26}\orcidlink{0000-0003-2053-0749},
    S. Udry\inst{1}\orcidlink{0000-0001-7576-6236},
    R. Vanderspek\inst{14}\orcidlink{0000-0001-6763-6562},
    V. Vaulato\inst{1}\orcidlink{0000-0001-7329-3471},
    J.N. Winn\inst{27}\orcidlink{0000-0002-4265-047X},
    \and
    C. Ziegler\inst{28}
    }
   \authorrunning{Y.G.C. Frensch et al.}

   \institute{Observatoire de Genève, 51 Ch. Pegasi, 1290 Versoix, Switzerland\\
              \url{yolanda.frensch@unige.ch}
         \and 
            European Southern Observatory, Karl-Schwarzschild-Strasse 3, 85748 Garching,
            Germany
        \and
            Leiden Observatory, Leiden University, Postbus 9513, 2300 RA Leiden, The Netherlands
        \and
            Instituto de Astrofisica e Ciencias do Espaco, Universidade do Porto, CAUP, Rua das Estrelas, 4150-762 Porto, Portugal
        \and 
            Departamento de Fisica e Astronomia, Faculdade de Ciencias, Universidade do Porto, Rua do Campo Alegre, 4169-007 Porto, Portugal 
        \and
            Department of Physics and Astronomy, Vanderbilt University, Nashville, TN 37235, USA 
        \and 
            Center for Astrophysics \textbar \ Harvard \& Smithsonian, 60 Garden Street, Cambridge, MA 02138, USA 
        \and 
            Astrobiology Research Unit, University of Li\`ege, All\'ee du 6 ao\^ut, 19, 4000 Li\`ege (Sart-Tilman), Belgium 
        \and 
            Department of Earth, Atmospheric and Planetary Science, Massachusetts Institute of Technology, 77 Massachusetts Avenue, Cambridge, MA 02139, USA 
        \and 
            Instituto de Astrof\'isica de Canarias (IAC), Calle V\'ia L\'actea s/n, 38200, La Laguna, Tenerife, Spain 
        \and 
            Astronomy Unit, Queen Mary University of London, G.O. Jones Building, Bethnal Green, London E1 4NS, United Kingdom 
        \and 
            Department of Physics and McDonnell Center for the Space Sciences, Washington University, St. Louis, MO 63130, USA 
        \and 
            El Sauce Observatory, Coquimbo Province, Chile 
        \and 
            Department of Physics and Kavli Institute for Astrophysics and Space Research, Massachusetts Institute of Technology, Cambridge, MA 02139, USA 
        \and
            Astrophysics Group, Keele University, ST5 5BG, U.K. 
        \and 
            STAR Institute, University of Li\`ege, All\'ee du 6 ao\^ut, 19, 4000 Li\`ege (Sart-Tilman), Belgium 
        \and 
            NASA Ames Research Center, Moffett Field, CA 94035, USA
        \and 
            Department of Physics and Astronomy, The University of North Carolina at Chapel Hill, Chapel Hill, NC 27599-3255, USA 
        \and 
            Departamento de Astrof\'isica, Universidad de La Laguna (ULL), E-38206 La Laguna, Tenerife, Spain 
        \and 
            Silesian University of Technology, Akademicka 16, 44-100 Gliwice, Poland 
        \and
            Instituto de Astrof\'isica de Andaluc\'ia (IAA-CSIC), Glorieta de la Astronom\'ia s/n, 18008 Granada, Spain 
        \and 
            Brierfield Observatory, Bowral, NSW Australia 
        \and 
            Department of Aeronautics and Astronautics, MIT, 77 Massachusetts Avenue, Cambridge, MA 02139, USA
        \and
            Sternberg Astronomical Institute Lomonosov Moscow State University 
        \and 
            Perth Exoplanet Survey Telescope, Perth, Western Australia 
        \and
            SETI Institute, Mountain View, CA 94043 USA/NASA Ames Research Center, Moffett Field, CA 94035 USA 
        \and
            Department of Astrophysical Sciences, Princeton University, 4 Ivy Lane, Princeton, NJ 08544, USA
        \and 
            Department of Physics, Engineering and Astronomy, Stephen F. Austin State University, 1936 North St, Nacogdoches, TX 75962, USA
        }

   \date{Received 24 January 2025; accepted 3 June 2025}
   
  \abstract
   {Albeit at a lower frequency than around hotter stars, short-period gas giants around low-mass stars ($T_\mathrm{eff} < 4965$ K) do exist, despite predictions from planetary population synthesis models that such systems should be exceedingly rare.}
  {By combining data from the Transiting Exoplanet Survey Satellite (\textit{TESS}) and ground-based follow-up observations, we seek to confirm and characterize giant planets transiting K dwarfs, particularly mid/late K dwarfs.}
  {Photometric data were obtained from the \textit{TESS} mission, supplemented by ground-based imaging- and photometric observations, as well as high-resolution spectroscopic data from the CORALIE spectrograph. Radial velocity (RV) measurements were analyzed to confirm the presence of companions.}
   {We report the confirmation and characterization of three giants transiting mid-K dwarfs. Within the TOI-2969 system, a giant planet of $1.16\pm 0.04$ $\mjup$ and a radius of 1.10 $\pm$ 0.08 $\rjup$ revolves around its K3V host in 1.82 days. The system of TOI-2989 contains a 3.0 $\pm$ 0.2 $\mjup$ giant with a radius of 1.12 $\pm$ 0.05 $\rjup$, which orbits its K4V host in 3.12 days. The K4V TOI-5300 hosts a giant of 0.6 $\pm$ 0.1 $\mjup$ with a radius of 0.88 $\pm$ 0.08 $\rjup$ and an orbital period of 2.3 days. The equilibrium temperatures of the companions range from 1001 to 1186 K, classifying them as Hot Jupiters. However, they do not present radius inflation. The estimated heavy element masses in their interior, inferred from the mass, radius, and evolutionary models, are $90 \pm 30 M_\oplus$, $114 \pm 30 M_\oplus$, and $84 \pm 21 M_\oplus$, respectively. The heavy element masses are significantly higher than most reported heavy elements for K-dwarf Hot Jupiters.}
   {These mass characterizations contribute to the poorly explored population of massive companions around low-mass stars.}

   \maketitle
\section{Introduction}
With the ever-increasing number of exoplanets, currently totaling 5819\footnote{See the NASA Exoplanet Archive \url{https://exoplanetarchive.ipac.caltech.edu}, accessed 20 January 2025}, our understanding of planetary systems continues to grow, even extending to the rarest of configurations. One such rare category is that of massive companions orbiting low-mass stars (mid-K to late-M). Planetary population synthesis models predict a very low occurrence rate for these systems, suggesting that the rate of planets with masses above 0.3 $\mjup$ decreases below 0.7 $M_\odot$ and drops to zero around stars with masses below 0.5 $M_{\odot}$ \citep{burn_new_2021}. However, these companions do exist, albeit at lower rates than around higher-mass stars. There are currently 19 well-characterized\footnote{Mass precision $<$25\%.} gas giants orbiting mid-K stars (K3 to K5), 13 orbiting late K stars (K6 to K9), and 20 orbiting M stars (versus 32 orbiting K0 to K2 and 136 orbiting G-type stars)\footnote{The spectral types are defined by the effective temperatures, as listed in Table 5 of \citet{pecaut_intrinsic_2013}.}. From the Transiting Exoplanet Survey Satellite (\textit{TESS}; \citeauthor{ricker_transiting_2014} \citeyear{ricker_transiting_2014}) photometry, \citet{gan_occurrence_2023} obtained a Hot Jupiter occurrence rate of $0.27 \pm 0.09 \%$ for early-type M stars with stellar masses ranging from $0.45 - 0.65 M_\odot$. For a wider range of $0.088 - 0.71 M_\odot$, \citet{bryant_occurrence_2023} measures an occurrence rate of $0.194 \pm 0.072 \%$, showing the occurrence rate being non-zero for stars with $M_\star \leq 0.4 M_\odot$. Results from radial-velocity surveys agree with short-period ($1-10$ d) gas giants $(0.3 - 3\mjup)$ being rare around low-mass stars. For M dwarfs, \citet{ribas_carmenes_2023} reports an occurrence rate from CARMENES data of $<0.6\%$, \citet{bonfils_harps_2013} from HARPS $<1\%$, \citet{pinamonti_hades_2022} from HARPS-North $<2\%$, and \citet{pass_mid--late_2023} from TRES, CHIRON, and MAROON-X $<1.5\%$. These results confirm that while gas giants around low-mass stars are rare, they do exist. For comparison, Hot Jupiter occurrence rates are higher around more massive stars, with \citet{wright_frequency_2012} reporting $1.2 \pm 0.38\%$ for F, G, and K dwarfs, and \citet{mayor_harps_2011}, including M dwarfs, finding $0.89 \pm 0.36\%$. While these values align with the upper limits from the various RV surveys, they are higher than the occurrence rates derived from transit data for low-mass stars.

Giants are considered to form either via gravitational instability \citep{boss_giant_1997} or core accretion \citep{pollack_formation_1996}, with Hot Jupiters likely forming ex-situ - at large orbital separations where the conditions are more favorable for both mechanisms - and subsequently migrating inwards \citep{fortney_hot_2021}. The observed relative paucity of giant planets around low-mass stars compared to more massive ones aligns with key predictions of both formation models. For core accretion, this scarcity is attributed to the insufficient mass surface density and longer orbital timescales associated with low-mass stars \citep{laughlin_core_2004, ida_toward_2005}. For gravitational instability, it is due to the requirement of massive, cold disks, which are uncommon around low-mass stars. Testing the predictions of formation and synthesis models and identifying where exactly the decrease in formation starts remains challenging and incomplete. Characterizing giants provides more insight into the poorly explored population of rare low-mass star companions.

To expand the known sample and to bridge the gap between heavier stars and the very low-mass star regime, an ongoing follow-up program on CORALIE aims to characterize giant planets identified by \textit{TESS} around K dwarfs. In this paper, we confirm and characterize three Hot Jupiters transiting mid-K dwarfs, contributing mass measurements to this still relatively unexplored population.

\section{Ongoing CORALIE program}
\textit{TESS} has identified 301 \textit{TESS} objects of interest (TOIs) orbiting low-mass stars, defined by an effective temperature of $T_\mathrm{eff} \leq 4965$K (inclusive up to K3) and a radius of $R_{\star} \leq 0.8 R_{\odot}$. These 301 TOIs have radii ranging from 7.5 to 16 $\rearth$ (0.67 to 1.43 $\rjup$), strongly suggesting that the potential companions are giant planets or brown dwarfs\footnote{This radii range also includes M dwarfs, but we vet the data to exclude stellar companions.}. Until now, 37 have been classified as false positives, and 26 have been confirmed according to the \textit{TESS} Follow-up Observing program \citep[TFOP;][]{collins_tess_2019}\footnote{\url{https://tess.mit.edu/followup}}. Of these 26, 7 have effective temperatures in the mid/late K range (3890 - 4965 K) \citep[e.g.][]{vines_ngts-6b_2019, hartman_hats-47b_2020, huang_tess_2020, martin_toi-1259ab_2021, jordan_hats-74ab_2022, kanodia_toi-3757_2022}. Motivated by the relatively few confirmed TOIs, the CORALIE spectrograph has an ongoing program for \textit{TESS} follow-up observations of giant planets and brown dwarf candidates around mid/late K dwarfs. The sample includes both mid- and late K dwarfs, probing the transition between higher-mass stars and very low-mass stars, to more precisely determine where the occurrence rate of gas giants exactly decreases. 

\begin{figure}[ht!]
        \includegraphics[width=\linewidth]{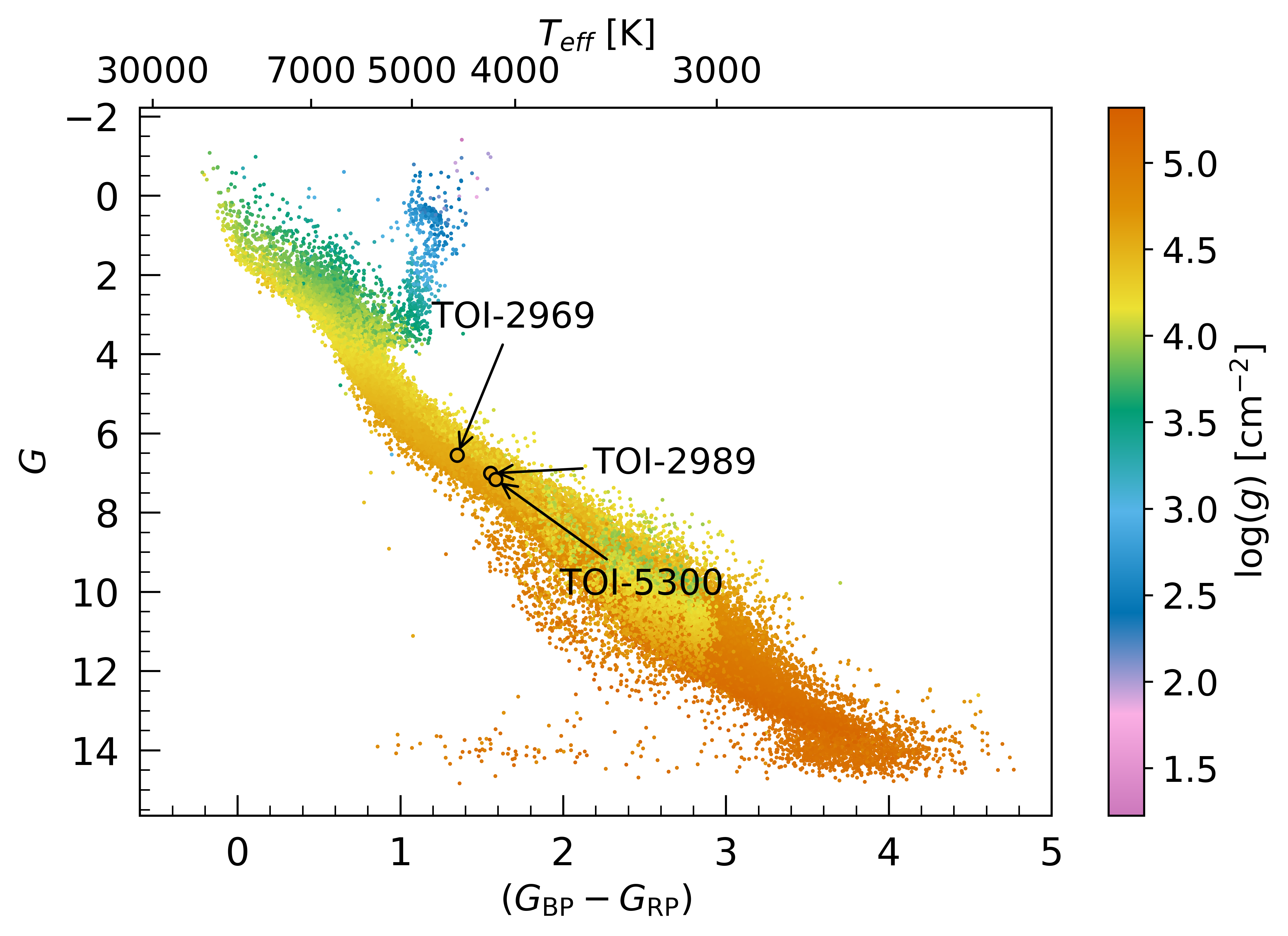}
        \caption{HR diagram of all \textit{Gaia} DR3 nearby stars with a parallax $\pi \geq 10$ mas, with the colours indicating $\log(g)$. The three stars presented in this work are overplotted and visible on the main sequence.}
        \label{fig:HRgaia}
\end{figure}

After selecting targets based on the planetary- (7.5-16 $\rearth$) and stellar radius ($\leq 0.8 R_\odot$), effective temperature (K9V to K4V, 3890K to 4600K, respectively) and the observability of CORALIE in the southern hemisphere ($V \,\mathrm{mag}<14$ and declination $< +20^{\circ}$), we plot the chosen stars on an HR diagram alongside \textit{Gaia} DR3 data \citep{gaia_collaboration_gaia_2023} for nearby stars ($\pi \geq 10\,\mathrm{mas}$, $d \leq 100\,\mathrm{pc}$). This approach enables us to exclude stars that are not on the single main sequence. In Figure \ref{fig:HRgaia}, the three stars presented in this paper are visible on the main sequence, illustrating the approach. 

Supplementary to the \textit{TESS} vetting, manual lightcurve vetting is performed before including targets in the program. During this process, we check for secondary eclipse events, odd-even depth differences, sector depth differences, and spurious events. V-shaped transits are included in the sample, as the likelihood of a grazing transit increases with larger companions around small-sized stars.

The observation strategy of the program continuously evolves. Initially, we commence with two measurements for spectroscopic vetting. These initial observations serve to eliminate eclipsing binaries, identifiable by significant RV variations and/or the presence of two components instead of one in the cross-correlation function (CCF). Note that our selection criteria exclude potential transiting giant planets in binary star systems where the two stars have comparable masses. Some of the TOIs excluded based on these criteria may still host planets. In the absence of clear indications of binarity, the observation frequency is increased and continuously monitored to obtain optimal phase coverage. This includes observations conducted at various phases outside the transit, using the \textit{TESS}-derived transit time and period to guide the timing of observations. Follow-up observations stop when a mass precision of at least 5$\sigma$ is reached. 

\section{Observations}
The photometric data for the targets were acquired from the \textit{TESS} mission (Section \ref{subsec:TESS}). Data from SOAR and SAI were employed in speckle interferometry to search for potential stellar companions (Section \ref{subsec:Interferometry}). Subsequent follow-up photometric observations involved El Sauce, PEST, LCO-CTIO, LCO-SAAO, LCO-HAL, TRAPPIST-South, Brierfield, and SUTO (Section \ref{subsec:SG1}). Ground-based high-resolution spectroscopic data were obtained using the CORALIE spectrograph (Section \ref{subsec:CORALIE}).

\subsection{\textit{TESS} photometry}
\label{subsec:TESS}
Table \ref{table:TESS-SPOC} shows the sectors in which the three systems presented in this paper were observed by \textit{TESS}, including the years and exposure times. For the analyses of these systems, we used the Presearch Data Conditioned Simple Aperture Photometry \citep[PDCSAP][]{stumpe_kepler_2012, stumpe_multiscale_2014, smith_kepler_2012} fluxes and corresponding errors, which were produced by the Science Processing Operation Center (SPOC; \citeauthor{jenkins_tess_2016} \citeyear{jenkins_tess_2016}). If no \textit{TESS}-SPOC data is available, we instead use the Quick Look Pipeline \citep[QLP][]{huang_photometry_2020, huang_tess_2020} photometry. Any data flagged for quality issues (e.g., scattered light, bad calibration, or insufficient targets for systematic error correction), recognizable by a quality parameter larger than 0, are excluded. Cosmic rays are mitigated on the satellite before downlink. The \textit{TESS} data was accessed via the \texttt{lightkurve} Python package \citep{lightkurve_collaboration_lightkurve_2018}. 

        \begin{table}
            \centering
        \caption{Properties of the \textit{TESS}-SPOC and QLP lightcurves.}
        \label{table:TESS-SPOC}
            \begin{tabular}{l c c c c c} \hline\hline
            TOI & Sector & Start date & End date & $t_\mathrm{exp}$ & $\sigma_\mathrm{OOT}$\\
            & & & & [s] & [ppm]\\ \hline 
2969 & 9$^*$ & 28-02-2019 & 25-03-2019 & 1800 & 2580 \\
 & 10$^*$ & 26-03-2019 & 22-04-2019 & 1800 & 25970 \\ 
 & 36$^*$ & 07-03-2021 & 01-04-2021 & 600 & 3280 \\ 
 & 63 & 10-03-2023 & 06-04-2023 & 200 & 3260 \\ 
 & 89$^*$ & 11-02-2025 & 12-03-2025 & 200 & 4710 \\ 
2989 & 9 & 28-02-2019 & 25-03-2019 & 1800 & 1460 \\ 
 & 36 & 07-03-2021 & 01-04-2021 & 600 & 3630 \\ 
5300 & 42 & 21-08-2021 & 15-09-2021 & 600 & 2000 \\
& 70 & 20-09-2023 & 16-10-2023 & 200 & 4720 \\ \hline
        \end{tabular}\begin{flushleft}
        \textbf{Notes:} Here, $\sigma_\mathrm{OOT}$ represents the standard deviation of the out-of-transit flux in the activity-filtered light curves. Sectors marked with $^*$ refer to QLP light curves; all others are \textit{TESS}-SPOC. \end{flushleft}
    \end{table}   

For TOI-2969, the \textit{TESS} sector 10 data are impacted by instrumental noise, as reflected in the notably higher $\sigma_\mathrm{OOT}$ relative to the other sectors. The two affected intervals of the light curve have been excluded from the analysis. For TOI-5300, the \textit{TESS} sector 70 light curve exhibits significantly more red noise than sector 42, with an out-of-transit jitter of 4720 ppm. These fluctuations occur on timescales comparable to the transit duration, compromising the transit signal's reliability. Consequently, sector 70 is excluded from the analysis to ensure data quality and robustness. 

\subsubsection{TOI-2969 - TIC 36452991}
TOI-2969 was alerted on 2021-06-04 \citep{guerrero_tess_2021} by the \textit{TESS} Science Office (TSO) after detection by the FAINT transit search pipeline \citep{kunimoto_searching_2021} using QLP Full Frame Image (FFI) data from sectors 9, 10, and 36. The SPOC transit search pipeline \citep{jenkins_impact_2002, jenkins_transiting_2010, jenkins_kepler_2020} also detected the signature in 2-min cadence data from sector 63. A difference image centroiding analysis located the host star within $0.95 \pm 2.5\arcsec$ of the transit source \citep{twicken_kepler_2018}.
 
\subsubsection{TOI-2989 – TIC 97825640}
TOI-2989 was detected by the FAINT pipeline using the QLP FFI light curve from sector 9. After vetting, the TSO issued a TOI alert on 2021-06-04. The SPOC transit planet search pipeline also detected the transit signal in sector 36 FFI light curve \citep{caldwell_tess_2020}, with difference image centroiding locating the host star within $1.2 \pm 2.5\arcsec$ of the transit source.
 
\subsubsection{TOI-5300 – TIC 267215820}
TOI-5300 was detected by the FAINT pipeline using the QLP FFI light curve from sector 42. The TSO reviewed the vetting information and issued a TOI alert on 2022-02-28. The SPOC transit planet search pipeline also identified the transit signal in sectors 42 (FFI, 200-sec cadence), and 70 (2-min cadence). A difference image centroiding analysis placed the host star within $0.683 \pm 2.5 \arcsec$ of the transit source.

\subsection{Speckle Interferometry}
\label{subsec:Interferometry}
\subsubsection{SOAR}
All stars in this paper were observed with the High-Resolution Camera (HRCAM) installed at the 4.1m Southern Astrophysical Research (SOAR) telescope, located at Cerro Pachón, Coquimbo, Chile. \cite{tokovinin_ten_2018} describes the method for identifying stellar companions. In summary, the presence of binaries is determined from spatial Fourier transform images obtained through speckle observations. A companion star will appear as fringes in these images. The auto-correlation function images, as shown in Figure \ref{fig:SOAR}, are reconstructed images that include a companion for TOI-2969 (with a separation of $3.3\arcsec$, \textit{Gaia} DR3 5407977460534995840), and a mirrored counterpart resulting from the imaging process. How this influences the RV and photometry data is discussed in combination with the orbital solution in Section \ref{sec:TOI2969}. The true position of the binary is determined using shift-and-added lucky imaging, which has a lower contrast sensitivity compared to speckle imaging. For TOI-2989 and TOI-5300, no stellar companions were detected, with detection limits of $\Delta I = 5.5^m$ and $\Delta I = 6.0^m$, respectively at $1.0\arcsec$. 

\subsubsection{SAI}
TOI-5300 was observed on UTC 2023 September 30 with the speckle polarimeter on the 2.5-m telescope at the Caucasian Observatory of Sternberg Astronomical Institute (SAI) of Lomonosov Moscow State University. The image is visible in Figure \ref{fig:SAI}. A low-noise CMOS detector Hamamatsu ORCA-quest \citep{safonov_speckle_2017} was used as a detector. The atmospheric dispersion compensator was active, which allowed using the $I_\mathrm{c}$ band. The respective angular resolution is $0.083\arcsec$. A total of 2500 frames with 60 ms exposure have been accumulated. The atmospheric conditions were exceptionally good at the time of observation; the long-exposure full width at half maximum (FWHM) was $0.54\arcsec$. We did not detect any stellar companions; detection limits are $\Delta I_\mathrm{c}=4.7^m$ and $6.2^m$ at distances $0.25\arcsec$ and $1.0\arcsec$ from the star, respectively.

\subsection{Ground-based photometry}
\label{subsec:SG1}
The \textit{TESS} pixel scale is $\sim 21\arcsec$ pixel$^{-1}$ and photometric apertures typically extend out to roughly $1\arcmin$, generally causing multiple stars to blend in the \textit{TESS} photometric aperture. The SPOC uses difference image centroiding to localize the transit source to typically $2.5\arcsec$ ($\sim$0.1 pixels). To verify the true source of the \textit{TESS} detection and to check for wavelength-dependent transit depth, we acquired ground-based time-series follow-up photometry of the fields around TOI-2969, TOI-2989, and TOI-5300 as part of TFOP. We used the {\tt TESS Transit Finder}, which is a customized version of the {\tt Tapir} software package \citep{jensen_tapir_2013}, to schedule our transit observations. All light curve data are available under each host star's web page on the Exoplanet Follow-up Observing Program (ExoFOP) website\footnote{\label{foot:EXOFOP}\url{https://exofop.ipac.caltech.edu/tess}} and are included in the global modeling described in section \ref{sec:Orbital_Solutions}. 

\subsubsection{TOI-2969}
\label{sec:SG1-TOI2969}
We observed a full transit window of TOI-2969 b on UTC 2021 June 12 in the Cousins R band from the Perth Exoplanet Survey Telescope (PEST) located near Perth, Australia. The 0.3 m PEST telescope has a $5544\times3694$ QHY183M camera. Images are binned $2\times2$ in software giving an image scale of $0.7\arcsec$ pixel$^{-1}$ resulting in a $32\arcmin\times21\arcmin$ field of view. A custom pipeline based on {\tt C-Munipack}\footnote{\url{http://c-munipack.sourceforge.net}} was used to calibrate the images and extract the differential photometry. We used circular photometric apertures of $7.1\arcsec$ that included all of the flux from the nearest known neighbor in the \textit{Gaia} DR3 catalog (\textit{Gaia} DR3 5407977460534995840), which is $3.4\arcsec$ northeast of TOI-2969 and 3 magnitudes fainter in \textit{TESS} band.

Two full transit windows were observed on 2021 December 19 and 2022 March 05 in $i'$ and $g'$ bands, respectively, from the Las Cumbres Observatory Global Telescope (LCOGT) 0.4\,m network nodes at Cerro Tololo Inter-American Observatory (CTIO), and South Africa Astronomical Observatory near Sutherland, South Africa (SAAO). We used circular photometric apertures of $3.9\arcsec$ and $3.5\arcsec$, respectively, that are $\sim 50\%$ contaminated with the $3.4\arcsec$ companion.

We observed a full transit window on UTC 2022 March 02 in the Johnson/Cousins R band from the Evans 0.36\,m telescope at El Sauce Observatory.  We used a circular photometric aperture of $4.3\arcsec$ that includes part of the flux from the $3.4\arcsec$ neighbor. We also used a circular photometric aperture of $1.6\arcsec$ that excludes most of the flux of the $3.4\arcsec$ neighbor, showing that the event occurs in TOI-2969. We used the larger aperture lightcurve in the global modeling since blending from the neighbor is only $\sim 3\%$, and because the smaller aperture lightcurve has much larger noise.

One full transit window was observed on 2022 November 24 in the B band from the Silesian University of Technology (SUTO) 0.3\,m telescope located in Pyskowice, Poland. The SUTO telescope is equipped with a $4656\times3520$ pixel Atik 11000M camera with an image scale of $0.712\arcsec$ pixel$^{-1}$, resulting in a $38\arcmin\times26\arcmin$ field of view. The differential photometric data were extracted using {\tt AstroImageJ}, using circular photometric apertures of $3.9\arcsec$.

We observed two more full transit windows on UTC 2023 February 25 and 2023 March 28 in $z'$ and B bands from the TRAnsiting Planets and PlanetesImals Small Telescope (TRAPPIST) South 0.6\,m telescope located at La Silla Observatory (Chile) \citep{jehin_trappist_2011, gillon_trappist_2011}. TRAPPIST-South is equipped with an FLI camera with an image scale of $0.6\arcsec$ pixel$^{-1}$, resulting in a $22\arcmin\times22\arcmin$ field of view. The image data were calibrated, and photometric data were extracted using a dedicated pipeline that uses the {\tt prose} framework described in \citet{garcia_prose_2022}. We used circular photometric apertures of $3.5\arcsec$ and $5.0\arcsec$ that included the flux from the $3.4\arcsec$ neighbor. 

An $\sim$on-time $\sim$23 ppt event was detected in all seven observations.

\subsubsection{TOI-2989}
We observed a full transit window of TOI-2989 b on UTC 2024 January 16 in the Johnson/Cousins R band from the Evans 0.36\,m telescope at El Sauce Observatory. We used circular photometric apertures of $5.4\arcsec$ that excluded all of the flux from the nearest known neighbor in the \textit{Gaia} DR3 catalog (\textit{Gaia} DR3 3531594171179942528), which is $\sim37\arcsec$ southeast of TOI-2989. An $\sim$on-time $\sim$28 ppt event was detected on-target. 

A partial and a full transit window were also observed from TRAPPIST-South on UTC 2022 April 25 and 2022 May 17 in NIR 700$\,$nm long pass band, and B bands using circular photometric apertures of $5.1\arcsec$ and $3.8\arcsec$, respectively. An $\sim$on-time $\sim$28 ppt event was detected on-target in both observations. 

\subsubsection{TOI-5300}
We observed a full transit window on UTC 2022 June 22 in the Cousins R band from Brierfield Observatory near Bowral, New S. Wales, Australia. The 0.36\,m telescope is equipped with a 4096 $\times$ 4096 Moravian 16803 camera. The image scale after binning 2 $\times$ 2 is $1.47\arcsec$ pixel$^{-1}$, resulting in a $50\arcmin\times50\arcmin$ field of view. The differential photometric data were extracted using {\tt AstroImageJ} and used a circular $8.8\arcsec$ photometric aperture that excluded all of the flux from the nearest known neighbor in the \textit{Gaia} DR3 catalog (\textit{Gaia} DR3 2642761924907362560), which is $\sim52\arcsec$ south of TOI-5300.

Two full transit windows were observed on UTC 2022 July 08 and 2020 October 18 in Sloan $i$ and $g$ bands, respectively, from the LCOGT 0.4\,m network node at Haleakala Observatory on Maui, Hawai'i (HAL). The photometric data were extracted using circular apertures of $6.6\arcsec$ for Sloan $i$ and $8.8\arcsec$ for Sloan $g$. Another full transit window was observed on UTC 2022 August 09 in the Sloan $g$ band from the LCOGT 0.4\,m network node at CTIO. The photometric data were extracted using circular $8.8\arcsec$ photometric apertures. 

We observed one full transit window on UTC 2022 August 26 in the Johnson/Cousins V band from TRAPPIST-South. The photometric data were extracted using circular $3.8\arcsec$ photometric apertures. A $\sim$23 ppt event was detected on-target in all observations.

\subsection{CORALIE spectroscopy}
\label{subsec:CORALIE}
Spectroscopic vetting and RV observations were performed with the CORALIE echelle spectrograph at the Swiss 1.2-meter Leonhard Euler telescope at La Silla Observatory (Chile) \citep{queloz_coralie_2001}. All observations were conducted out of transit, ensuring that the Rossiter-McLaughlin effect did not affect the data. The CORALIE data was accessed using \texttt{dace-query}, a Python package from the Data \& Analysis Center for Exoplanets\footnote{\url{https://dace.unige.ch/dashboard/}}. The RVs are derived by version 3.8 of the CORALIE Data Reduction System (DRS), which employs the CCF with numerical stellar templates closely matched to the spectral types of each TOI (in this case, K5). The DRS provides the RVs, the FWHM, the bisector span, and the contrast. Additionally, the pipeline provides data on the activity indices derived from the Na, Ca, and H$\mathrm{\alpha}$ lines. Table \ref{table:CORALIE} provides an overview of the CORALIE data we obtained for our analysis.

To exclude potential diluted binaries, the data was also reduced using other stellar masks with spectral types further from our stars (A0, F0, G5, and M2). In binary systems with components of different spectral types, applying different stellar masks can enhance the contribution of the companion to the CCF, potentially shifting the measured RV, as demonstrated in the case of HD 41004 \citep{santos_coralie_2002}. Our analysis did not reveal any significant mask effects. Furthermore, all RV observations were checked for strong correlations with stellar activity indicators. However, short-period stellar activity, such as starspots or flares, typically causes small RV variations \citep[root mean square (RMS) $\sim 2-10$ m/s][]{cretignier_measuring_2020}. The significant RV variations (RMS ranging from 100 - 400 m/s) observed in our three stars suggest an external influence, likely from a companion, rather than intrinsic stellar activity.

            \begin{table}
                \centering
                \caption{Characteristics of the CORALIE observations.}
                \label{table:CORALIE}
                \begin{tabular}{l c c c c c} \hline\hline
                TOI & $N_\mathrm{meas}$ & Span & $\langle t_\mathrm{exp}\rangle$ & $\mathrm{med}(\sigma_\mathrm{RV})$ & RMS(RV)\\
                & & [d] & [min] & [m s$^{-1}$] & [m s$^{-1}$] \\ \hline 
2969 & 24 &  114.9 & 45 & 34 & 179 \\ 
2989 & 26 &  435.9 & 40 & 133 & 379 \\ 
5300 & 33 &  423.0 & 45 & 100 & 118 \\ \hline
                \end{tabular}
                
            \end{table}

\section{Stellar properties}
\label{sec:stellar_properties}
Table \ref{table:stellar_parameters} provides an overview of the stellar properties. The spectral type is derived from the effective temperature, using Table 5 of \cite{pecaut_intrinsic_2013}. The magnitudes $H$, $K$, $V$, and $B$ originate from the TIC working group \citep{paegert_tess_2021, stassun_revised_2019}, as the stars are too faint to have been observed by \textit{Tycho} \citep{esa_hipparcos_1997}. The $\log R'_\mathrm{HK}$ values are not reported, as the flux in the $H$ and $K$ bands are insufficient in the CORALIE data. The right ascension $\alpha$, declination $\delta$, proper motions in both directions ($\mu_{\alpha^*}, \mu_\delta$), parallax $\pi$ and the derived distance $d$ originate from \textit{Gaia} DR3 \citep{gaia_collaboration_gaia_2023}. The effective temperature $T_\mathrm{eff}$, the microturbulence $v_\mathrm{tur}$, and the metallicity $\mathrm{[Fe/H]}$, are results of the spectral analysis, described in more detail in Section \ref{sec:Spectral_Analysis}. The extinction $A_\mathrm{V}$, the bolometric luminosity $L_\mathrm{bol}$, the stellar radius $R_{\star}$, and the stellar mass $M_{\star}$ follow from the spectral energy distribution (SED) analysis, described in Section \ref{sec:SED_Analysis}. The surface gravity, $\log(g)$, is computed using the stellar density derived in section \ref{sec:Orbital_Solutions} and the stellar radius.

The reported FWHM represents the average of the FWHM values derived from the CCF, obtained with the correlation of the CORALIE stellar spectra with a K5 mask. The standard deviation of these values is used as the error. A higher standard deviation in the FWHM may indicate greater stellar activity, as it reflects variations in spectral line widths. The rotational velocity $v \sin i$ is an approximation based on the FWHM adapted from \cite{santos_coralie_2002}. The rotational period is approximated from the public \textit{Gaia} DR3 photometric data (section \ref{sec:Prot_Gaia}), and/or the WASP transit survey (section \ref{sec:Prot_Wasp}). 

\begin{table*}
\caption{Stellar parameters of the stars presented in this paper.}
\label{table:stellar_parameters}

\centering
\begin{tabular}{l l c c c c c}
\hline\hline
&  & TOI-2969 & TOI-2989 & TOI-5300 & &\\ 
\multicolumn{2}{l}{Parameter} & \footnotesize{TIC 36452991} & \footnotesize{TIC 97825640} & \footnotesize{TIC 267215820} & Source & Sec. \\ \hline
SpType & & K3V & K4V & K4V & $T_\mathrm{eff}$ & \ref{sec:stellar_properties}\\
$B$ & [mag] & $14.01\pm 0.02$  & $15.23\pm 0.04$  & $15.02\pm 0.03$  & \textit{TESS}\\
$V$ & [mag] & $12.80\pm 0.09$  & $13.86\pm 0.06$  & $13.9\pm 0.3$  & \textit{TESS}\\
$J$ & [mag] & $10.96\pm 0.03$  & $11.59\pm 0.03$  & $11.33\pm 0.02$  & \textit{TESS}\\
$H$ & [mag] & $10.38\pm 0.03$  & $10.99\pm 0.02$  & $10.71\pm 0.02$  & \textit{TESS} \\
$K$ & [mag] & $10.25\pm 0.02$  & $10.86\pm 0.02$  & $10.57\pm 0.02$  & \textit{TESS}\\
$\alpha$ & [deg] & $150.063\pm 0.008$  & $171.232\pm 0.010$  & $356.48\pm 0.01$  & \textit{Gaia} DR3\\
$\delta$ & [deg] & $-47.445\pm 0.008$  & $-27.06\pm 0.01$  & $0.46\pm 0.01$  & \textit{Gaia} DR3\\
$\mu_\alpha^*$ & [mas/yr] & $12.01\pm 0.01$  & $-118.09\pm 0.01$  & $16.12\pm 0.02$  & \textit{Gaia} DR3\\
$\mu_\delta$ & [mas/yr] & $-0.54\pm 0.01$  & $-15.52\pm 0.01$  & $-44.07\pm 0.01$  & \textit{Gaia} DR3\\
$\pi$ & [mas] & $6.153\pm 0.010$  & $5.13\pm 0.02$  & $6.15\pm 0.02$  & \textit{Gaia} DR3\\
$d$ & [pc] & $162.5\pm 0.3$  & $194.9\pm 0.6$  & $162.7\pm 0.4$  & $\pi$ & \ref{sec:stellar_properties}\\
$T_{\mathrm{eff}}$ & [K] & $4738\pm 100$  & $4672\pm 170$  & $4610\pm 207$  & Spectra & \ref{sec:Spectral_Analysis}\\
$v_\mathrm{tur}$ & [km s$^{-1}$] & $0.8\pm 0.2$  & $1.4\pm 0.3$  & $1.5\pm 0.4$  & Spectra & \ref{sec:Spectral_Analysis}\\
$\mathrm{[Fe/H]}$ & [dex] & $0.08\pm 0.05$  & $-0.04\pm 0.07$  & $-0.17\pm 0.07$  & Spectra & \ref{sec:Spectral_Analysis}\\
$A_V$ & [mag] & $0.17\pm 0.02$  & $0.32\pm 0.05$  & $0.05\pm 0.03$  & SED & \ref{sec:SED_Analysis}\\
$L_\mathrm{bol}$ & [$L_{\odot}$] & $0.260\pm 0.006$  & $0.183\pm 0.004$  & $0.173\pm 0.004$  & SED & \ref{sec:SED_Analysis}\\
$R_{\star}$ & [$R_{\odot}$] & $0.70\pm 0.05$  & $0.76\pm 0.03$  & $0.65\pm 0.06$  & SED & \ref{sec:SED_Analysis}\\
$\log(g)$ & [cm s$^{-2}$] & $4.56 \pm 0.04$ & $4.58 \pm 0.04$ & $4.64 \pm 0.08$ & $\rho_\star$, $R_\star$ & \ref{sec:stellar_properties}\\
$M_{\star}$ & [$M_{\odot}$] & $0.71\pm 0.02$  & $0.77\pm 0.02$  & $0.67\pm 0.02$  & SED & \ref{sec:SED_Analysis}\\
FWHM & [km s$^{-1}$] & $8.12\pm 0.06$  & $8.0\pm 0.4$  & $8.0\pm 0.2$  & CCF & \ref{sec:stellar_properties}\\
$v\sin i$ & [km s$^{-1}$] & $< 2$ & $< 2$ & $< 2$ & FWHM & \ref{sec:stellar_properties} \\
$P_{\mathrm{rot}}$ & [d] & $26.8\pm 2.0$  & $29.2\pm 1.5$  &  & \textit{Gaia} & \ref{sec:Prot_Gaia}\\
$P_{\mathrm{rot}}$ & [d] &  & $30\pm 1$  & $31\pm 1$  & WASP & \ref{sec:Prot_Wasp}\\
\hline
\end{tabular}
\end{table*}

\subsection{Spectral analysis}
\label{sec:Spectral_Analysis}
The stellar spectroscopic parameters ($T_{\mathrm{eff}}$, $v_\mathrm{tur}$, [Fe/H]) were derived using the ARES+MOOG methodology, which is described in detail in \citet[][]{sousa_sweet-cat_2021, sousa_aresmoog_2014, santos_sweet-cat_2013}. To consistently measure the equivalent widths (EW), we used the ARES code\footnote{The last version, ARES v2, can be downloaded at \url{https://github.com/sousasag/ARES}} \citep{sousa_new_2007, sousa_ares_2015}. The spectral analysis was done using the combined spectrum obtained by shifting to the measured RV and taking the mean of the individual exposures for each star. In this analysis, we used the list of lines presented in \citet[][]{tsantaki_deriving_2013}, which is suitable for stars with $T_\mathrm{eff} < 5200\,\mathrm{K}$. The best set of spectroscopic parameters for each spectrum was found by using a minimization process to find the ionization and excitation equilibrium. This process uses a grid of Kurucz model atmospheres \citep{kurucz_synthe_1993} and the latest version of the radiative transfer code MOOG \citep{sneden_carbon_1973}. 

\subsection{Effective temperatures}
\begin{table}
    \centering
    \caption{The effective temperatures from different sources.}
    \begin{tabular}{l l c c c c}
    \hline \hline
             & & TOI-2969 & TOI-2989 & TOI-5300\\ \hline
         TIC v8 & [K] & $4581 \pm 128$ & $4262 \pm 123$ & $4219 \pm 126$ \\
         \textit{Gaia} DR3 & [K] & $4539 \pm 42$ & $4497\pm3$ & $4462\pm3$ \\
         Spectra & [K] & $4738\pm100$ & $4672\pm170$ & $4610\pm207$ \\ \hline
    \end{tabular}
    \begin{flushleft}
        \textbf{Notes:} TIC v8 is the \textit{TESS} Input Catalog, and \textit{Gaia} DR3 temperatures follow from the GSP-Phot library.
    \end{flushleft}
    \label{tab:Teff_differences}
\end{table}

Given the significant discrepancy ($200-400$ K) between the effective temperatures obtained from our spectral analysis and those in the \textit{TESS} Input Catalog \citep[TICv8][]{stassun_revised_2019}, caution is advised when relying on TICv8 values for effective temperature estimates. Our analysis indicates that all the stars in our sample exceed the selection criterion of $T_\mathrm{eff} < 4600$ K and are generally more consistent with effective temperatures from the General Stellar Parametrizer from Photometry (GSP-Phot) library of \textit{Gaia} DR3, except for TOI-2969 (see Table \ref{tab:Teff_differences}). We adopt the spectroscopic values, which are directly derived from our observations, ensuring consistency.

\subsection{SED analysis}
\label{sec:SED_Analysis}
As an independent determination of the basic stellar parameters, we performed an analysis of the broadband SEDs of the stars together with the \textit{Gaia} DR3 parallaxes \citep[with no systematic offset applied; see, e.g.,][]{stassun_parallax_2021}. This analysis was conducted to determine an empirical measurement of the stellar radii, following the procedures described in \citet{stassun_eclipsing_2016,stassun_accurate_2017,stassun_empirical_2018}.
We obtained the $JHK_S$ magnitudes from 2MASS, the W1--W3 magnitudes from WISE, the $G_{\rm BP}$ and $G_{\rm RP}$ magnitudes, as well as the absolute flux-calibrated spectrophotometry, from \textit{Gaia}. Together, the available photometry spans the full stellar SED over the wavelength range 0.4-10 $\upmu$m (see Figure~\ref{fig:sed}). 
 
We performed a fit using PHOENIX stellar atmosphere models \citep{husser_new_2013}, with $T_{\rm eff}$, $\log g$, and [Fe/H] set to the earlier determined values. The extinction $A_\mathrm{V}$ was limited to the maximum line-of-sight value from the Galactic dust maps of \citet{schlegel_maps_1998}. Integrating the (unreddened) model SED gives the bolometric flux at Earth, $F_{\rm bol}$. Taking the $F_{\rm bol}$ and \textit{Gaia} parallax directly gives the bolometric luminosity, $L_{\rm bol}$. The Stefan-Boltzmann relation then yields the stellar radius, $R_\star$. Finally, we can estimate the stellar mass, $M_\star$, from appropriate empirical relations depending on the stellar mass \citep[i.e.,][]{torres_accurate_2010}. 

\subsection{Rotational period}
\label{sec:Prot_Gaia_Wasp}
\subsubsection{\textit{Gaia}}
\label{sec:Prot_Gaia}
\begin{figure}[ht!]
    \includegraphics[width=\linewidth]{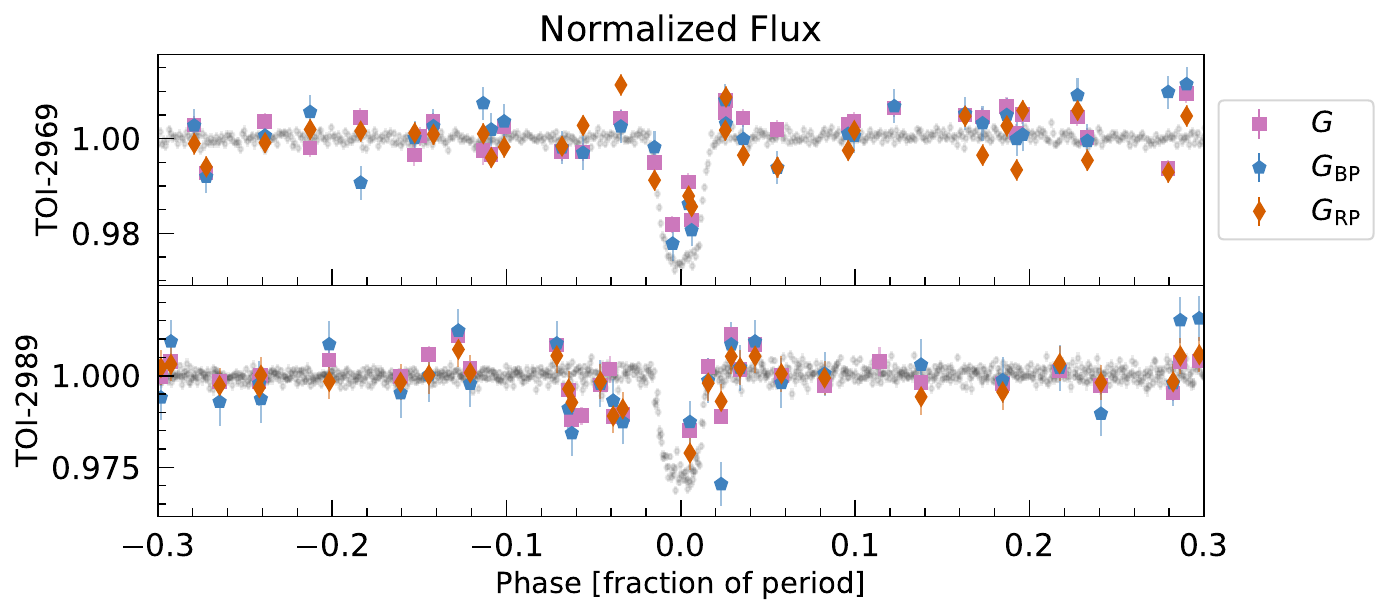}
    \caption{The phase-folded \textit{Gaia} photometric data. Coloured points indicate the \textit{Gaia} different wave bands. The black points are the \textit{TESS} binned data, with a binning of 1/1000 of the period. The errors of the \textit{Gaia} data have been corrected as suggested in \citep{evans_gaia_2023} by adding an error in quadrature as a function of the magnitude. The long-period signals above 0.01\% FAP have been subtracted and the errors are scaled by the ratio of the standard deviation before and after signal subtraction.}
    \label{fig:GaiaPhasefolded}
\end{figure}
TOI-2969 and TOI-2989 are flagged as variable in \textit{Gaia} DR3. Variability in stars can arise from multiple origins. The classification of the variability processing \citep{eyer_gaia_2023} can be found in the variability summary database \verb|gaiadr3.vari_summary|. For both stars, the variability flag arises from solar-like variability, which can originate from flares, stellar spots, and/or chromospheric variability. When \textit{Gaia} flags a star as variable, the corresponding epoch photometry becomes publicly available. Thus, we could access the \textit{Gaia} photometric data for these two stars. 

Upon performing a Lomb-Scargle periodogram on the photometric data (combining the three available magnitudes $G$, $G_\mathrm{BP}$, and $G_\mathrm{RP}$), we identified potential rotational periods (see Appendix \ref{app:GaiaPhotometry}). TOI-2969 has a potential rotational period of $26.8 \pm 2.0$ days, visible in $G_\mathrm{BP}$ and $G_\mathrm{RP}$. In the $G$ band, a period of $16 \pm 10$ days is visible, and in addition, a similar period of $19 \pm 10$ days also appears in all three magnitudes when using a window function. Since both potential rotational periods fall within the range identified by the window function, the rotational period remains uncertain, but the 26.8-day signal is more probable. TOI-2989 has a rotational period of $29.2 \pm 1.5$ days visible in all three magnitudes. This is not due to data sampling, as it does not appear in the window function.

\textit{Gaia} photometry errors are underestimated due to uncalibrated systemic errors \citep{evans_gaia_2023}. When adding 1\% to the noise, the Lomb-Scargle periodogram still identifies the rotational periods as significant above 1\% False Alarm Probability (FAP). These rotational periods are not detected in the \textit{TESS} data (from which the short-period companion transits are excluded) when using a Box-fitting Least Squares (BLS) \citep{kovacs_box-fitting_2002}, nor when computing the Lomb-Scargle periodogram. This is due to the absence of consecutive \textit{TESS} sectors, which limits the data to a 27-day span. While BLS is not expected to detect stellar rotation periods from spot-modulated light curves due to its focus on identifying low-duty-cycle transit-like features, the absence of a signal in the BLS search suggests that the modulation is unlikely to be caused by a transit. This is more challenging to confirm with the lower cadence \textit{Gaia} data.

After removing the rotational period variation, the transit periods do not appear when a BLS is performed on the resulting \textit{Gaia} data. When phase-folding the \textit{Gaia} photometric data to the period and epoch from the orbital solution presented in Section \ref{sec:Orbital_Solutions}, TOI-2969 and TOI-2989 show a change in flux at the time of transit. This shows that, in principle, \textit{Gaia} has detected the transit. Although the data exhibit significant scatter, see Figure \ref{fig:GaiaPhasefolded}.

\subsubsection{WASP}
\label{sec:Prot_Wasp}
We obtained data from the WASP transit survey \citep{pollacco_wasp_2006} to look for rotational modulations of the host stars.  WASP data is gathered using Canon 200-mm, f/1.8 lenses backed by 2048$\times$2048 CCDs, observing with a 400--700 nm passband, and producing photometry from extraction apertures with a radius of $48\arcsec$ centered on each star \citep{pollacco_wasp_2006}. We searched the accumulated lightcurves for periodicities in the range of 1 to 130 days using methods discussed in \citet{maxted_wasp-41b_2011}; see Section \ref{app:wasp_period} for the WASP lightcurves periodograms.

For TOI-2969, WASP-South recorded 12\,840 data points between 2006 and 2012. No significant rotational modulation was detected, given that the amplitude of the \textit{Gaia} modulation is three times less for TOI-2969 ($\sim5$ mmag) than for TOI-2989 ($\sim15$ mmag), WASP likely does not have the sensitivity for the rotation seen by \textit{Gaia}. Plus, the WASP extraction aperture contains multiple stars of similar brightness. Therefore, any detected modulation could not be attributed to a specific star. However, while TOI-2969\,b was never a WASP candidate, knowing the TESS ephemeris, we find that the standard WASP transit-search algorithm \citep{collier_cameron_efficient_2007} detects a tentative transit, giving an ephemeris of,

\begin{multline*} {\rm Transit [TDB(JD)]} = 245\,5830.9392 \pm\ 0.0035 + \\ N \times 1.823808 \pm\ 0.000052.
\end{multline*}
This likely represents the earliest recorded data of the TOI-2969 b transit. However, given that the WASP precision is insufficient to improve the ephemeris and may introduce noise without significantly enhancing the results, the data were not included in the joint fit.

TOI-2989 was observed by WASP-South over the span of $\sim$\,165 nights in 2009 and 2010, obtaining 15\,000 data points.  The 2010 data show a significant modulation at a period of 15.0 $\pm$ 0.5 d with an amplitude of 5 mmag and a false-alarm likelihood below 2\%. The 2009 data show a significant periodicity compatible with twice this period (30 $\pm$ 3 d), and an amplitude of 7 mmag.  To check whether this might be caused by moonlight, we made a similar analysis of several nearby stars in the field of view, but did not find the modulation. We are likely detecting a rotational modulation of TOI-2989 at a period of 30 $\pm$ 1 d, with its first harmonic present in the data from 2010. This is consistent with the rotational period observed in the \textit{Gaia} variability data.

TOI-5300 was observed by WASP-South over the span of $\sim$\,140 nights in 2008 and 2009, producing 9500 data points. The 2009 data show a significant modulation at a period of 15.6 $\pm$ 0.6 d, with an amplitude of 6 mmag and a false-alarm likelihood below 1\%. The 2009 data show a marginal detection (10\%\ false-alarm likelihood) of a modulation compatible with twice this period (32 $\pm$ 3 d). We also checked for moonlight interference by performing a similar analysis of nearby field stars, but these don't show the modulation. We are likely detecting a rotational modulation at a period of 31.2 $\pm$ 1.2 d, with its first harmonic present in the data from 2010. Given the 6 mmag amplitude, the signal should be visible in the \textit{Gaia} G-band photometry, which has a median uncertainty of 0.2 mmag for a $G$ magnitude of $\sim$13 \citep{riello_gaia_2021}.

\section{Orbital solutions}
\label{sec:Orbital_Solutions}
\begin{table*}[ht!]
\caption{Fitted and derived parameters for the companions presented in this paper.}
\label{table:orbital_solution}
\centering
\begin{tabular}{l l l c c c}
\hline\hline
\multicolumn{2}{l}{Parameter} &  & TOI-2969 b & TOI-2989 b & TOI-5300 b \\ \hline
Fitted parameters & \\
\footnotesize{Orbital period} & $P$ & [days]  &\begin{tabular}{@{}c@{}}$1.8237146$\\ $\pm 0.0000002$\end{tabular} &\begin{tabular}{@{}c@{}}$3.122832$\\ $\pm 0.000001$\end{tabular} &\begin{tabular}{@{}c@{}}$2.262196$\\ $\pm 0.000003$\end{tabular}\\
\footnotesize{Time of transit} & $T_0$ & [rBJD]$^{(a)}$  &\begin{tabular}{@{}c@{}}$9303.30015$\\ $_{-0.00008}^{+0.00009}$\end{tabular} &\begin{tabular}{@{}c@{}}$9302.2242$\\ $\pm 0.0002$\end{tabular} &\begin{tabular}{@{}c@{}}$9470.4174$\\ $\pm 0.0002$\end{tabular}\\
\footnotesize{Radius ratio} & $R_\mathrm{pl}/R_{\star}$ &  &\begin{tabular}{@{}c@{}}$0.1613$\\ $\pm 0.0008$\end{tabular} &\begin{tabular}{@{}c@{}}$0.152$\\ $_{-0.001}^{+0.002}$\end{tabular} &\begin{tabular}{@{}c@{}}$0.1389$\\ $_{-0.0010}^{+0.0011}$\end{tabular}\\
\footnotesize{Impact parameter} & $b$ &  & $0.713\pm 0.010$  & $0.36_{-0.07}^{+0.05}$  & $0.13_{-0.09}^{+0.11}$ \\
\footnotesize{Stellar density} & $\rho_{\star}$ & [$\rho_\odot$]  & $1.93\pm 0.07$  & $1.8\pm 0.1$  & $2.13_{-0.12}^{+0.07}$ \\
\footnotesize{Eccentricity (fixed)} & $e$ & & 0& 0& 0\\
\footnotesize{RV semi-amplitude} & $K$ & [m s$^{-1}$]  & $243\pm 8$  & $503_{-37}^{+35}$  & $121\pm 22$ \\ \hline
Derived parameters & \\
\footnotesize{Planetary radius} & $R_\mathrm{pl}$ & [$R_\mathrm{Jup}$]  & $1.10\pm 0.08$  & $1.12\pm 0.05$  & $0.88\pm 0.08$ \\
\footnotesize{Planetary mass} & $M_\mathrm{pl}$ & [$M_\mathrm{Jup}$]  & $1.16\pm 0.04$  & $3.0\pm 0.2$  & $0.6\pm 0.1$  \\
\footnotesize{Planetary bulk density} & $\rho_\mathrm{pl}$ & [g cm$^{-3}$]  & $1.1_{-0.2}^{+0.3}$  & $2.7\pm 0.4$  & $1.1_{-0.3}^{+0.4}$  \\
\footnotesize{Inclination} & $i$ & [$^{\circ}$]  & $84.9\pm 0.4$  & $88.1_{-0.3}^{+0.4}$  & $89.2_{-0.6}^{+0.5}$  \\
\footnotesize{Semi-major axis} & $a_\mathrm{pl}$ & [mAU]  & $26.1 \pm 0.2$ & $38.4 \pm 0.3$ & $29.5 \pm 0.3$ \\
\footnotesize{Transit duration} & $T_{14}$ & [hours]  & $1.846_{-0.011}^{+0.010}$  & $2.68\pm 0.02$  & $2.36\pm 0.02$ \\
\footnotesize{Equilibrium temperature}$^{(b)}$ & $T_\mathrm{eq}$ & [K]  & $1186\pm 52$  & $1001_{-42}^{+43}$  & $1043_{-66}^{+67}$  \\ 
\footnotesize{Insolation} & $S_\mathrm{pl}$ & [$S_{\oplus}$]  & $382_{-11}^{+12}$  & $124\pm 4$  & $198\pm 6$  \\
\hline
CORALIE parameters & \\
Systemic RV & $\gamma_\mathrm{CORALIE}$ & [km s$^{-1}$]  & $52.043\pm 0.005$  & $-9.68\pm 0.02$  & $-66.61\pm 0.02$  \\
Jitter & $\sigma_\mathrm{CORALIE}$ & [m s$^{-1}$]  & $8_{-3}^{+5}$  & $3_{-3}^{+16}$  & $2_{-2}^{+10}$  \\ 
Residual noise & $\mathrm{RMS}\left(\mbox{O-C}\right)$ & [m s$^{-1}$]  & 23 & 112 & 68 \\ \hline
\end{tabular}
\begin{flushleft}
\textbf{Notes:} The limb darkening and photometric instrumental parameters can be found in Appendix \ref{app:limb_instrument_table}. The values assumed for the solar and planetary constants are the IAU 2015 Resolution B 3 values from \citep{prsa_nominal_2016}.\\
$^{(a)}$ The reduced Barycentric Julian Date in Barycentric Dynamical Time ($\mathrm{rBJD_{TDB}}$), obtained by subtracting 2 450 000 from the $\mathrm{BJD_{TDB}}$.\\
$^{(b)}$ Assuming a Bond albedo $A=0$.\\
\end{flushleft}
\end{table*}
To derive the orbital solution, a joint-fit analysis, combining RV and photometric data, was performed using the Python software \texttt{Juliet} \citep{espinoza_juliet_2019}. The Dynamic Nested Sampling package \texttt{dynesty} \citep{speagle_dynesty_2020} is used as a sampler for estimating Bayesian posteriors and evidence. Given that the RV variations align well with the photometric data, no Gaussian Process (GP) model was added to account for stellar activity. For CORALIE, accidental off-target observations are excluded from the analysis (e.g., no star in fiber, verifiable via the integrated guiding frame images).

Table \ref{table:priors} presents the priors used for the joint modeling. The estimated values for the orbital period $P$, time of transit $T_0$, and stellar density $\rho_{\star}$, are obtained from the ExoFOP website\footref{foot:EXOFOP}. The \textit{TESS} limb darkening $q_1$ and $q_2$ parameters are calculated by the quadratic law, as described by \cite{kipping_efficient_2013}, via \texttt{LDCU}\footnote{\url{https://github.com/delinea/LDCU}}, a modified version of the Python code \texttt{limb-darkening} \citep{espinoza_limb_2015}. The eccentricity $e$ and the argument of periastron $\omega$ are treated in two ways: they are either both fixed ($e=0$ and $\omega=90^{\circ}$), or $e$ follows a beta prior with parameters as defined by \cite{kipping_parametrizing_2013}, while $\omega$ is assigned a uniform prior ranging from 0 to 180 degrees. The log-evidence is used to compare the models and determine whether having 
$e$ and $\omega$ free or fixed fits the data best. The radius ratio $R_\mathrm{pl}/R_\star$ and impact parameter $b$ are assigned a uniform prior ranging from 0 to 1.
Based on the absence of V-shaped transit features, the transits are not grazing; thus, $b$ larger than 1 is not considered. All other priors use values as suggested by the \texttt{Juliet} documentation. The resulting posterior distributions are given in the form of \texttt{corner} plots in Appendix \ref{app:Corner} \citep{foreman-mackey_cornerpy_2016}.

To evaluate the presence of contaminating sources with a threshold of six in magnitude difference, we utilized \texttt{tpfplotter} to display the average image of the target pixel files generated by the \textit{TESS}-SPOC \citep{aller_planetary_2020}. The sectors where our three TOIs were observed and checked for contamination are outlined in Table \ref{table:TESS-SPOC}. For TOI-2989 and TOI-5300, across all sectors, the apertures used by \textit{TESS} for light curve extraction are uncontaminated by neighboring stars. However, TOI-2969 is affected by contamination, we expect maximally between $\sim 30 \mbox{ to } 35\%$ contamination from nearby stars. A dilution factor is included as a uniform prior from 0 to 1 for the ground-based photometry and the QLP light curves of TOI-2969. The QLP and \textit{TESS}-SPOC data are treated as separate instruments with their own instrumental parameters in the joint fit, except for the limb-darkening, which is shared due to the identical wavelength range and thus must have the same value. Since the PDCSAP light curves account for contamination, a dilution factor is not included for \textit{TESS}-SPOC data. All ground-based follow-up photometry is detrended for airmass using a linear regression model.

Table \ref{table:orbital_solution} presents the orbital solutions derived using \texttt{Juliet}. The fitted- and instrumental parameters are taken from the model, with the errors corresponding to the $1\sigma$ Monte Carlo uncertainties, while the derived parameters were subsequently computed. Specifically, the planetary radius $R_\mathrm{pl}$ is obtained from the stellar radius and the radius ratio; the planetary mass $M_\mathrm{pl}$ is determined using the RV equation (for transiting planets, $M_\mathrm{pl}\sin i \approx M_\mathrm{pl}$, assuming $i\approx 90^{\circ}$); and the bulk planetary density $\rho_\mathrm{pl}$ is calculated from $R_\mathrm{pl}$ and $M_\mathrm{pl}$. The semi-major axis $a_\mathrm{pl}$ is derived from Kepler's third law. The inclination $i$ and the transit duration $T_{14}$ are computed following the method detailed in \cite{seager_unique_2003} and \cite{kipping_characterizing_2014}, respectively. The planet equilibrium temperature $T_\mathrm{eq}$ is determined from $T_\mathrm{eff}$, $R_\star$, and $a_\mathrm{pl}$, assuming a Bond albedo $A=0$. The insolation flux $S_\mathrm{pl}$ is calculated from $L$ and $a_\mathrm{pl}$. The limb darkening coefficients and photometric instrumental parameters are listed in Appendix \ref{app:limb_instrument_table}.

\begin{figure}[H]
	\centering
	\includegraphics[width=.9\linewidth]{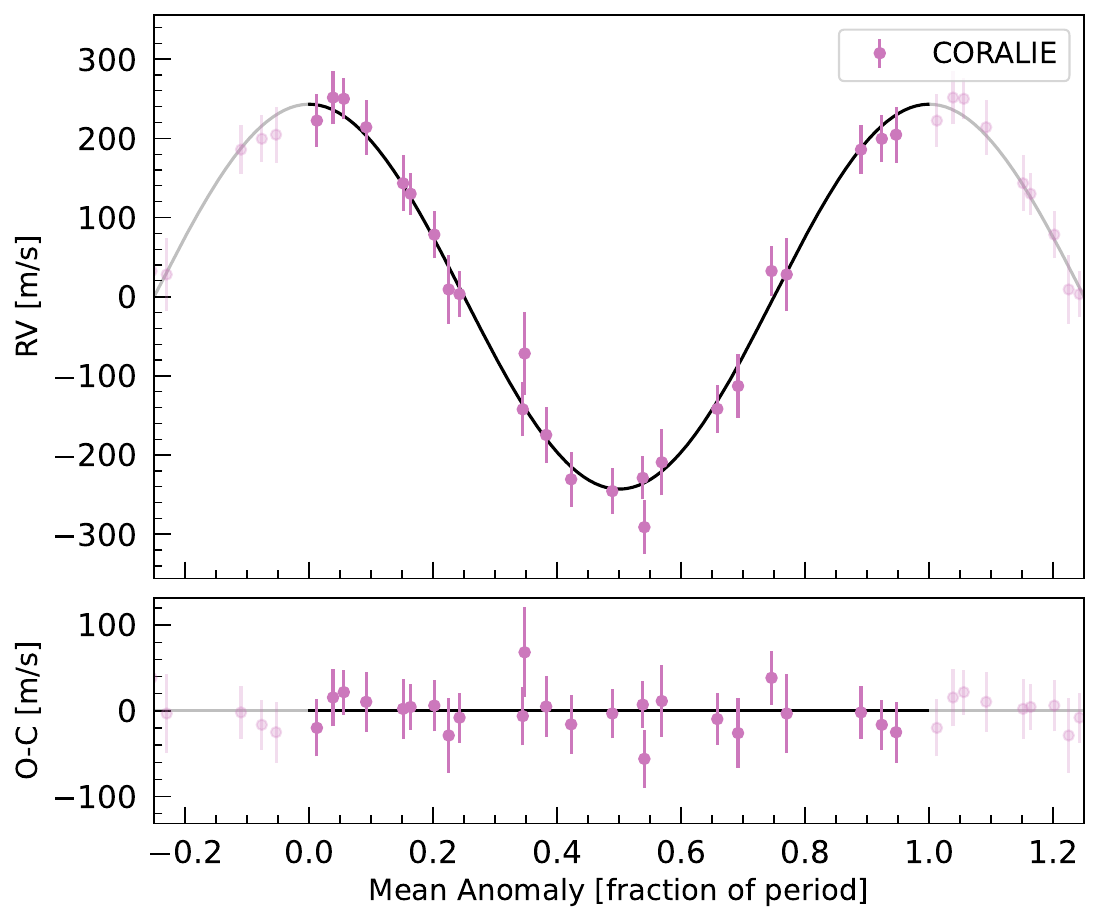}
	\caption{Overlay of TOI-2969's orbital solution with CORALIE RV observations, accompanied by residuals in the lower panel (with an RMS of 23 m/s).}
	\label{fig:TOI2969_CORALIE}
\end{figure}
\vfill
\subsection{TOI-2969 b}
\label{sec:TOI2969}
TOI-2969 b is a Hot Jupiter orbiting its K3V host with an orbital period of just 1.82 days. At its proximity of 0.0261 AU, the equilibrium temperature is 1186 K, and it receives an insolation of 382 $\Searth$. The planet is more massive than Jupiter, with a mass of 1.16 $\mjup$ and a radius of 1.10 $\rjup$. 

TOI-2969 has a stellar companion separated by $3.34\arcsec$, as observed by HRCam with SOAR \citep{tokovinin_ten_2018}. \textit{Gaia} DR3 \citep{gaia_collaboration_gaia_2023} also shows two sources within $3.4\arcsec$ (\textit{Gaia} DR3 5407977460534995840 and \textit{Gaia} DR3 5407977460540294784, the latter being TOI-2969); however, their parallaxes differ significantly (0.106 mas and 6.153 mas respectively), indicating that these stars do not belong to the same system. RV contamination from the nearby star is negligible, as it is 3.1 mag fainter in $G$ mag, located outside the $2\arcsec$ CORALIE fiber, and resolved given that we do not observe under seeing conditions worse than $1.8\arcsec$. All fitted photometry contamination values are smaller than our maximally expected $30 - 35\,\%$ (see Fig. \ref{fig:TOI2969_TESS}). With respect to the ground-based photometry QLP has the highest contamination ($21\,\%$), which can be expected as \textit{TESS} uses the largest photometric window. As the transit is confirmed on target (see section \ref{sec:SG1-TOI2969}), the RV variation seen for TOI-2969 can be concluded as caused by a giant companion with a planetary mass of 1.16 $\mjup$. The \texttt{Juliet} fit does not include an eccentricity or argument of periastron, as they do not improve the model. The resulting model versus the RV and photometric data are visible in Figures \ref{fig:TOI2969_CORALIE} and \ref{fig:TOI2969_TESS}. No significant signs of additional planets were detected; the CORALIE jitter aligns with zero.

\subsection{TOI-2989 b}
\label{sec:TOI2989}
TOI-2989 is a high proper motion star, which may indicate a different evolutionary pathway or membership in a Galactic kinematic population. A quick approximation combining the proper motion and parallax shows that its tangential velocity is $\sim110$ km/s. When combined with the systemic velocity measured by CORALIE, the total velocity is also around 110 km/s. This would potentially place the star in the thick disk population \citep{nissen_thin_2003}. It hosts a Hot Jupiter with an orbital period of 3.12 days and is situated at $a_\mathrm{pl} = $ 0.0384 AU, resulting in $T_\mathrm{eq} = 1001$ K and receiving insolation 124 times that of Earth. The planet's radius of 1.12 $\rjup$ and a mass of 3.0 $\mjup$ suggest it has a massive gaseous envelope.

\begin{figure}[ht!]
	\centering
	\includegraphics[width=.9\linewidth]{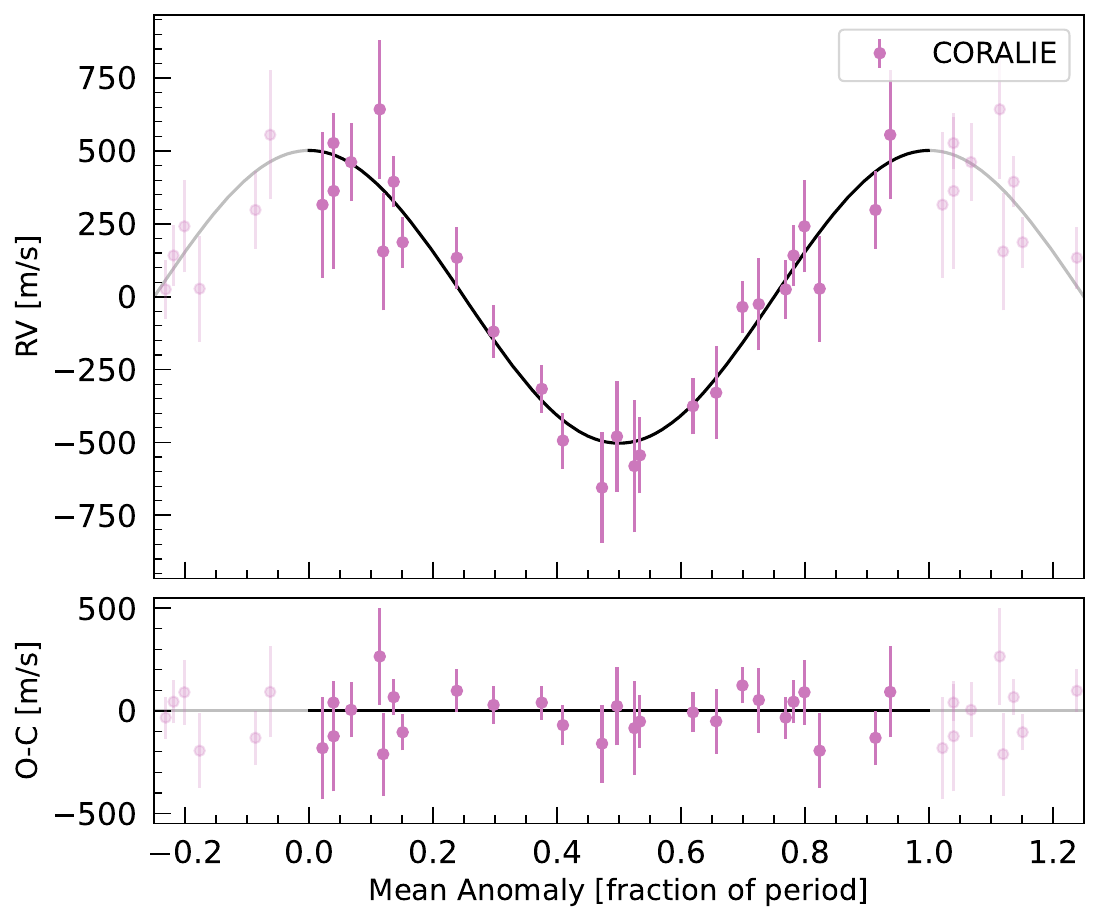}
	\caption{TOI-2989's orbital solution alongside its CORALIE RV data. Residuals, exhibiting an RMS of 112 m/s, are visualized in the lower panel.}
	\label{fig:TOI2989_CORALIE}
\end{figure}

The \texttt{Juliet} fit did not improve with the addition of eccentricity and argument of periastron and was thus not included. The resulting models, along with the RV residual with an RMS value of 112 m/s, can be found in Figure \ref{fig:TOI2989_CORALIE} and the photometric observations in Figure \ref{fig:TOI2989_TESS}. Note that a fluctuation is visible at the ingress in the El Sauce data. The \textit{TESS} light curves were checked for similar depth fluctuations, and the BLS of sector 9 found an 11.8-day period, caused by a dip at the edge of the sector. No corresponding signal was found in sector 36. Given that the El Sauce data are ground-based and no matching signal appears in the \textit{TESS} light curves, the fluctuation is most likely due to instrumental systematics or external factors such as weather variability. The data show no clear indications of further planets; the CORALIE jitter is consistent with zero.

\subsection{TOI-5300 b}
\label{sec:TOI5300}
\begin{figure}[ht!]
    \centering
    \includegraphics[width=.9\linewidth]{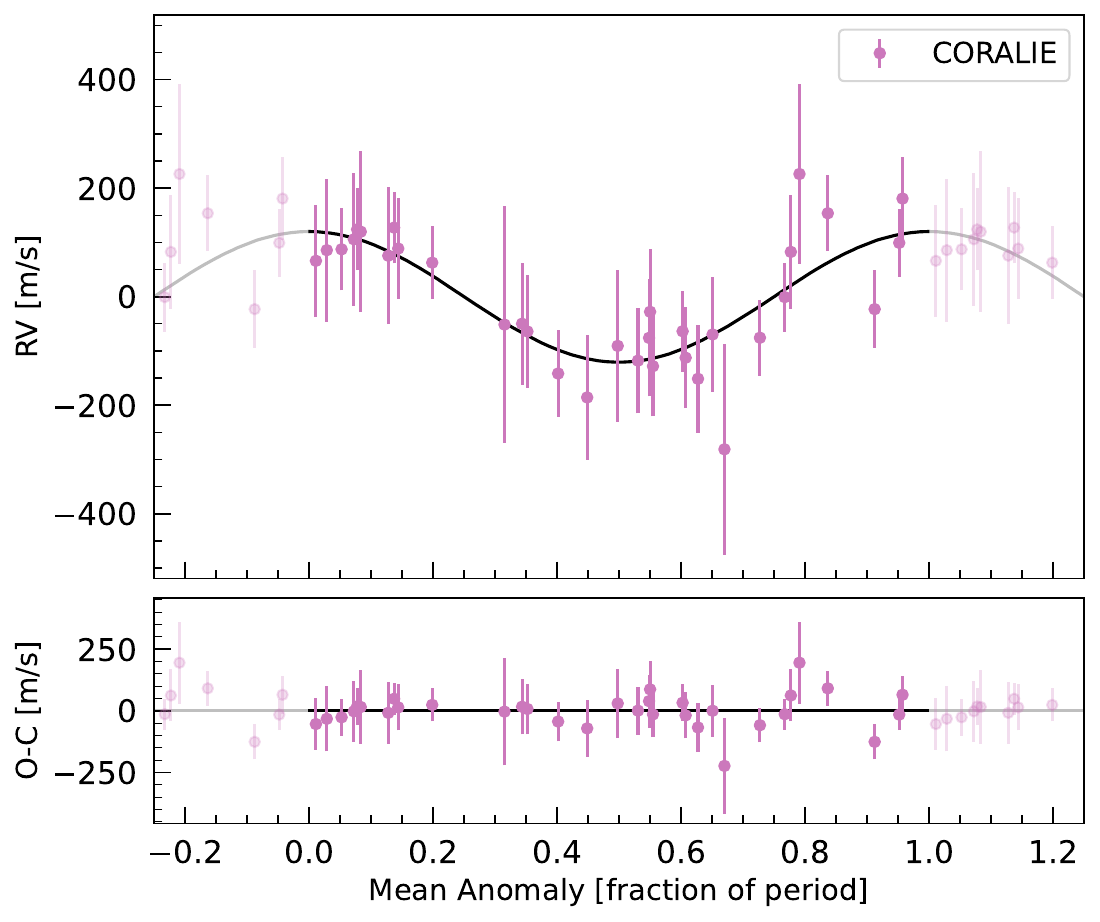}
    \caption{TOI-5300's orbital solution superimposed on CORALIE RV data, with the residuals plotted in the lower panel (with an RMS of 68 m/s).}
    \label{fig:TOI5300_CORALIE}
\end{figure}
TOI-5300 b orbits its K4V host in $2.26$ days. The planet's radius $R_\mathrm{pl} = 0.88 \rjup$ and mass $M_\mathrm{pl} = 0.6 \mjup$ are smaller than those of Jupiter, giving it a bulk density of $1.1$ g cm$^{-3}$. At its proximity of 0.0236 AU, it receives an insolation of 198 $S_\oplus$, resulting in $T_\mathrm{eq} = 1043$ K. The Hot Jupiter TOI-5300 appears to be the only planet in its system, as there is no robust evidence for additional planets; the CORALIE jitter is consistent with zero. Figure \ref{fig:TOI5300_CORALIE} and \ref{fig:TOI5300_TESS} show the resulting orbital solution. As with the other Hot Jupiters presented in this paper, the \texttt{Juliet} fit did not improve when including eccentricity and argument of periastron as free parameters, so these were fixed at $e = 0$ and $\omega = 90^{\circ}$.

\begin{figure}[ht!]
	\centering
	\includegraphics[width=.9\linewidth]{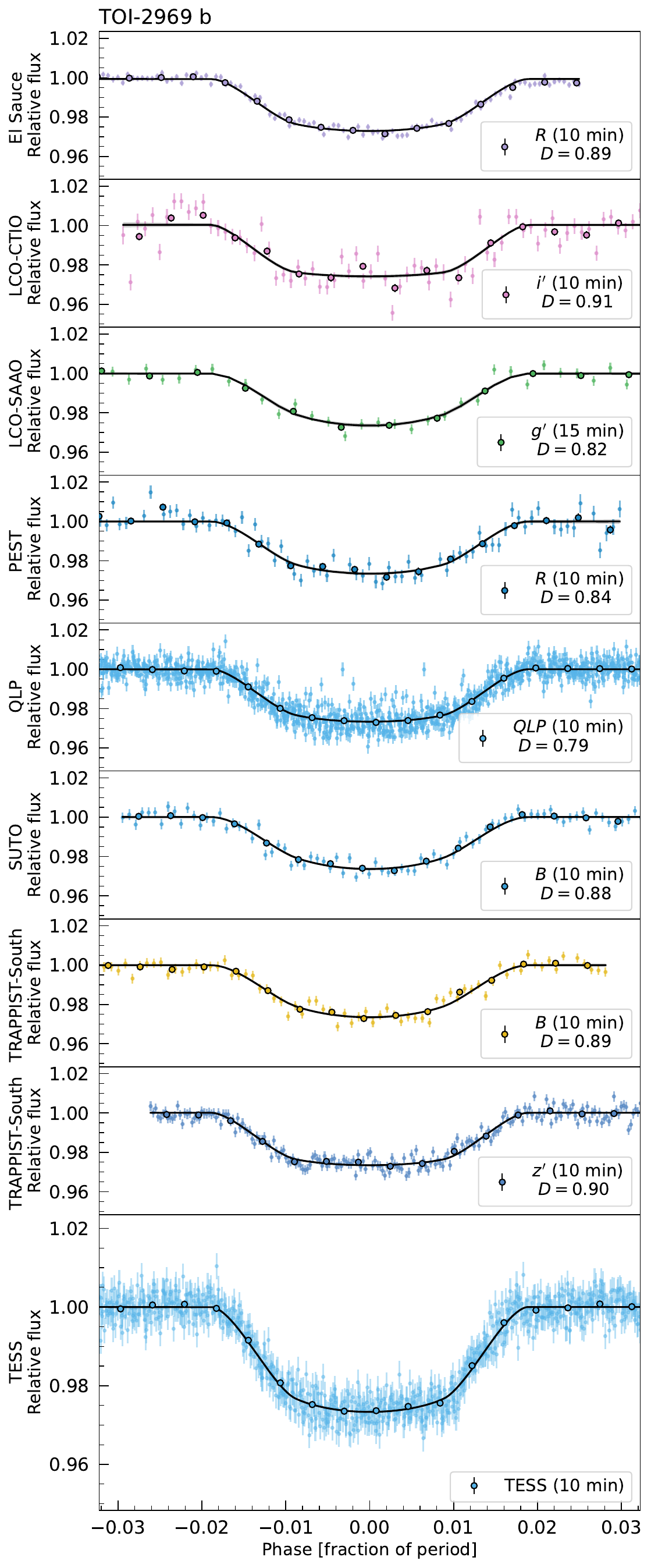}
	\caption{The phase-folded \textit{TESS} light curve of TOI-2969. The \texttt{Juliet} fit is shown as a black line, while \textit{TESS} data is displayed in light blue in the bottom panel. The upper panels feature ground-based follow-up photometric observations from LCO-SAAO, LCO-CTIO, TRAPPIST-South, El Sauce, SUTO, and PEST. Markers with black edges denote 10-minute binned data points, except for LCO-SAAO ($g'$) observations, where the data is binned by 15 minutes. If the dilution $D$ is fitted, it's indicated per light curve, otherwise it's set to 1.}
	\label{fig:TOI2969_TESS}
\end{figure}
\begin{figure}[ht!]
	\centering
	\includegraphics[width=.9\linewidth]{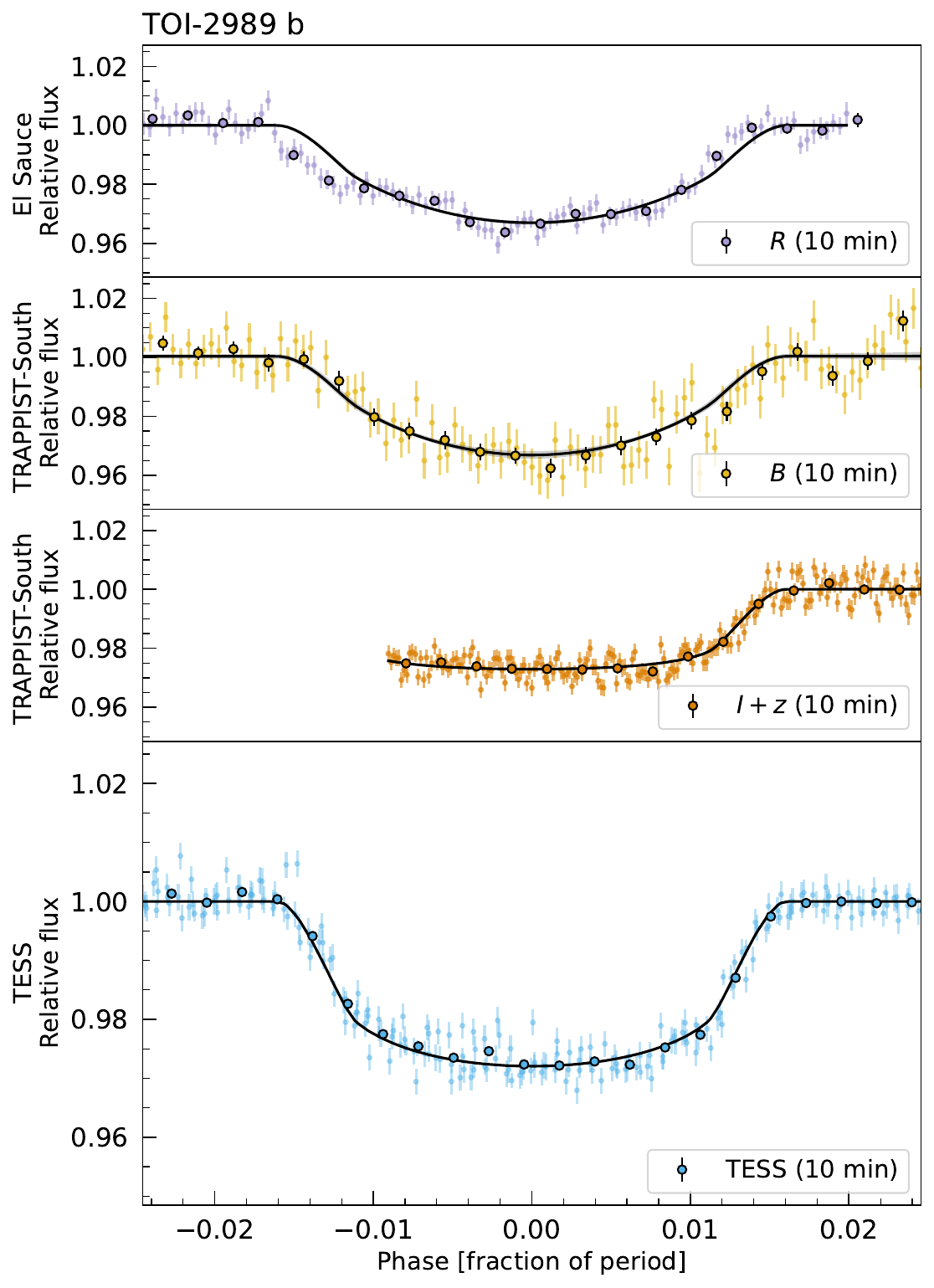}
	\caption{The phase-folded \textit{TESS} light curve of TOI-2989. The \texttt{Juliet} fit is shown as a black line, while \textit{TESS} data is displayed in light blue in the bottom panel. The upper panels feature ground-based follow-up photometric observations from El Sauce and TRAPPIST-South. Markers with black edges denote 10-minute binned data points.}
	\label{fig:TOI2989_TESS}
\end{figure}
\begin{figure}[ht!]
    \centering
    \includegraphics[width=.9\linewidth]{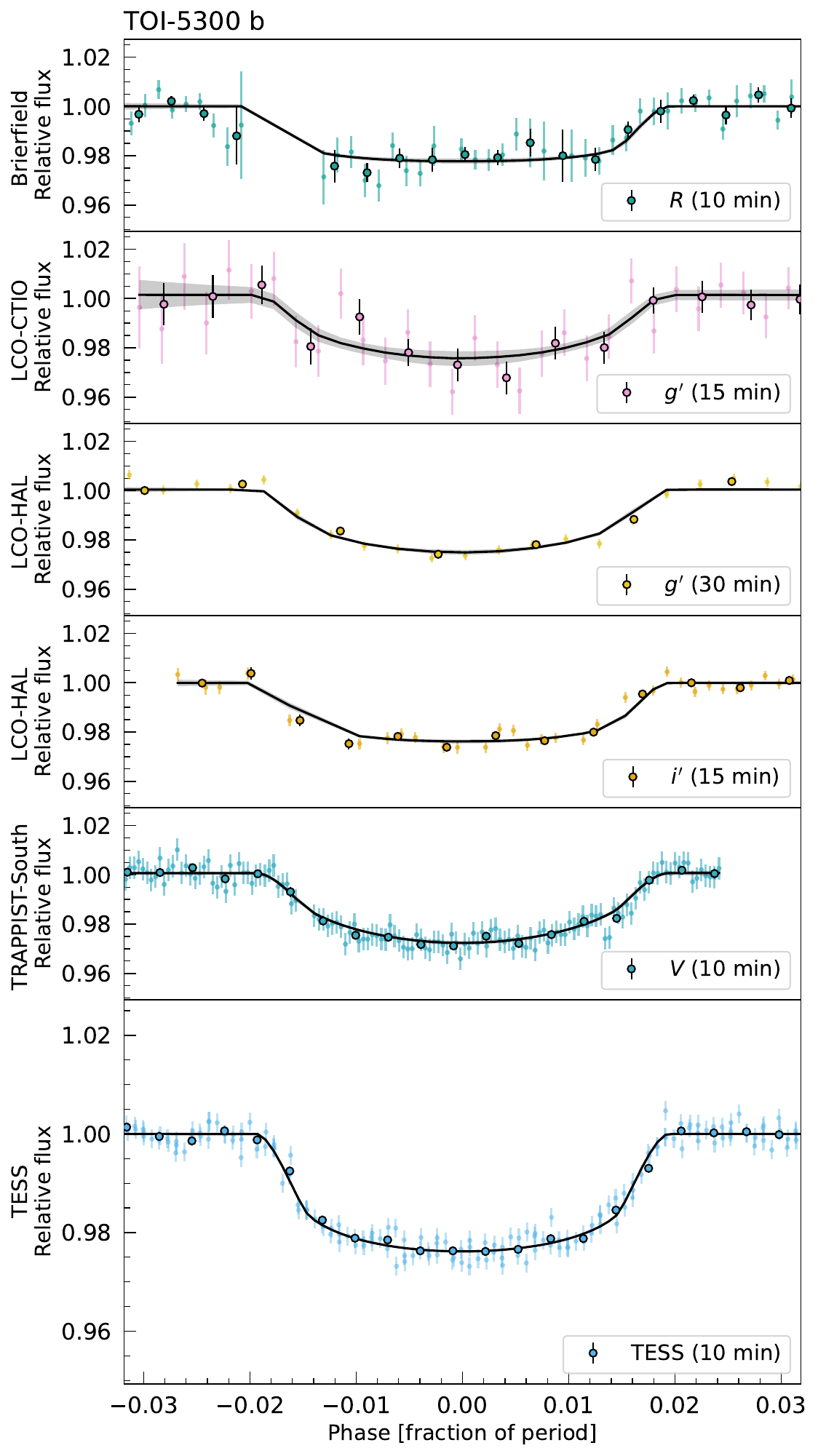}
    \caption{The phase-folded \textit{TESS} light curve for TOI-5300. The \texttt{Juliet} model fit is depicted as a black line, with the \textit{TESS} data shown in light blue in the lower panel. The upper panels include ground-based photometric observations from LCO-CTIO, LCO-HAL, TRAPPIST-South, and Brierfield. Black-edged markers represent data points binned in 10-minute intervals for \textit{TESS}, Brierfield, and TRAPPIST-South, 15-minute intervals for LCO-CTIO and LCO-HAL ($i'$), and 30-minute intervals for LCO-HAL ($g'$).}
    \label{fig:TOI5300_TESS}
\end{figure}

\section{Discussion}
\label{sec:Discussion}
\begin{figure*}
	\includegraphics[width=\linewidth]{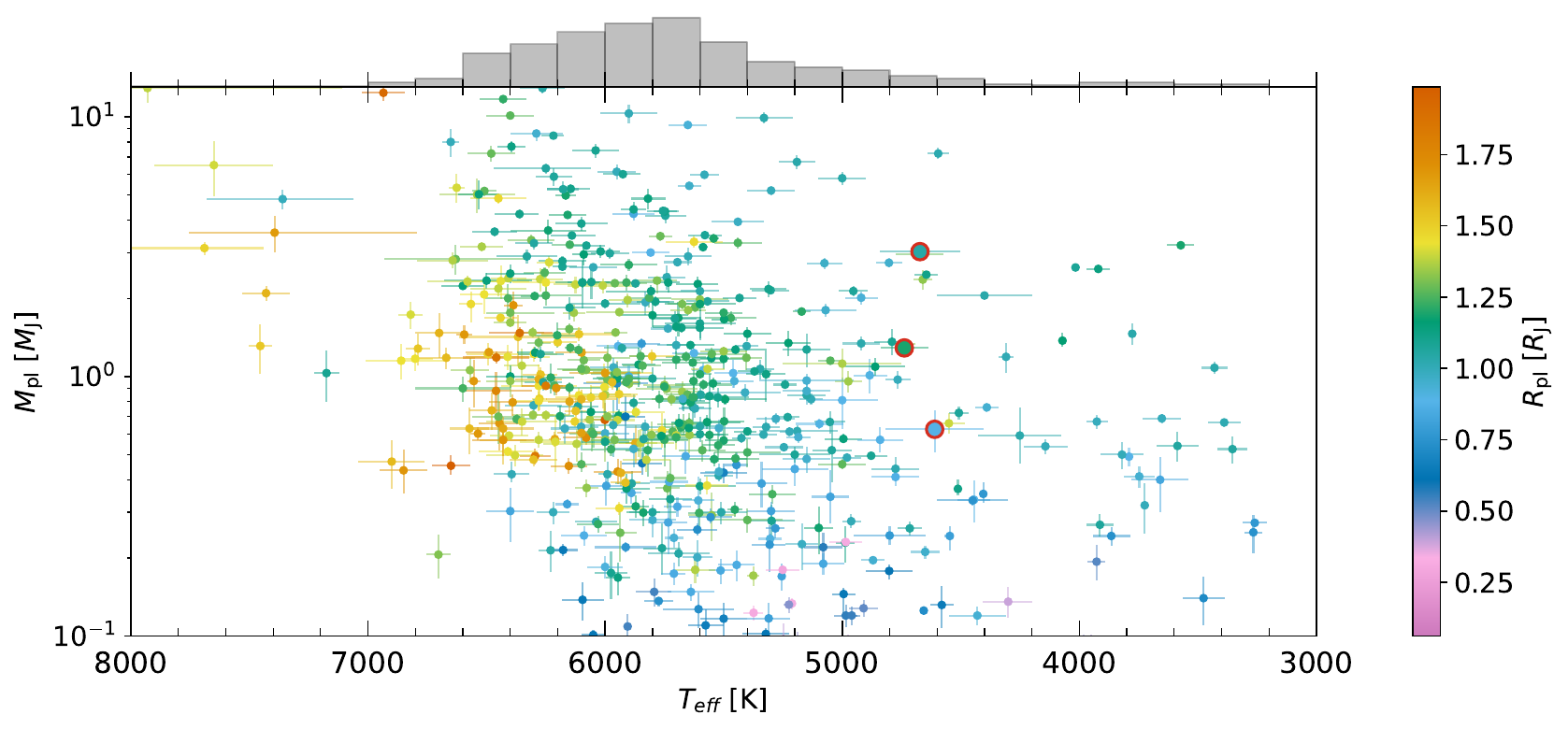}
	\caption{Overview of the presented companions (red encircled) compared to known planets from the PlanetS catalog (extended from \cite{otegi_revisited_2020, parc_super-earths_2024}, updated August 2024). The data is color-coded by planetary radii. The histogram at the top displays the relative occurrence of the transiting gas giants with masses ranging from 0.1 to 13 $M_\mathrm{Jup}$. The low-mass star regime remains relatively poorly populated, and our three mass characterizations contribute to this population.}
	\label{fig:overview}
\end{figure*}

We confirm the presence of three Hot Jupiters orbiting mid-K dwarfs: TOI-2969 b, TOI-2989 b, and TOI-5300 b. Figure \ref{fig:overview} shows these companions compared to known transiting exoplanets, highlighting their location in the diagram's relatively poorly populated region. Characterizing these objects contributes to filling the low-mass star ends of the exoplanet distribution, enriching our understanding of planetary demographics.

\subsection{Heavy element masses}
The heavy element content of the three Hot Jupiters, TOI-2969 b, TOI-2989 b, and TOI-5300 b, is estimated using interior models. 

According to the empirical formulas of \citep{sestovic_investigating_2018}, TOI-2969 b, TOI-2989 b, and TOI-5300 b should not be inflated, with the insolation not exceeding the threshold in incident flux. Indeed, when applying the \cite{fortney_planetary_2007} models that do not include inflation, TOI-2969 b can be explained by a 50 $M_\oplus$ core mass at an approximate age of 4.5 Gyr. Similarly, TOI-2989 b is consistent with a $100 M_\oplus$ core mass at 1 Gyr, and TOI-5300 b with a $100 M_\oplus$ core mass at 4.5 Gyr. 

However, the three planets can also be described by models including inflation. Applying the models from \cite{baraffe_structure_2008} with typical Hot Jupiter irradiation (equivalent to solar exposure at 0.045 AU), and using approximate values for age and heavy element fraction (derived from the fitted interior models presented later in this section), we find the following theoretical radii: For TOI-2969 b, assuming an age of around 3 Gyr and a heavy element mass fraction of $Z = 0.10$, the radius would be close to $1.06 \rjup$. For TOI-2989 b, with an assumed age of about 1 Gyr and the same heavy element fraction, the radius is predicted to be near $1.09 \rjup$. For TOI-5300 b, assuming an age of roughly 3 Gyr and a higher metallicity of $Z = 0.50$, the expected radius is about $0.77 \rjup$. These estimated radii correspond well with the observed values, which are $1.10 \pm 0.08 \rjup$, $1.12 \pm 0.05 \rjup$, and $0.83 \pm 0.07 \rjup$, respectively. 
To test if these planets can also be fitted when including heating efficiency leading to radius inflation, we utilize the grid of interior models for Hot Jupiters published by \cite{sarkis_evidence_2021}. These models, based on the planetary evolution code \texttt{completo} \citep{mordasini_characterization_2012}, assume a composition of an H/He envelope, without a central core \citep[using the SCvH equation of state (EoS);][]{saumon_equation_1995}, where heavy elements are modeled as water \cite[ANEOS equation of state][]{thompson_grain_1990} and assumed to be homogeneously mixed. The envelope is coupled with a fully non-gray atmospheric model from \texttt{petitCODE} \citep{molliere_model_2015, molliere_observing_2017}. This grid of interior models does not trace the planet's evolution over time, but relies on the internal luminosity value to estimate the planetary internal structure. We choose a uniform prior on the internal luminosity; as noted in \cite{sarkis_evidence_2021}, the choice of the prior impacts the derived luminosity and heating efficiency coefficient. However, we find that the heavy element fractions are compatible at $1\,\sigma$ when a log-uniform prior is used. The fraction of heavy elements in the interior is $0.24 \pm 0.08$, $0.12 \pm 0.03$, and $0.44 \pm 0.08$ for TOI-2969 b, TOI-2989 b, and TOI-5300 b, respectively. From the fraction of heavy elements, the heavy element mass can be derived \citep[e.g.][]{ulmer-moll_two_2022}, and we find that TOI-2969 b, TOI-2989 b, and TOI-5300 b contain a significant amount of heavy element mass of $88 \pm 30 M_\oplus$, $114 \pm 30 M_\oplus$, and $84 \pm 21 M_\oplus$, respectively. Compared with other studies discussing the heavy-element mass of companions orbiting K dwarfs \citep[e.g.][]{hartman_hat-p-12b_2009, hartman_hat-p-18b_2011, grunblatt_seeing_2017, torres_improved_2008, hacker_toi-2374_2024, delamer_toi-4201_2024, hellier_wasp-29b_2010}; our three targets have a relatively high heavy-element mass, which is linked to their higher densities. 

The inclusion of heating efficiency does not significantly impact the radii of TOI-2989 b and TOI-5300 b. For these two planets, the inferred heavy element content is consistent between models with and without inflation. However, for TOI-2969 b, including inflation impacts the planetary radius, resulting in a larger estimated heavy element content when modeled with inflation. TOI-2969 b may be more affected by additional heating efficiency, as it has the lowest density in the sample and receives the highest stellar irradiation, leading to an equilibrium temperature of 1186 K. This highlights the degeneracy between incorporating additional heating efficiency, which increases the planetary radius, and adding heavy elements to the interior, which decreases the overall radius.

In conclusion, all planets can be modeled without including additional heating efficiency and are found to contain a significant amount of heavy elements. We note that the models used in this work rely on the SCvH equation of state for H/He and that more recent EoS from \cite{chabrier_new_2021} usually lead to smaller planetary radii and a lower amount of heavy elements \citep[e.g.][]{muller_theoretical_2020}.

\subsection{Planet-metallicity correlation}
It is established that metal-rich stars are more likely to host giant exoplanets, as described by the planet-metallicity correlation for FGK stars \citep{ida_toward_2004, santos_spectroscopic_2004, fischer_planetmetallicity_2005}. This correlation appears particularly true for low-mass stars, where the expected lower disk mass of M dwarfs \citep{vorobyov_mass_2008, alibert_extrasolar_2011} can be compensated with a higher metallicity (and vice versa) \citep{thommes_gas_2008, mordasini_extrasolar_2012}. Our stars, with masses between 0.67 and 0.77 $M_\odot$, exhibit [Fe/H] metallicities of 0.08 $\pm$ 0.05, -0.04 $\pm$ 0.07, and -0.17 $\pm$ 0.07, which align with, or slightly exceed, the average metallicity expected for K dwarfs of similar mass \citep[e.g. Fig. 9 from][]{fischer_planetmetallicity_2005}, where [M/H] $\approx$ -0.15 is typical. Since [M/H] represents the total metal abundance and is greater than or equal to [Fe/H], this allows for direct comparison. \cite{guillot_correlation_2006, fortney_atmosphere_2006} have shown that the metal content of a planet correlates with that of its host star. Given the significant uncertainty in metallicity and heavy element masses, concluding whether these three targets agree with the correlation is complex.

\subsection{Eccentricity}
All orbital solutions presented in this paper have the eccentricity and argument of periastron fixed to constant values ($e=0$ and $\omega=90^{\circ}$). When including the eccentricity and argument of periastron as free parameters, we determine $3\,\sigma$ upper limits on the eccentricity of $0.01$ for TOI-2969 b, $0.03$ for TOI-2989 b, and $0.05$ for TOI-5300 b. The argument of periastron does not converge. We find that TOI-2969 b, TOI-2989 b, and TOI-5300 b are best described by models with fixed circular orbits ($e=0$) based on the \texttt{Juliet} fits' log-evidence values. This is consistent with the expectation that most Hot Jupiters, especially those with orbital periods up to approximately 3 days, have circularized orbits. If these planets initially had elliptical orbits, they would have quickly circularized due to the strong tidal dissipation caused by their proximity to their host stars \citep{hut_tidal_1981}. 

\subsection{Multiplicity}
Hot Jupiters typically lack nearby planetary companions, likely because their inward migration clears out planets in close orbits. The planetary systems of WASP-132 \citep{grieves_refining_2024} and TOI-1130 \citep{borsato_characterisation_2024} are notable exceptions among K dwarfs, each hosting an inner short-period Super-Earth alongside a Hot Jupiter. Within the precision of our CORALIE data, there are no indications of additional companions with periods up to $\sim$ 1 year, as all jitter values are consistent with zero. Nonetheless, these targets remain relevant for additional RV follow-up observations as $52 \pm 5\%$ of Hot Jupiters have additional, longer period companions as shown by \cite{bryan_statistics_2016}.

\subsection{Atmospheric characterization}
With the orbital periods shorter than 4 days, the three planets presented in this paper can also be considered for atmospheric characterization. The indicators most commonly used for the expected signal-to-noise (S/N) of transmission and emission spectroscopy are the transmission spectroscopy metric (TSM) and emission spectroscopy metric (ESM), as defined by \citet{kempton_framework_2018}. TOI-2969 has an elevated ESM of 149, a TSM of 90, and a relatively large scale height of 205 km, suggesting it is promising for atmospheric studies. In contrast, TOI-2989, with an ESM of 71, a TSM of 20, and a scale height of 70 km, is less favorable for atmospheric studies. TOI-5300, with an ESM of 76, a TSM of 77, and a large scale height of 224 km, also shows potential for atmospheric characterization, though it is not as favorable as TOI-2969.

\section{Conclusions}
\label{sec:Conclusions}
We confirm and characterize three non-inflated Hot Jupiters - TOI-2969 b, TOI-2989 b, and TOI-5300 b - orbiting mid-K dwarfs. These mass measurements highlight the importance of spectroscopic follow-up, as they contribute to the growing well-characterized catalog of gas giants around low-mass stars. These stars are part of an ongoing CORALIE program, which should provide further characterizations in the future. The unique characteristics of the discovered objects, e.g., non-inflated Hot Jupiters with a significant amount of heavy elements, offer valuable opportunities for future research in planetary and substellar science, such as atmospheric studies and low-mass star formation. With TOI-2969 b standing out as the most promising target for emission spectroscopy. 

\bibliographystyle{aa}
\bibliography{references.bib}

\begin{thebibliography}{117}
\expandafter\ifx\csname natexlab\endcsname\relax\def\natexlab#1{#1}\fi

\bibitem[{Alibert {et~al.}(2011)Alibert, Mordasini, \&
  Benz}]{alibert_extrasolar_2011}
Alibert, Y., Mordasini, C., \& Benz, W. 2011, \aap, 526, A63

\bibitem[{Aller {et~al.}(2020)Aller, Lillo-Box, Jones, Miranda, \&
  Barceló~Forteza}]{aller_planetary_2020}
Aller, A., Lillo-Box, J., Jones, D., Miranda, L.~F., \& Barceló~Forteza, S.
  2020, \aap, 635, A128

\bibitem[{Baraffe {et~al.}(2008)Baraffe, Chabrier, \&
  Barman}]{baraffe_structure_2008}
Baraffe, I., Chabrier, G., \& Barman, T. 2008, \aap, 482, 315

\bibitem[{Bonfils {et~al.}(2013)Bonfils, Delfosse, Udry, Forveille, Mayor,
  Perrier, Bouchy, Gillon, Lovis, Pepe, Queloz, Santos, Ségransan, \&
  Bertaux}]{bonfils_harps_2013}
Bonfils, X., Delfosse, X., Udry, S., {et~al.} 2013, \aap, 549, A109

\bibitem[{Borsato {et~al.}(2024)Borsato, Degen, Leleu, Hooton, Egger,
  Bekkelien, Brandeker, Collier~Cameron, Günther, Nascimbeni, Persson,
  Bonfanti, Wilson, Correia, Zingales, Guillot, Triaud, Piotto, Gandolfi, Abe,
  Alibert, Alonso, Bárczy, Navascues, Barros, Baumjohann, Beck, Bendjoya,
  Benz, Billot, Broeg, Busch, Csizmadia, Cubillos, Davies, Deleuil, Deline,
  Delrez, Demangeon, Demory, Derekas, Edwards, Ehrenreich, Erikson, Fortier,
  Fossati, Fridlund, Gazeas, Gillon, Güdel, Heitzmann, Helling, Hoyer, Isaak,
  Kiss, Korth, Lam, Laskar, Lecavelier Des~Etangs, Lendl, Magrin, Marafatto,
  Maxted, Mecina, Mékarnia, Mordasini, Mura, Olofsson, Ottensamer, Pagano,
  Pallé, Peter, Pollacco, Queloz, Ragazzoni, Rando, Ratti, Rauer, Ribas,
  Salmon, Santos, Scandariato, Ségransan, Simon, Smith, Sousa, Stalport,
  Suarez, Sulis, Szabó, Udry, Van~Grootel, Venturini, Villaver, Walton, \&
  Wolter}]{borsato_characterisation_2024}
Borsato, L., Degen, D., Leleu, A., {et~al.} 2024, \aap, 689, A52

\bibitem[{Boss(1997)}]{boss_giant_1997}
Boss, A.~P. 1997, Science, 276, 1836

\bibitem[{Bryan {et~al.}(2016)Bryan, Knutson, Howard, Ngo, Batygin, Crepp,
  Fulton, Hinkley, Isaacson, Johnson, Marcy, \& Wright}]{bryan_statistics_2016}
Bryan, M.~L., Knutson, H.~A., Howard, A.~W., {et~al.} 2016, \aj, 821, 89

\bibitem[{Bryant {et~al.}(2023)Bryant, Bayliss, \&
  Van Eylen}]{bryant_occurrence_2023}
Bryant, E.~M., Bayliss, D., \& Van Eylen, V. 2023, \mnras, 521, 3663

\bibitem[{Burn {et~al.}(2021)Burn, Schlecker, Mordasini, Emsenhuber, Alibert,
  Henning, Klahr, \& Benz}]{burn_new_2021}
Burn, R., Schlecker, M., Mordasini, C., {et~al.} 2021, \aap, 656, A72

\bibitem[{Caldwell {et~al.}(2020)Caldwell, Tenenbaum, Twicken, Jenkins, Ting,
  Smith, Hedges, Fausnaugh, Rose, \& Burke}]{caldwell_tess_2020}
Caldwell, D.~A., Tenenbaum, P., Twicken, J.~D., {et~al.} 2020, Research Notes
  of the AAS, 4, 201

\bibitem[{Chabrier \& Debras(2021)}]{chabrier_new_2021}
Chabrier, G. \& Debras, F. 2021, \aj, 917, 4

\bibitem[{Collier~Cameron {et~al.}(2007)Collier~Cameron, Wilson, West, Hebb,
  Wang, Aigrain, Bouchy, Christian, Clarkson, Enoch, Esposito, Guenther,
  Haswell, Hébrard, Hellier, Horne, Irwin, Kane, Loeillet, Lister, Maxted,
  Mayor, Moutou, Parley, Pollacco, Pont, Queloz, Ryans, Skillen, Street, Udry,
  \& Wheatley}]{collier_cameron_efficient_2007}
Collier~Cameron, A., Wilson, D.~M., West, R.~G., {et~al.} 2007, \mnras, 380,
  1230

\bibitem[{Collins(2019)}]{collins_tess_2019}
Collins, K. 2019, in American {Astronomical} {Society} {Meeting} {Abstracts},
  Vol. 233, American {Astronomical} {Society} {Meeting} {Abstracts} \#233,
  140.05

\bibitem[{Cretignier {et~al.}(2020)Cretignier, Dumusque, Allart, Pepe, \&
  Lovis}]{cretignier_measuring_2020}
Cretignier, M., Dumusque, X., Allart, R., Pepe, F., \& Lovis, C. 2020, \aap,
  633, A76

\bibitem[{Delamer {et~al.}(2024)Delamer, Kanodia, Cañas, Müller, Helled, Lin,
  Libby-Roberts, Gupta, Mahadevan, Teske, Butler, Yee, Crane, Shectman, Osip,
  Beletsky, Monson, Hebb, Powers, Wisniewski, Alvarado-Montes, Bender, Dong,
  Han, Ninan, Robertson, Roy, Schwab, Stefánsson, \&
  Wright}]{delamer_toi-4201_2024}
Delamer, M., Kanodia, S., Cañas, C.~I., {et~al.} 2024, \aj$\,$Letters, 962,
  L22

\bibitem[{{ESA}(1997)}]{esa_hipparcos_1997}
{ESA}. 1997, ESA Special Publication, 1200

\bibitem[{Espinoza \& Jordán(2015)}]{espinoza_limb_2015}
Espinoza, N. \& Jordán, A. 2015, \mnras, 450, 1879

\bibitem[{Espinoza {et~al.}(2019)Espinoza, Kossakowski, \&
  Brahm}]{espinoza_juliet_2019}
Espinoza, N., Kossakowski, D., \& Brahm, R. 2019, \mnras, 490, 2262

\bibitem[{Evans {et~al.}(2023)Evans, Eyer, Busso, Riello, De~Angeli, Burgess,
  Audard, Clementini, Garofalo, Holl, Jevardat De~Fombelle, Lanzafame,
  Lecoeur-Taibi, Mowlavi, Nienartowicz, Palaversa, \&
  Rimoldini}]{evans_gaia_2023}
Evans, D.~W., Eyer, L., Busso, G., {et~al.} 2023, \aap, 674, A4

\bibitem[{Eyer {et~al.}(2023)Eyer, Audard, Holl, Rimoldini, Carnerero,
  Clementini, De~Ridder, Distefano, Evans, Gavras, Gomel, Lebzelter, Marton,
  Mowlavi, Panahi, Ripepi, Wyrzykowski, Nienartowicz, Jevardat De~Fombelle,
  Lecoeur-Taibi, Rohrbasser, Riello, García-Lario, Lanzafame, Mazeh, Raiteri,
  Zucker, Ábrahám, Aerts, Aguado, Anderson, Bashi, Binnenfeld, Faigler,
  Garofalo, Karbevska, Kóspál, Kruszyńska, Kun, Lanza, Leccia, Marconi,
  Messina, Molinaro, Molnár, Muraveva, Musella, Nagy, Pagano, Palaversa,
  Plachy, Prša, Rybicki, Shahaf, Szabados, Szegedi-Elek, Trabucchi, Barblan,
  Grenon, Roelens, \& Süveges}]{eyer_gaia_2023}
Eyer, L., Audard, M., Holl, B., {et~al.} 2023, \aap, 674, A13

\bibitem[{Fischer \& Valenti(2005)}]{fischer_planetmetallicity_2005}
Fischer, D.~A. \& Valenti, J. 2005, \aj, 622, 1102

\bibitem[{Foreman-Mackey(2016)}]{foreman-mackey_cornerpy_2016}
Foreman-Mackey, D. 2016, The Journal of Open Source Software, 1, 24

\bibitem[{Fortney {et~al.}(2021)Fortney, Dawson, \& Komacek}]{fortney_hot_2021}
Fortney, J.~J., Dawson, R.~I., \& Komacek, T.~D. 2021, \jgr: Planets, 126,
  e2020JE006629

\bibitem[{Fortney {et~al.}(2007)Fortney, Marley, \&
  Barnes}]{fortney_planetary_2007}
Fortney, J.~J., Marley, M.~S., \& Barnes, J.~W. 2007, \aj, 659, 1661

\bibitem[{Fortney {et~al.}(2006)Fortney, Saumon, Marley, Lodders, \&
  Freedman}]{fortney_atmosphere_2006}
Fortney, J.~J., Saumon, D., Marley, M.~S., Lodders, K., \& Freedman, R.~S.
  2006, \aj, 642, 495

\bibitem[{{Gaia Collaboration} {et~al.}(2023){Gaia Collaboration}, Vallenari,
  Brown, Prusti, De~Bruijne, Arenou, Babusiaux, Biermann, Creevey, Ducourant,
  Evans, Eyer, Guerra, Hutton, Jordi, Klioner, Lammers, Lindegren, Luri,
  Mignard, Panem, Pourbaix, Randich, Sartoretti, Soubiran, Tanga, Walton,
  Bailer-Jones, Bastian, Drimmel, Jansen, Katz, Lattanzi, Van~Leeuwen, Bakker,
  Cacciari, Castañeda, De~Angeli, Fabricius, Fouesneau, Frémat, Galluccio,
  Guerrier, Heiter, Masana, Messineo, Mowlavi, Nicolas, Nienartowicz, Pailler,
  Panuzzo, Riclet, Roux, Seabroke, Sordo, Thévenin, Gracia-Abril, Portell,
  Teyssier, Altmann, Andrae, Audard, Bellas-Velidis, Benson, Berthier, Blomme,
  Burgess, Busonero, Busso, Cánovas, Carry, Cellino, Cheek, Clementini,
  Damerdji, Davidson, De~Teodoro, Nuñez~Campos, Delchambre, Dell’Oro,
  Esquej, Fernández-Hernández, Fraile, Garabato, García-Lario, Gosset,
  Haigron, Halbwachs, Hambly, Harrison, Hernández, Hestroffer, Hodgkin, Holl,
  Janßen, Jevardat De~Fombelle, Jordan, Krone-Martins, Lanzafame, Löffler,
  Marchal, Marrese, Moitinho, Muinonen, Osborne, Pancino, Pauwels,
  Recio-Blanco, Reylé, Riello, Rimoldini, Roegiers, Rybizki, Sarro, Siopis,
  Smith, Sozzetti, Utrilla, Van~Leeuwen, Abbas, Ábrahám, Abreu~Aramburu,
  Aerts, Aguado, Ajaj, Aldea-Montero, Altavilla, Álvarez, Alves, Anders,
  Anderson, Anglada~Varela, Antoja, Baines, Baker, Balaguer-Núñez, Balbinot,
  Balog, Barache, Barbato, Barros, Barstow, Bartolomé, Bassilana, Bauchet,
  Becciani, Bellazzini, Berihuete, Bernet, Bertone, Bianchi, Binnenfeld,
  Blanco-Cuaresma, Blazere, Boch, Bombrun, Bossini, Bouquillon, Bragaglia,
  Bramante, Breedt, Bressan, Brouillet, Brugaletta, Bucciarelli, Burlacu,
  Butkevich, Buzzi, Caffau, Cancelliere, Cantat-Gaudin, Carballo, Carlucci,
  Carnerero, Carrasco, Casamiquela, Castellani, Castro-Ginard, Chaoul, Charlot,
  Chemin, Chiaramida, Chiavassa, Chornay, Comoretto, Contursi, Cooper, Cornez,
  Cowell, Crifo, Cropper, Crosta, Crowley, Dafonte, Dapergolas, David, David,
  De~Laverny, De~Luise, De~March, De~Ridder, De~Souza, De~Torres, Del~Peloso,
  Del~Pozo, Delbo, Delgado, Delisle, Demouchy, Dharmawardena, Di~Matteo,
  Diakite, Diener, Distefano, Dolding, Edvardsson, Enke, Fabre, Fabrizio,
  Faigler, Fedorets, Fernique, Fienga, Figueras, Fournier, Fouron, Fragkoudi,
  Gai, Garcia-Gutierrez, Garcia-Reinaldos, García-Torres, Garofalo, Gavel,
  Gavras, Gerlach, Geyer, Giacobbe, Gilmore, Girona, Giuffrida, Gomel, Gomez,
  González-Núñez, González-Santamaría, González-Vidal, Granvik, Guillout,
  Guiraud, Gutiérrez-Sánchez, Guy, Hatzidimitriou, Hauser, Haywood, Helmer,
  Helmi, Sarmiento, Hidalgo, Hilger, Hładczuk, Hobbs, Holland, Huckle,
  Jardine, Jasniewicz, Jean-Antoine~Piccolo, Jiménez-Arranz, Jorissen,
  Juaristi~Campillo, Julbe, Karbevska, Kervella, Khanna, Kontizas, Kordopatis,
  Korn, Kóspál, Kostrzewa-Rutkowska, Kruszyńska, Kun, Laizeau, Lambert,
  Lanza, Lasne, Le~Campion, Lebreton, Lebzelter, Leccia, Leclerc,
  Lecoeur-Taibi, Liao, Licata, Lindstrøm, Lister, Livanou, Lobel, Lorca, Loup,
  Madrero~Pardo, Magdaleno~Romeo, Managau, Mann, Manteiga, Marchant, Marconi,
  Marcos, Marcos~Santos, Marín~Pina, Marinoni, Marocco, Marshall, Martin~Polo,
  Martín-Fleitas, Marton, Mary, Masip, Massari, Mastrobuono-Battisti, Mazeh,
  McMillan, Messina, Michalik, Millar, Mints, Molina, Molinaro, Molnár,
  Monari, Monguió, Montegriffo, Montero, Mor, Mora, Morbidelli, Morel, Morris,
  Muraveva, Murphy, Musella, Nagy, Noval, Ocaña, Ogden, Ordenovic, Osinde,
  Pagani, Pagano, Palaversa, Palicio, Pallas-Quintela, Panahi, Payne-Wardenaar,
  Peñalosa~Esteller, Penttilä, Pichon, Piersimoni, Pineau, Plachy, Plum,
  Poggio, Prša, Pulone, Racero, Ragaini, Rainer, Raiteri, Rambaux, Ramos,
  Ramos-Lerate, Re~Fiorentin, Regibo, Richards, Rios~Diaz, Ripepi, Riva, Rix,
  Rixon, Robichon, Robin, Robin, Roelens, Rogues, Rohrbasser, Romero-Gómez,
  Rowell, Royer, Ruz~Mieres, Rybicki, Sadowski, Sáez~Núñez,
  Sagristà~Sellés, Sahlmann, Salguero, Samaras, Sanchez~Gimenez, Sanna,
  Santoveña, Sarasso, Schultheis, Sciacca, Segol, Segovia, Ségransan, Semeux,
  Shahaf, Siddiqui, Siebert, Siltala, Silvelo, Slezak, Slezak, Smart, Snaith,
  Solano, Solitro, Souami, Souchay, Spagna, Spina, Spoto, Steele,
  Steidelmüller, Stephenson, Süveges, Surdej, Szabados, Szegedi-Elek, Taris,
  Taylor, Teixeira, Tolomei, Tonello, Torra, Torra, Torralba~Elipe, Trabucchi,
  Tsounis, Turon, Ulla, Unger, Vaillant, Van~Dillen, Van~Reeven, Vanel,
  Vecchiato, Viala, Vicente, Voutsinas, Weiler, Wevers, Wyrzykowski, Yoldas,
  Yvard, Zhao, Zorec, Zucker, \& Zwitter}]{gaia_collaboration_gaia_2023}
{Gaia Collaboration}, Vallenari, A., Brown, A. G.~A., {et~al.} 2023, \aap, 674,
  A1

\bibitem[{Gan {et~al.}(2023)Gan, Wang, Wang, Mao, Huang, Collins, Stassun,
  Shporer, Zhu, Ricker, Vanderspek, Latham, Seager, Winn, Jenkins, Barkaoui,
  Belinski, Ciardi, Evans, Girardin, Maslennikova, Mazeh, Panahi, Pozuelos,
  Radford, Schwarz, Twicken, Wünsche, \& Zucker}]{gan_occurrence_2023}
Gan, T., Wang, S.~X., Wang, S., {et~al.} 2023, \aj, 165, 17

\bibitem[{Garcia {et~al.}(2022)Garcia, Timmermans, Pozuelos, Ducrot, Gillon,
  Delrez, Wells, \& Jehin}]{garcia_prose_2022}
Garcia, L.~J., Timmermans, M., Pozuelos, F.~J., {et~al.} 2022, \mnras, 509,
  4817

\bibitem[{{Gillon} {et~al.}(2011){Gillon}, {Jehin}, {Magain}, {Chantry},
  {Hutsem{\'e}kers}, {Manfroid}, {Queloz}, \& {Udry}}]{gillon_trappist_2011}
{Gillon}, M., {Jehin}, E., {Magain}, P., {et~al.} 2011, in European Physical
  Journal Web of Conferences, Vol.~11, European Physical Journal Web of
  Conferences, 06002

\bibitem[{{Grieves} {et~al.}(2025){Grieves}, {Bouchy}, {Armstrong},
  {Akinsanmi}, {Psaridi}, {Ulmer-Moll}, {Frensch}, {Helled}, {M{\"u}ller},
  {Knierim}, {Santos}, {Adibekyan}, {Parc}, {Lendl}, {Battley}, {Unger},
  {Chaverot}, {Bayliss}, {Dumusque}, {Hawthorn}, {Figueira}, {Fetzner Keniger},
  {Lillo-Box}, {Dyregaard Nielsen}, {Osborn}, {Sousa}, {Str{\o}m}, \&
  {Udry}}]{grieves_refining_2024}
{Grieves}, N., {Bouchy}, F., {Armstrong}, D.~J., {et~al.} 2025, \aap, 693, A144

\bibitem[{Grunblatt {et~al.}(2017)Grunblatt, Huber, Gaidos, Lopez, Howard,
  Isaacson, Sinukoff, Vanderburg, Nofi, Yu, North, Chaplin, Foreman-Mackey,
  Petigura, Ansdell, Weiss, Fulton, \& Lin}]{grunblatt_seeing_2017}
Grunblatt, S.~K., Huber, D., Gaidos, E., {et~al.} 2017, \aj, 154, 254

\bibitem[{Guerrero {et~al.}(2021)Guerrero, Seager, Huang, Vanderburg, Soto,
  Mireles, Hesse, Fong, Glidden, Shporer, Latham, Collins, Quinn, Burt,
  Dragomir, Crossfield, Vanderspek, Fausnaugh, Burke, Ricker, Daylan, Essack,
  Günther, Osborn, Pepper, Rowden, Sha, Villanueva~Jr., Yahalomi, Yu, Ballard,
  Batalha, Berardo, Chontos, Dittmann, Esquerdo, Mikal-Evans, Jayaraman,
  Krishnamurthy, Louie, Mehrle, Niraula, Rackham, Rodriguez, Rowden,
  Sousa-Silva, Watanabe, Wong, Zhan, Zivanovic, Christiansen, Ciardi, Swain,
  Lund, Mullally, Fleming, Rodriguez, Boyd, Quintana, Barclay, Colón,
  Rinehart, Schlieder, Clampin, Jenkins, Twicken, Caldwell, Coughlin, Henze,
  Lissauer, Morris, Rose, Smith, Tenenbaum, Ting, Wohler, Bakos, Bean,
  Berta-Thompson, Bieryla, Bouma, Buchhave, Butler, Charbonneau, Doty, Ge,
  Holman, Howard, Kaltenegger, Kane, Kjeldsen, Kreidberg, Lin, Minsky, Narita,
  Paegert, Pál, Palle, Sasselov, Spencer, Sozzetti, Stassun, Torres, Udry, \&
  Winn}]{guerrero_tess_2021}
Guerrero, N.~M., Seager, S., Huang, C.~X., {et~al.} 2021, \apjs, 254, 39

\bibitem[{Guillot {et~al.}(2006)Guillot, Santos, Pont, Iro, Melo, \&
  Ribas}]{guillot_correlation_2006}
Guillot, T., Santos, N.~C., Pont, F., {et~al.} 2006, \aap, 453, L21

\bibitem[{Hacker {et~al.}(2024)Hacker, Díaz, Armstrong,
  Fernández Fernández, Müller, Delgado-Mena, Sousa, Adibekyan, Stassun,
  Collins, Yee, Bayliss, Bieryla, Bouchy, Butler, Crane, Dumusque, Hartman,
  Helled, Jenkins, Keniger, Lewis, Lillo-Box, Lund, Nielsen, Osborn, Osip,
  Paegert, Radford, Santos, Seager, Shectman, Srdoc, Strøm, Tan, Teske, Vezie,
  Watanabe, Watkins, Wheatley, Winn, Wohler, \& Ziegler}]{hacker_toi-2374_2024}
Hacker, A., Díaz, R.~F., Armstrong, D.~J., {et~al.} 2024, \mnras, 532, 1612

\bibitem[{Hartman {et~al.}(2011)Hartman, Bakos, Sato, Torres, Noyes, Latham,
  Kovács, Fischer, Howard, Johnson, Marcy, Buchhave, Füresz, Perumpilly,
  Béky, Stefanik, Sasselov, Esquerdo, Everett, Csubry, Lázár, Papp, \&
  Sári}]{hartman_hat-p-18b_2011}
Hartman, J.~D., Bakos, G.~{\'A}., Sato, B., {et~al.} 2011, \aj, 726, 52

\bibitem[{Hartman {et~al.}(2009)Hartman, Bakos, Torres, Kovács, Noyes, Pál,
  Latham, Sipőcz, Fischer, Johnson, Marcy, Butler, Howard, Esquerdo, Sasselov,
  Kovács, Stefanik, Fernandez, Lázár, Papp, \&
  Sári}]{hartman_hat-p-12b_2009}
Hartman, J.~D., Bakos, G.~{\'A}., Torres, G., {et~al.} 2009, \aj, 706, 785

\bibitem[{Hartman {et~al.}(2020)Hartman, Jordán, Bayliss, Bakos, Bento,
  Bhatti, Brahm, Csubry, Espinoza, Henning, Mancini, Penev, Rabus, Sarkis, Suc,
  De~Val-Borro, Zhou, Crane, Shectman, Teske, Wang, Butler, Lázár, Papp,
  Sári, Anderson, Hellier, West, Barkaoui, Pozuelos, Jehin, Gillon, Nielsen,
  Lendl, Udry, Ricker, Vanderspek, Latham, Seager, Winn, Christiansen,
  Crossfield, Henze, Jenkins, Smith, \& Ting}]{hartman_hats-47b_2020}
Hartman, J.~D., Jordán, A., Bayliss, D., {et~al.} 2020, \aj, 159, 173

\bibitem[{Hellier {et~al.}(2010)Hellier, Anderson, Collier~Cameron, Gillon,
  Lendl, Maxted, Queloz, Smalley, Triaud, West, Brown, Enoch, Lister, Pepe,
  Pollacco, Ségransan, \& Udry}]{hellier_wasp-29b_2010}
Hellier, C., Anderson, D.~R., Collier~Cameron, A., {et~al.} 2010, \aj, 723, L60

\bibitem[{Huang {et~al.}(2020{\natexlab{a}})Huang, Quinn, Vanderburg, Becker,
  Rodriguez, Pozuelos, Gandolfi, Zhou, Mann, Collins, Crossfield, Barkaoui,
  Collins, Fridlund, Gillon, Gonzales, Günther, Henry, Howell, James, Jao,
  Jehin, Jensen, Kane, Lissauer, Matthews, Matson, Paredes, Schlieder, Stassun,
  Shporer, Sha, Tan, Georgieva, Mathur, Palle, Persson, Eylen, Ricker,
  Vanderspek, Latham, Winn, Seager, Jenkins, Burke, Goeke, Rinehart, Rose,
  Ting, Torres, \& Wong}]{huang_tess_2020}
Huang, C.~X., Quinn, S.~N., Vanderburg, A., {et~al.} 2020{\natexlab{a}},
  \aj$\,$Letters, 892, L7

\bibitem[{Huang {et~al.}(2020{\natexlab{b}})Huang, Vanderburg, Pál, Sha, Yu,
  Fong, Fausnaugh, Shporer, Guerrero, Vanderspek, \&
  Ricker}]{huang_photometry_2020}
Huang, C.~X., Vanderburg, A., Pál, A., {et~al.} 2020{\natexlab{b}}, Research
  Notes of the AAS, 4, 206

\bibitem[{Husser {et~al.}(2013)Husser, Wende-von Berg, Dreizler, Homeier,
  Reiners, Barman, \& Hauschildt}]{husser_new_2013}
Husser, T.-O., Wende-von Berg, S., Dreizler, S., {et~al.} 2013, \aap, 553, A6

\bibitem[{Hut(1981)}]{hut_tidal_1981}
Hut, P. 1981, \aap, 99, 126

\bibitem[{Ida \& Lin(2004)}]{ida_toward_2004}
Ida, S. \& Lin, D. N.~C. 2004, \aj, 616, 567

\bibitem[{Ida \& Lin(2005)}]{ida_toward_2005}
Ida, S. \& Lin, D. N.~C. 2005, \aj, 626, 1045

\bibitem[{Jehin {et~al.}(2011)Jehin, Gillon, Queloz, Magain, Manfroid, Chantry,
  Lendl, Hutsemékers, \& Udry}]{jehin_trappist_2011}
Jehin, E., Gillon, M., Queloz, D., {et~al.} 2011, The Messenger, 145, 2

\bibitem[{Jenkins(2002)}]{jenkins_impact_2002}
Jenkins, J.~M. 2002, \aj, 575, 493

\bibitem[{{Jenkins} {et~al.}(2010){Jenkins}, {Chandrasekaran}, {McCauliff},
  {Caldwell}, {Tenenbaum}, {Li}, {Klaus}, {Cote}, \&
  {Middour}}]{jenkins_transiting_2010}
{Jenkins}, J.~M., {Chandrasekaran}, H., {McCauliff}, S.~D., {et~al.} 2010, in
  Society of Photo-Optical Instrumentation Engineers (SPIE) Conference Series,
  Vol. 7740, Software and Cyberinfrastructure for Astronomy, ed. N.~M.
  {Radziwill} \& A.~{Bridger}, 77400D

\bibitem[{{Jenkins} {et~al.}(2020){Jenkins}, {Tenenbaum}, {Seader}, {Burke},
  {McCauliff}, {Smith}, {Twicken}, \& {Chandrasekaran}}]{jenkins_kepler_2020}
{Jenkins}, J.~M., {Tenenbaum}, P., {Seader}, S., {et~al.} 2020, {Kepler Data
  Processing Handbook: Transiting Planet Search}, Kepler Science Document
  KSCI-19081-003, id. 9. Edited by Jon M. Jenkins.

\bibitem[{Jenkins {et~al.}(2016)Jenkins, Twicken, McCauliff, Campbell,
  Sanderfer, Lung, Mansouri-Samani, Girouard, Tenenbaum, Klaus, Smith,
  Caldwell, Chacon, Henze, Heiges, Latham, Morgan, Swade, Rinehart, \&
  Vanderspek}]{jenkins_tess_2016}
Jenkins, J.~M., Twicken, J.~D., McCauliff, S., {et~al.} 2016, in Society of
  {Photo}-{Optical} {Instrumentation} {Engineers} ({SPIE}) {Conference}
  {Series}, Vol. 9913, Software and {Cyberinfrastructure} for {Astronomy} {IV},
  ed. G.~Chiozzi \& J.~C. Guzman, 99133E

\bibitem[{{Jensen}(2013)}]{jensen_tapir_2013}
{Jensen}, E. 2013, {Tapir: A web interface for transit/eclipse observability},
  Astrophysics Source Code Library, record ascl:1306.007

\bibitem[{Jordán {et~al.}(2022)Jordán, Hartman, Bayliss, Bakos, Brahm,
  Bryant, Csubry, Henning, Hobson, Mancini, Penev, Rabus, Suc, De~Val-Borro,
  Wallace, Barkaoui, Ciardi, Collins, Esparza-Borges, Furlan, Gan, Benkhaldoun,
  Ghachoui, Gillon, Howell, Jehin, Fukui, Kawauchi, Livingston, Luque, Matson,
  Matthews, Osborn, Murgas, Narita, Palle, Parvianen, \&
  Waalkes}]{jordan_hats-74ab_2022}
Jordán, A., Hartman, J.~D., Bayliss, D., {et~al.} 2022, \aj, 163, 125

\bibitem[{Kanodia {et~al.}(2022)Kanodia, Libby-Roberts, Cañas, Ninan,
  Mahadevan, Stefansson, Lin, Jones, Monson, Parker, Kobulnicky, Swaby, Powers,
  Beard, Bender, Blake, Cochran, Dong, Diddams, Fredrick, Gupta, Halverson,
  Hearty, Logsdon, Metcalf, McElwain, Morley, Rajagopal, Ramsey, Robertson,
  Roy, Schwab, Terrien, Wisniewski, \& Wright}]{kanodia_toi-3757_2022}
Kanodia, S., Libby-Roberts, J., Cañas, C.~I., {et~al.} 2022, \aj, 164, 81

\bibitem[{Kempton {et~al.}(2018)Kempton, Bean, Louie, Deming, Koll, Mansfield,
  Christiansen, López-Morales, Swain, Zellem, Ballard, Barclay, Barstow,
  Batalha, Beatty, Berta-Thompson, Birkby, Buchhave, Charbonneau, Cowan,
  Crossfield, Val-Borro, Doyon, Dragomir, Gaidos, Heng, Hu, Kane, Kreidberg,
  Mallonn, Morley, Narita, Nascimbeni, Pallé, Quintana, Rauscher, Seager,
  Shkolnik, Sing, Sozzetti, Stassun, Valenti, \&
  Essen}]{kempton_framework_2018}
Kempton, E. M.-R., Bean, J.~L., Louie, D.~R., {et~al.} 2018, \pasp, 130, 114401

\bibitem[{Kipping(2013{\natexlab{a}})}]{kipping_efficient_2013}
Kipping, D.~M. 2013{\natexlab{a}}, \mnras, 435, 2152

\bibitem[{Kipping(2013{\natexlab{b}})}]{kipping_parametrizing_2013}
Kipping, D.~M. 2013{\natexlab{b}}, \mnras: Letters, 434, L51

\bibitem[{Kipping(2014)}]{kipping_characterizing_2014}
Kipping, D.~M. 2014, \mnras, 440, 2164

\bibitem[{Kovács {et~al.}(2002)Kovács, Zucker, \&
  Mazeh}]{kovacs_box-fitting_2002}
Kovács, G., Zucker, S., \& Mazeh, T. 2002, \aap, 391, 369

\bibitem[{Kunimoto \& Daylan(2021)}]{kunimoto_searching_2021}
Kunimoto, M. \& Daylan, T. 2021, in Posters from the {TESS} {Science}
  {Conference} {II} ({TSC2}), 62

\bibitem[{{Kurucz}(1993)}]{kurucz_synthe_1993}
{Kurucz}, R.~L. 1993, {SYNTHE spectrum synthesis programs and line data}

\bibitem[{Laughlin {et~al.}(2004)Laughlin, Bodenheimer, \&
  Adams}]{laughlin_core_2004}
Laughlin, G., Bodenheimer, P., \& Adams, F.~C. 2004, \aj, 612, L73

\bibitem[{{Lightkurve Collaboration} {et~al.}(2018){Lightkurve Collaboration},
  Cardoso, Hedges, Gully-Santiago, Saunders, Cody, Barclay, Hall, Sagear,
  Turtelboom, Zhang, Tzanidakis, Mighell, Coughlin, Bell, Berta-Thompson,
  Williams, Dotson, \& Barentsen}]{lightkurve_collaboration_lightkurve_2018}
{Lightkurve Collaboration}, Cardoso, J. V. d.~M., Hedges, C., {et~al.} 2018,
  Astrophysics Source Code Library, ascl:1812.013

\bibitem[{Martin {et~al.}(2021)Martin, El-Badry, Hodžić, Triaud, Angus,
  Birky, Foreman-Mackey, Hedges, Montet, Murphy, Santerne, Stassun, Stephan,
  Wang, Benni, Krushinsky, Chazov, Mishevskiy, Ziegler, Soubkiou, Benkhaldoun,
  Boisse, Battley, Miller, Caldwell, Collins, Henze, Guerrero, Jenkins, Latham,
  Levine, McDermott, Mullally, Ricker, Seager, Shporer, Vanderburg, Vanderspek,
  \& Winn}]{martin_toi-1259ab_2021}
Martin, D.~V., El-Badry, K., Hodžić, V.~K., {et~al.} 2021, \mnras, 507, 4132

\bibitem[{Maxted {et~al.}(2011)Maxted, Anderson, Collier~Cameron, Hellier,
  Queloz, Smalley, Street, Triaud, West, Gillon, Lister, Pepe, Pollacco,
  Ségransan, Smith, \& Udry}]{maxted_wasp-41b_2011}
Maxted, P. F.~L., Anderson, D.~R., Collier~Cameron, A., {et~al.} 2011, \pasp,
  123, 547

\bibitem[{{Mayor} {et~al.}(2011){Mayor}, {Marmier}, {Lovis}, {Udry},
  {S{\'e}gransan}, {Pepe}, {Benz}, {Bertaux}, {Bouchy}, {Dumusque}, {Lo Curto},
  {Mordasini}, {Queloz}, \& {Santos}}]{mayor_harps_2011}
{Mayor}, M., {Marmier}, M., {Lovis}, C., {et~al.} 2011, ArXiv e-prints,
  arXiv:1109.2497

\bibitem[{Mollière {et~al.}(2017)Mollière, van Boekel, Bouwman, Henning,
  Lagage, \& Min}]{molliere_observing_2017}
Mollière, P., van Boekel, R., Bouwman, J., {et~al.} 2017, \aap, 600, A10

\bibitem[{Mollière {et~al.}(2015)Mollière, van Boekel, Dullemond, Henning, \&
  Mordasini}]{molliere_model_2015}
Mollière, P., van Boekel, R., Dullemond, C., Henning, T., \& Mordasini, C.
  2015, \aj, 813, 47

\bibitem[{Mordasini {et~al.}(2012{\natexlab{a}})Mordasini, Alibert, Benz,
  Klahr, \& Henning}]{mordasini_extrasolar_2012}
Mordasini, C., Alibert, Y., Benz, W., Klahr, H., \& Henning, T.
  2012{\natexlab{a}}, \aap, 541, A97

\bibitem[{Mordasini {et~al.}(2012{\natexlab{b}})Mordasini, Alibert, Klahr, \&
  Henning}]{mordasini_characterization_2012}
Mordasini, C., Alibert, Y., Klahr, H., \& Henning, T. 2012{\natexlab{b}}, \aap,
  547, A111

\bibitem[{Müller {et~al.}(2020)Müller, Ben-Yami, \&
  Helled}]{muller_theoretical_2020}
Müller, S., Ben-Yami, M., \& Helled, R. 2020, \aj, 903, 147

\bibitem[{{Nissen}(2004)}]{nissen_thin_2003}
{Nissen}, P.~E. 2004, in Origin and Evolution of the Elements, ed.
  A.~{McWilliam} \& M.~{Rauch}, 154

\bibitem[{Otegi {et~al.}(2020)Otegi, Bouchy, \& Helled}]{otegi_revisited_2020}
Otegi, J.~F., Bouchy, F., \& Helled, R. 2020, \aap, 634, A43

\bibitem[{{Paegert} {et~al.}(2021){Paegert}, {Stassun}, {Collins}, {Pepper},
  {Torres}, {Jenkins}, {Twicken}, \& {Latham}}]{paegert_tess_2021}
{Paegert}, M., {Stassun}, K.~G., {Collins}, K.~A., {et~al.} 2021, arXiv
  e-prints, arXiv:2108.04778

\bibitem[{Parc {et~al.}(2024)Parc, Bouchy, Venturini, Dorn, \&
  Helled}]{parc_super-earths_2024}
Parc, L., Bouchy, F., Venturini, J., Dorn, C., \& Helled, R. 2024, \aap, 688,
  A59

\bibitem[{Pass {et~al.}(2023)Pass, Winters, Charbonneau, Irwin, Latham,
  Berlind, Calkins, Esquerdo, \& Mink}]{pass_mid--late_2023}
Pass, E.~K., Winters, J.~G., Charbonneau, D., {et~al.} 2023, \aj, 166, 11

\bibitem[{Pecaut \& Mamajek(2013)}]{pecaut_intrinsic_2013}
Pecaut, M.~J. \& Mamajek, E.~E. 2013, \apjs, 208, 9

\bibitem[{Pinamonti {et~al.}(2022)Pinamonti, Sozzetti, Maldonado, Affer,
  Micela, Bonomo, Lanza, Perger, Ribas, González~Hernández, Bignamini,
  Claudi, Covino, Damasso, Desidera, Giacobbe, González-Álvarez, Herrero,
  Leto, Maggio, Molinari, Morales, Pagano, Petralia, Piotto, Poretti, Rebolo,
  Scandariato, Suárez~Mascareño, Toledo-Padrón, \&
  Zanmar~Sánchez}]{pinamonti_hades_2022}
Pinamonti, M., Sozzetti, A., Maldonado, J., {et~al.} 2022, \aap, 664, A65

\bibitem[{Pollacco {et~al.}(2006)Pollacco, Skillen, Cameron, Christian,
  Hellier, Irwin, Lister, Street, West, Anderson, Clarkson, Deeg, Enoch, Evans,
  Fitzsimmons, Haswell, Hodgkin, Horne, Kane, Keenan, Maxted, Norton, Osborne,
  Parley, Ryans, Smalley, Wheatley, \& Wilson}]{pollacco_wasp_2006}
Pollacco, D., Skillen, I., Cameron, A., {et~al.} 2006, \pasp, 118, 1407

\bibitem[{Pollack {et~al.}(1996)Pollack, Hubickyj, Bodenheimer, Lissauer,
  Podolak, \& Greenzweig}]{pollack_formation_1996}
Pollack, J.~B., Hubickyj, O., Bodenheimer, P., {et~al.} 1996, Icarus, 124, 62

\bibitem[{Prša {et~al.}(2016)Prša, Harmanec, Torres, Mamajek, Asplund,
  Capitaine, Christensen-Dalsgaard, Depagne, Haberreiter, Hekker, Hilton, Kopp,
  Kostov, Kurtz, Laskar, Mason, Milone, Montgomery, Richards, Schmutz, Schou,
  \& Stewart}]{prsa_nominal_2016}
Prša, A., Harmanec, P., Torres, G., {et~al.} 2016, \aj, 152, 41

\bibitem[{Queloz {et~al.}(2001)Queloz, Mayor, Udry, Burnet, Carrier,
  Eggenberger, Naef, Santos, Pepe, Rupprecht, Avila, Baeza, Benz, Bertaux,
  Bouchy, Cavadore, Delabre, Eckert, Fischer, Fleury, Gilliotte, Goyak, Guzman,
  Kohler, Lacroix, Lizon, Megevand, Sivan, Sosnowska, \&
  Weilenmann}]{queloz_coralie_2001}
Queloz, D., Mayor, M., Udry, S., {et~al.} 2001, The Messenger, 105, 1

\bibitem[{Ribas {et~al.}(2023)Ribas, Reiners, Zechmeister, Caballero, Morales,
  Sabotta, Baroch, Amado, Quirrenbach, Abril, Aceituno, Anglada-Escudé,
  Azzaro, Barrado, Béjar, Benítez De~Haro, Bergond, Bluhm, Calvo~Ortega,
  Cardona~Guillén, Chaturvedi, Cifuentes, Colomé, Cont, Cortés-Contreras,
  Czesla, Díez-Alonso, Dreizler, Duque-Arribas, Espinoza, Fernández,
  Fuhrmeister, Galadí-Enríquez, García-López, González-Álvarez,
  González~Hernández, Guenther, De~Guindos, Hatzes, Henning, Herrero, Hintz,
  Huelmo, Jeffers, Johnson, De~Juan, Kaminski, Kemmer, Khaimova, Khalafinejad,
  Kossakowski, Kürster, Labarga, Lafarga, Lalitha, Lampón, Lillo-Box, Lodieu,
  López~González, López-Puertas, Luque, Magán, Mancini, Marfil, Martín,
  Martín-Ruiz, Molaverdikhani, Montes, Nagel, Nortmann, Nowak, Pallé,
  Passegger, Pavlov, Pedraz, Perdelwitz, Perger, Ramón-Ballesta, Reffert,
  Revilla, Rodríguez, Rodríguez-López, Sadegi, Sánchez~Carrasco,
  Sánchez-López, Sanz-Forcada, Schäfer, Schlecker, Schmitt, Schöfer,
  Schweitzer, Seifert, Shan, Skrzypinski, Solano, Stahl, Stangret, Stock,
  Stürmer, Tabernero, Tal-Or, Trifonov, Vanaverbeke, Yan, \&
  Zapatero~Osorio}]{ribas_carmenes_2023}
Ribas, I., Reiners, A., Zechmeister, M., {et~al.} 2023, \aap, 670, A139

\bibitem[{Ricker {et~al.}(2014)Ricker, Winn, Vanderspek, Latham, Bakos, Bean,
  Berta-Thompson, Brown, Buchhave, Butler, Butler, Chaplin, Charbonneau,
  Christensen-Dalsgaard, Clampin, Deming, Doty, De~Lee, Dressing, Dunham, Endl,
  Fressin, Ge, Henning, Holman, Howard, Ida, Jenkins, Jernigan, Johnson,
  Kaltenegger, Kawai, Kjeldsen, Laughlin, Levine, Lin, Lissauer, MacQueen,
  Marcy, McCullough, Morton, Narita, Paegert, Palle, Pepe, Pepper, Quirrenbach,
  Rinehart, Sasselov, Sato, Seager, Sozzetti, Stassun, Sullivan, Szentgyorgyi,
  Torres, Udry, \& Villasenor}]{ricker_transiting_2014}
Ricker, G.~R., Winn, J.~N., Vanderspek, R., {et~al.} 2014, JATIS, 1, 014003

\bibitem[{Riello {et~al.}(2021)Riello, De~Angeli, Evans, Montegriffo, Carrasco,
  Busso, Palaversa, Burgess, Diener, Davidson, Rowell, Fabricius, Jordi,
  Bellazzini, Pancino, Harrison, Cacciari, Van~Leeuwen, Hambly, Hodgkin,
  Osborne, Altavilla, Barstow, Brown, Castellani, Cowell, De~Luise, Gilmore,
  Giuffrida, Hidalgo, Holland, Marinoni, Pagani, Piersimoni, Pulone, Ragaini,
  Rainer, Richards, Sanna, Walton, Weiler, \& Yoldas}]{riello_gaia_2021}
Riello, M., De~Angeli, F., Evans, D.~W., {et~al.} 2021, \aap, 649, A3

\bibitem[{Safonov {et~al.}(2017)Safonov, Lysenko, \&
  Dodin}]{safonov_speckle_2017}
Safonov, B.~S., Lysenko, P.~A., \& Dodin, A.~V. 2017, Astronomy Letters, 43,
  344

\bibitem[{Santos {et~al.}(2004)Santos, Israelian, \&
  Mayor}]{santos_spectroscopic_2004}
Santos, N.~C., Israelian, G., \& Mayor, M. 2004, \aap, 415, 1153

\bibitem[{Santos {et~al.}(2002)Santos, Mayor, Naef, Pepe, Queloz, Udry, Burnet,
  Clausen, Helt, Olsen, \& Pritchard}]{santos_coralie_2002}
Santos, N.~C., Mayor, M., Naef, D., {et~al.} 2002, \aap, 392, 215

\bibitem[{Santos {et~al.}(2013)Santos, Sousa, Mortier, Neves, Adibekyan,
  Tsantaki, Delgado~Mena, Bonfils, Israelian, Mayor, \&
  Udry}]{santos_sweet-cat_2013}
Santos, N.~C., Sousa, S.~G., Mortier, A., {et~al.} 2013, \aap, 556, A150

\bibitem[{Sarkis {et~al.}(2021)Sarkis, Mordasini, Henning, Marleau, \&
  Mollière}]{sarkis_evidence_2021}
Sarkis, P., Mordasini, C., Henning, T., Marleau, G.~D., \& Mollière, P. 2021,
  \aap, 645, A79

\bibitem[{Saumon {et~al.}(1995)Saumon, Chabrier, \& van
  Horn}]{saumon_equation_1995}
Saumon, D., Chabrier, G., \& van Horn, H.~M. 1995, \apjs, 99, 713

\bibitem[{Schlegel {et~al.}(1998)Schlegel, Finkbeiner, \&
  Davis}]{schlegel_maps_1998}
Schlegel, D.~J., Finkbeiner, D.~P., \& Davis, M. 1998, \aj, 500, 525

\bibitem[{Seager \& Mallen‐Ornelas(2003)}]{seager_unique_2003}
Seager, S. \& Mallen‐Ornelas, G. 2003, \aj, 585, 1038

\bibitem[{Sestovic {et~al.}(2018)Sestovic, Demory, \&
  Queloz}]{sestovic_investigating_2018}
Sestovic, M., Demory, B.-O., \& Queloz, D. 2018, \aap, 616, A76

\bibitem[{Smith {et~al.}(2012)Smith, Stumpe, Van~Cleve, Jenkins, Barclay,
  Fanelli, Girouard, Kolodziejczak, McCauliff, Morris, \&
  Twicken}]{smith_kepler_2012}
Smith, J.~C., Stumpe, M.~C., Van~Cleve, J.~E., {et~al.} 2012, \pasp, 124, 1000

\bibitem[{Sneden(1973)}]{sneden_carbon_1973}
Sneden, C.~A. 1973, {PhD} {Thesis}, University of Texas, Austin

\bibitem[{{Sousa}(2014)}]{sousa_aresmoog_2014}
{Sousa}, S.~G. 2014, in Determination of Atmospheric Parameters of B, ed.
  E.~{Niemczura}, B.~{Smalley}, \& W.~{Pych}, 297--310

\bibitem[{Sousa {et~al.}(2021)Sousa, Adibekyan, Delgado-Mena, Santos,
  Rojas-Ayala, Soares, Legoinha, Ulmer-Moll, Camacho, Barros, Demangeon, Hoyer,
  Israelian, Mortier, Tsantaki, \& Monteiro}]{sousa_sweet-cat_2021}
Sousa, S.~G., Adibekyan, V., Delgado-Mena, E., {et~al.} 2021, \aap, 656, A53

\bibitem[{Sousa {et~al.}(2015)Sousa, Santos, Adibekyan, Delgado-Mena, \&
  Israelian}]{sousa_ares_2015}
Sousa, S.~G., Santos, N.~C., Adibekyan, V., Delgado-Mena, E., \& Israelian, G.
  2015, \aap, 577, A67

\bibitem[{Sousa {et~al.}(2007)Sousa, Santos, Israelian, Mayor, \&
  Monteiro}]{sousa_new_2007}
Sousa, S.~G., Santos, N.~C., Israelian, G., Mayor, M., \& Monteiro, M. J. P.
  F.~G. 2007, \aap, 469, 783

\bibitem[{Speagle(2020)}]{speagle_dynesty_2020}
Speagle, J.~S. 2020, \mnras, 493, 3132

\bibitem[{Stassun {et~al.}(2017)Stassun, Collins, \&
  Gaudi}]{stassun_accurate_2017}
Stassun, K.~G., Collins, K.~A., \& Gaudi, B.~S. 2017, \aj, 153, 136

\bibitem[{Stassun {et~al.}(2018)Stassun, Corsaro, Pepper, \&
  Gaudi}]{stassun_empirical_2018}
Stassun, K.~G., Corsaro, E., Pepper, J.~A., \& Gaudi, B.~S. 2018, \aj, 155, 22

\bibitem[{Stassun {et~al.}(2019)Stassun, Oelkers, Paegert, Torres, Pepper, Lee,
  Collins, Latham, Muirhead, Chittidi, Rojas-Ayala, Fleming, Rose, Tenenbaum,
  Ting, Kane, Barclay, Bean, Brassuer, Charbonneau, Ge, Lissauer, Mann, McLean,
  Mullally, Narita, Plavchan, Ricker, Sasselov, Seager, Sharma, Shiao,
  Sozzetti, Stello, Vanderspek, Wallace, \& Winn}]{stassun_revised_2019}
Stassun, K.~G., Oelkers, R.~J., Paegert, M., {et~al.} 2019, \aj, 158, 138

\bibitem[{Stassun \& Torres(2016)}]{stassun_eclipsing_2016}
Stassun, K.~G. \& Torres, G. 2016, \aj, 152, 180

\bibitem[{Stassun \& Torres(2021)}]{stassun_parallax_2021}
Stassun, K.~G. \& Torres, G. 2021, \aj$\,$Letters, 907, L33

\bibitem[{Stumpe {et~al.}(2014)Stumpe, Smith, Catanzarite, Van~Cleve, Jenkins,
  Twicken, \& Girouard}]{stumpe_multiscale_2014}
Stumpe, M.~C., Smith, J.~C., Catanzarite, J.~H., {et~al.} 2014, \pasp, 126, 100

\bibitem[{Stumpe {et~al.}(2012)Stumpe, Smith, Van~Cleve, Twicken, Barclay,
  Fanelli, Girouard, Jenkins, Kolodziejczak, McCauliff, \&
  Morris}]{stumpe_kepler_2012}
Stumpe, M.~C., Smith, J.~C., Van~Cleve, J.~E., {et~al.} 2012, \pasp, 124, 985

\bibitem[{Thommes {et~al.}(2008)Thommes, Matsumura, \&
  Rasio}]{thommes_gas_2008}
Thommes, E.~W., Matsumura, S., \& Rasio, F.~A. 2008, Science, 321, 814

\bibitem[{Thompson(1990)}]{thompson_grain_1990}
Thompson, C.~V. 1990, Annual Review of Materials Research, 20, 245

\bibitem[{Tokovinin(2018)}]{tokovinin_ten_2018}
Tokovinin, A. 2018, \pasp, 130, 035002

\bibitem[{Torres {et~al.}(2010)Torres, Andersen, \&
  Giménez}]{torres_accurate_2010}
Torres, G., Andersen, J., \& Giménez, A. 2010, The \aap$\,$Review, 18, 67

\bibitem[{Torres {et~al.}(2008)Torres, Winn, \& Holman}]{torres_improved_2008}
Torres, G., Winn, J.~N., \& Holman, M.~J. 2008, \aj, 677, 1324

\bibitem[{Tsantaki {et~al.}(2013)Tsantaki, Sousa, Adibekyan, Santos, Mortier,
  \& Israelian}]{tsantaki_deriving_2013}
Tsantaki, M., Sousa, S.~G., Adibekyan, V.~Z., {et~al.} 2013, \aap, 555, A150

\bibitem[{Twicken {et~al.}(2018)Twicken, Catanzarite, Clarke, Girouard,
  Jenkins, Klaus, Li, McCauliff, Seader, Tenenbaum, Wohler, Bryson, Burke,
  Caldwell, Haas, Henze, \& Sanderfer}]{twicken_kepler_2018}
Twicken, J.~D., Catanzarite, J.~H., Clarke, B.~D., {et~al.} 2018, \pasp, 130,
  064502

\bibitem[{Ulmer-Moll {et~al.}(2022)Ulmer-Moll, Lendl, Gill, Villanueva, Hobson,
  Bouchy, Brahm, Dragomir, Grieves, Mordasini, Anderson, Acton, Bayliss,
  Bieryla, Burleigh, Casewell, Chaverot, Eigmüller, Feliz, Gaudi, Gillen,
  Goad, Gupta, Günther, Henderson, Henning, Jenkins, Jones, Jordán, Kendall,
  Latham, Mireles, Moyano, Nadol, Osborn, Pepper, Tala~Pinto, Psaridi, Queloz,
  Quinn, Rojas, Sarkis, Schlecker, Tilbrook, Torres, Trifonov, Udry, Vines,
  West, Wheatley, Yao, Zhao, \& Zhou}]{ulmer-moll_two_2022}
Ulmer-Moll, S., Lendl, M., Gill, S., {et~al.} 2022, \aap, 666, A46

\bibitem[{Vines {et~al.}(2019)Vines, Jenkins, Acton, Briegal, Bayliss, Bouchy,
  Belardi, Bryant, Burleigh, Cabrera, Casewell, Chaushev, Cooke, Csizmadia,
  Eigmüller, Erikson, Foxell, Gill, Gillen, Goad, Jackman, King, Louden,
  McCormac, Moyano, Nielsen, Pollacco, Queloz, Rauer, Raynard, Smith, Soto,
  Tilbrook, Titz-Weider, Turner, Udry, Walker, Watson, West, \&
  Wheatley}]{vines_ngts-6b_2019}
Vines, J.~I., Jenkins, J.~S., Acton, J.~S., {et~al.} 2019, \mnras, 489, 4125

\bibitem[{Vorobyov \& Basu(2008)}]{vorobyov_mass_2008}
Vorobyov, E.~I. \& Basu, S. 2008, \aj, 676, L139

\bibitem[{Wright {et~al.}(2012)Wright, Marcy, Howard, Johnson, Morton, \&
  Fischer}]{wright_frequency_2012}
Wright, J.~T., Marcy, G.~W., Howard, A.~W., {et~al.} 2012, \aj, 753, 160

\end{thebibliography}
\begin{acknowledgements}
      The authors thank the ESO staff at La Silla for operating and maintaining the instruments for so many years. A special thanks to Laurent Eyer for his assistance in accessing and inspecting the Gaia photometric data. This work has been carried out within the framework of the NCCR PlanetS supported by the Swiss National Science Foundation under grants 51NF40\_182901 and 51NF40\_205606. Funding for the \textit{TESS} mission is provided by NASA's Science Mission Directorate. We acknowledge the use of public \textit{TESS} data from pipelines at the \textit{TESS} Science Office and the \textit{TESS} Science Processing Operations Center. This research has made use of the Exoplanet Follow-up Observation program (ExoFOP; DOI: 10.26134/ExoFOP5) website, which is operated by the California Institute of Technology, under contract with the National Aeronautics and Space Administration under the Exoplanet Exploration program. Resources supporting this work were provided by the NASA High-End Computing (HEC) program through the NASA Advanced Supercomputing (NAS) Division at Ames Research Center for the production of the SPOC data products. This paper includes data collected by the \textit{TESS} mission that are publicly available from the Mikulski Archive for Space Telescopes (MAST). The research leading to these results has received funding from the ARC grant for Concerted Research Actions, financed by the Wallonia-Brussels Federation. TRAPPIST is funded by the Belgian Fund for Scientific Research (Fond National de la Recherche Scientifique, FNRS) under the grant PDR T.0120.21. Based in part on observations obtained at the Southern Astrophysical Research (SOAR) telescope, which is a joint project of the Minist\'{e}rio da Ci\^{e}ncia, Tecnologia e Inova\c{c}\~{o}es (MCTI/LNA) do Brasil, the US National Science Foundation’s NOIRLab, the University of North Carolina at Chapel Hill (UNC), and Michigan State University (MSU). KAC and CNW acknowledge support from the \textit{TESS} mission via subaward s3449 from MIT. SGS acknowledges the support from FCT through Investigador FCT contract nr. CEECIND/00826/2018 and POPH/FSE (EC). NCS acknowledges funding by the European Union (ERC, FIERCE, 101052347). Views and opinions expressed are however those of the author(s) only and do not necessarily reflect those of the European Union or the European Research Council. Neither the European Union nor the granting authority can be held responsible for them. This work was supported by FCT - Fundação para a Ciência e a Tecnologia through national funds by these grants: UIDB/04434/2020, UIDP/04434/2020. IAS acknowledges the support of M.V. Lomonosov Moscow State University program of Development. This work makes use of observations from the LCOGT network. The postdoctoral fellowship of KB is funded by F.R.S.-FNRS grant T.0109.20 and by the Francqui Foundation. MG and EJ are F.R.S.-FNRS Research Directors. This publication benefits from the support of the French Community of Belgium in the context of the FRIA Doctoral Grant awarded to MT. We acknowledge financial support from the Agencia Estatal de Investigaci\'on of the Ministerio de Ciencia e Innovaci\'on MCIN/AEI/10.13039/501100011033 and the ERDF “A way of making Europe” through project PID2021-125627OB-C32, and from the Centre of Excellence “Severo Ochoa” award to the Instituto de Astrofisica de Canarias. AP was financed by grants 02/140/RGJ24/0031 and BK 2025. FJP acknowledges financial support from the Severo Ochoa grant CEX2021-001131- S MICIU/AEI/10.13039/501100011033 and Ministerio de Ciencia e Innovación through the project PID2022-137241NB-C43. MS acknowledges financial support from the Swiss National Science Foundation (SNSF) for project 200021\_200726.
\end{acknowledgements}

\appendix	
\onecolumn
\section{Speckle Interferometry Images}
\label{subsec:DirectImaging}

\begin{figure}[H]
    \centering
    \begin{subfigure}[b]{0.45\linewidth}
        \includegraphics[width=\linewidth]{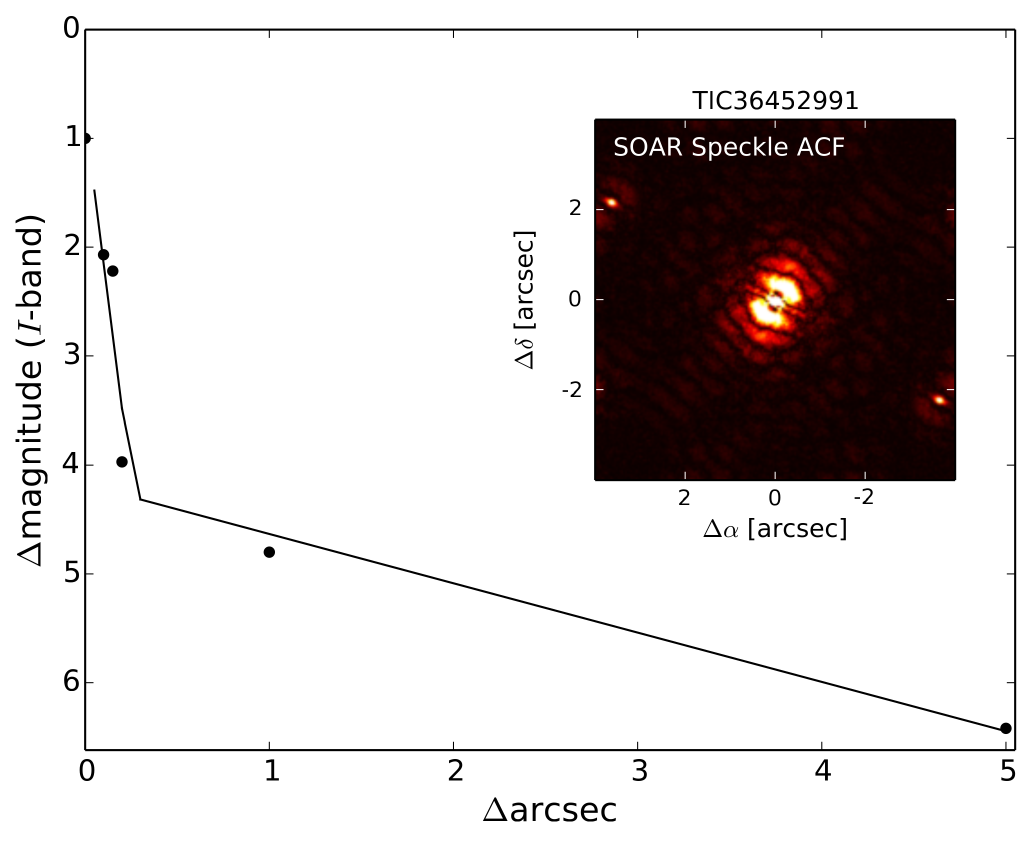}
        \caption{TOI-2969, UTC 2022 March 22}
    \end{subfigure}
    
    \begin{subfigure}[b]{0.45\linewidth}
        \includegraphics[width=\linewidth]{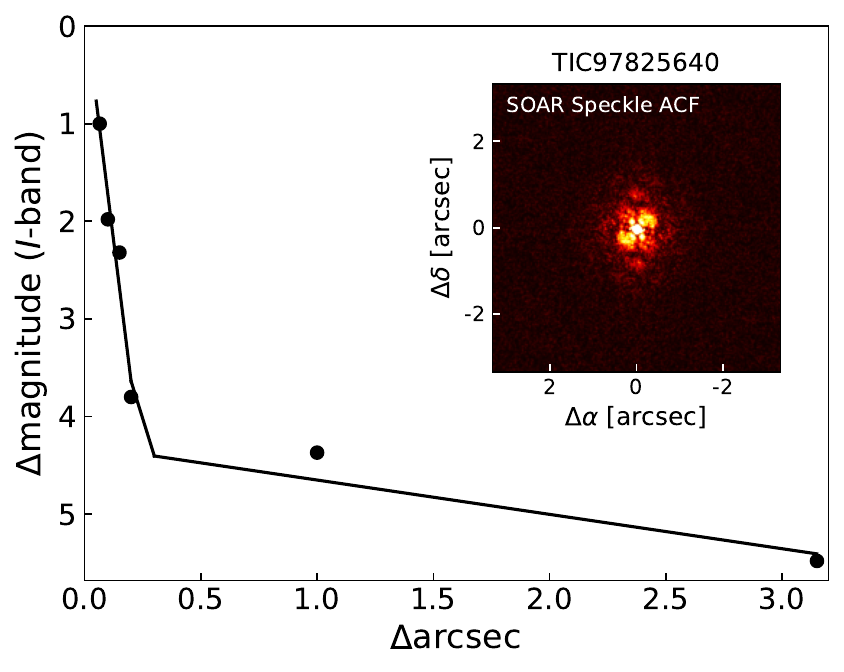}
        \caption{TOI-2989, UTC 2022 July 10}
    \end{subfigure}
    \begin{subfigure}[b]{0.45\linewidth}
        \includegraphics[width=\linewidth]{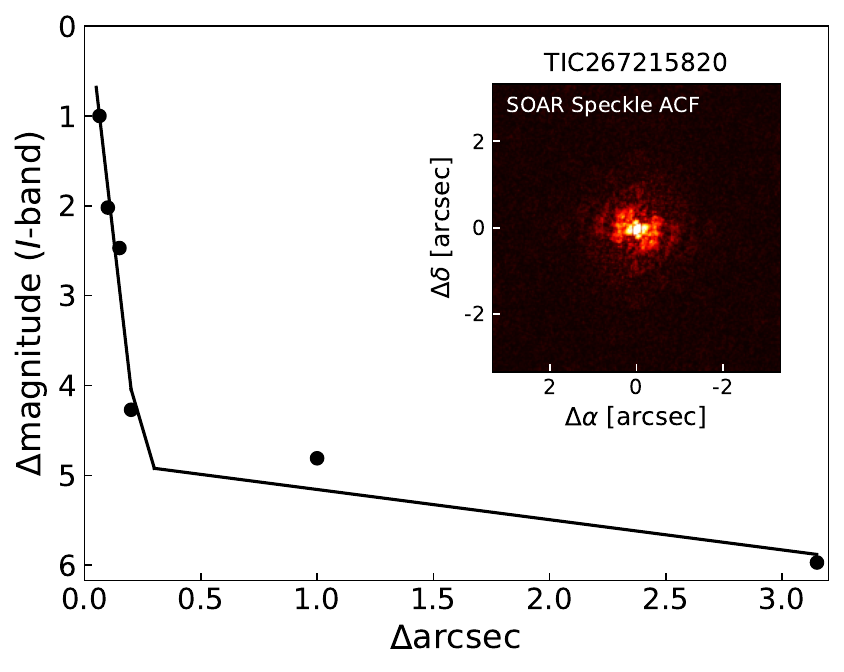}
        \caption{TOI-5300, UTC 2022 July 10}
    \end{subfigure}
    \caption{Speckle observations from SOAR.}
    \label{fig:SOAR}
\end{figure}

\begin{figure}[H]\centering
    \includegraphics[width=.49\linewidth]{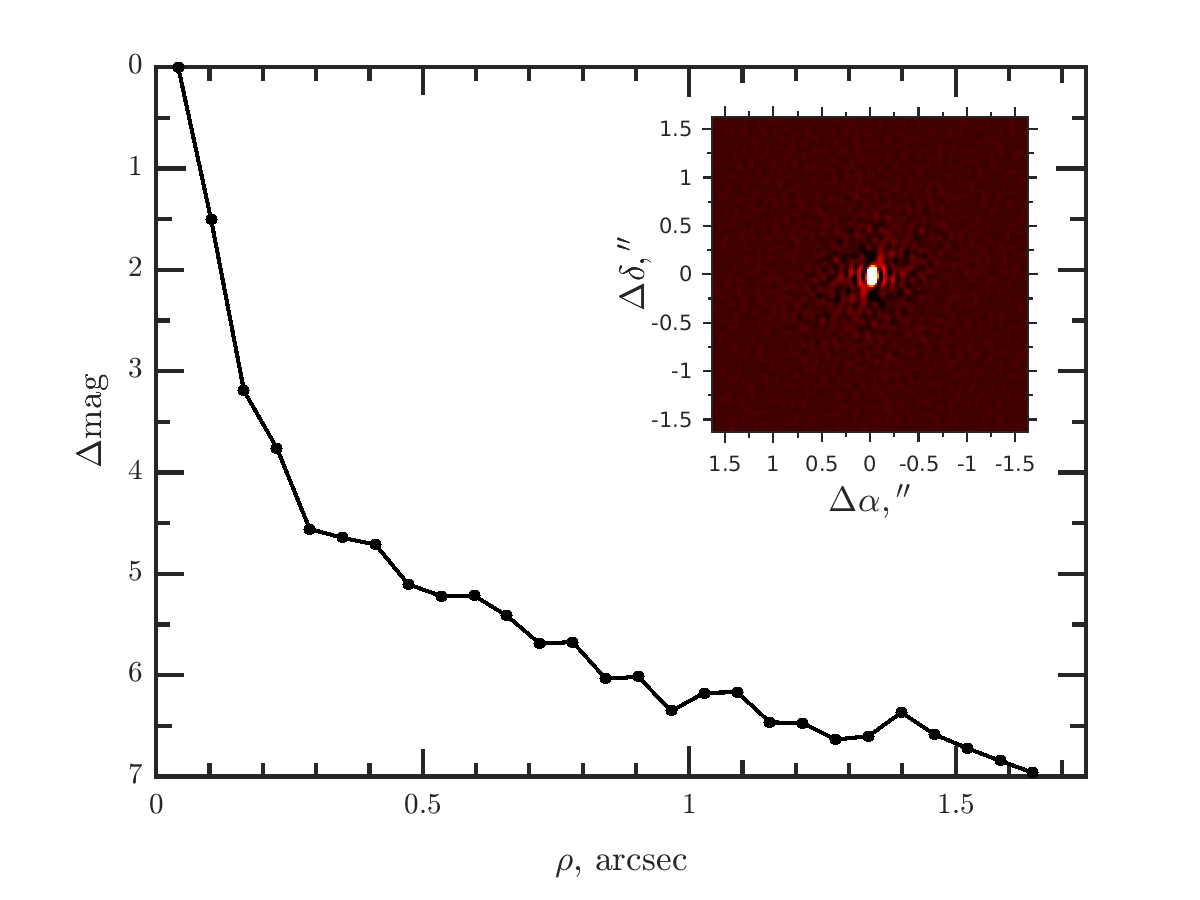}
    \caption{SAI speckle interferometry - TOI-5300, UTC 2023 September 30}
    \label{fig:SAI}
\end{figure}
\FloatBarrier

\newpage
\section{SED analysis}
\begin{figure}[H]
\centering
\begin{subfigure}[b]{0.49\linewidth}
    \includegraphics[width=\linewidth]{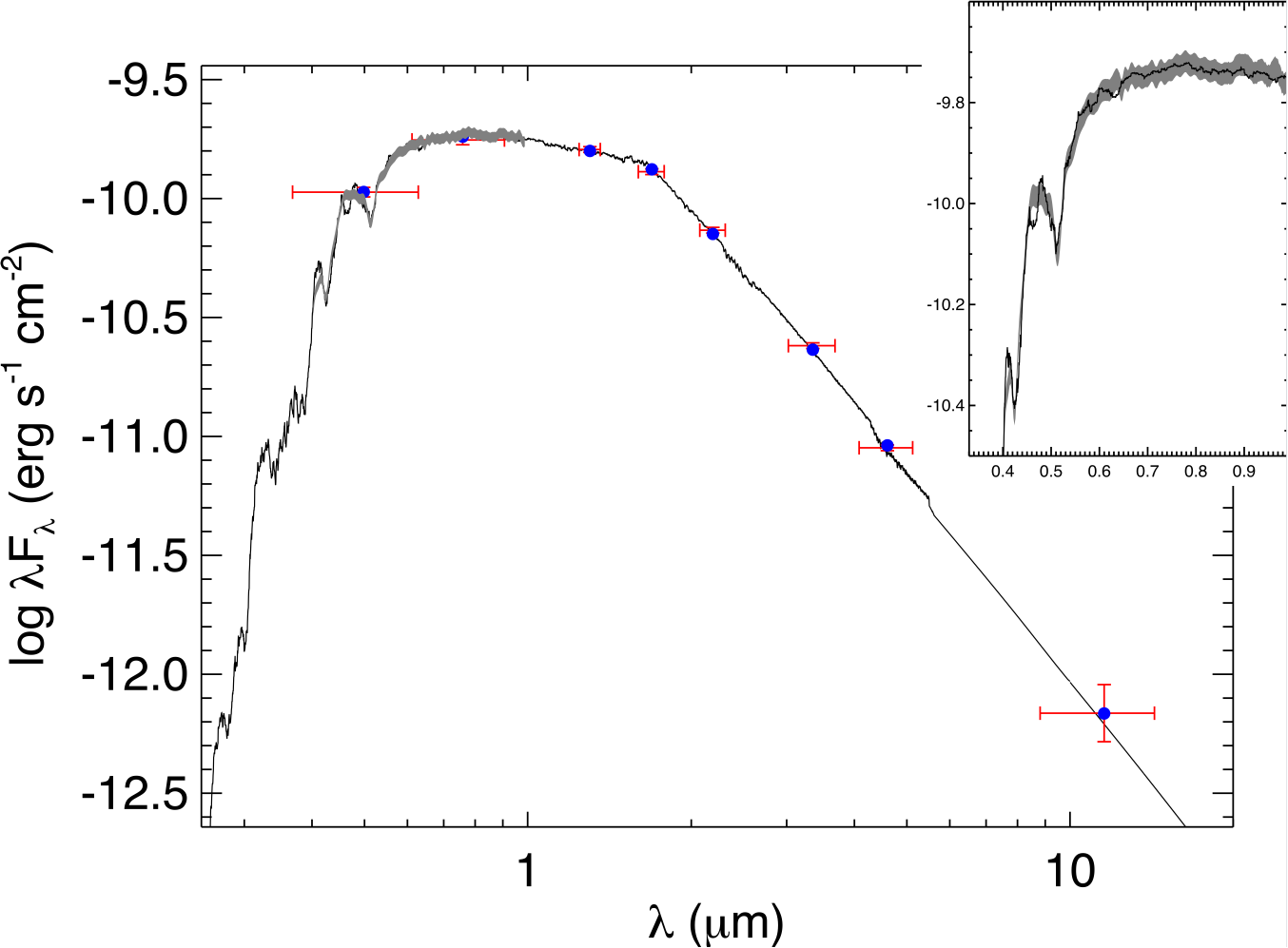}
    \caption{TOI-2969}
\end{subfigure}
\begin{subfigure}[b]{0.49\linewidth}
    \includegraphics[width=\linewidth]{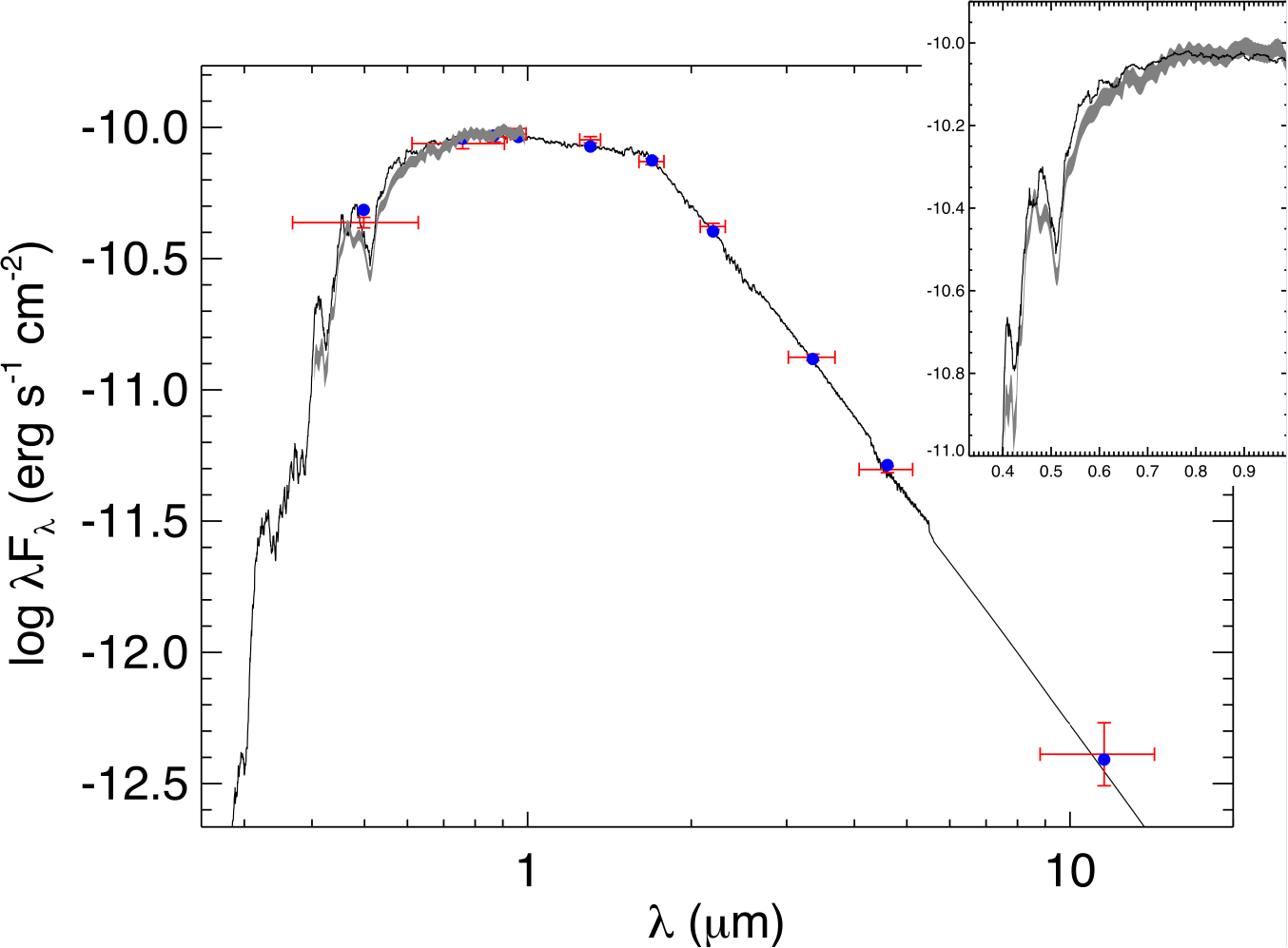}
    \caption{TOI-2989}
\end{subfigure}

\begin{subfigure}[b]{0.49\linewidth}
    \includegraphics[width=\linewidth]{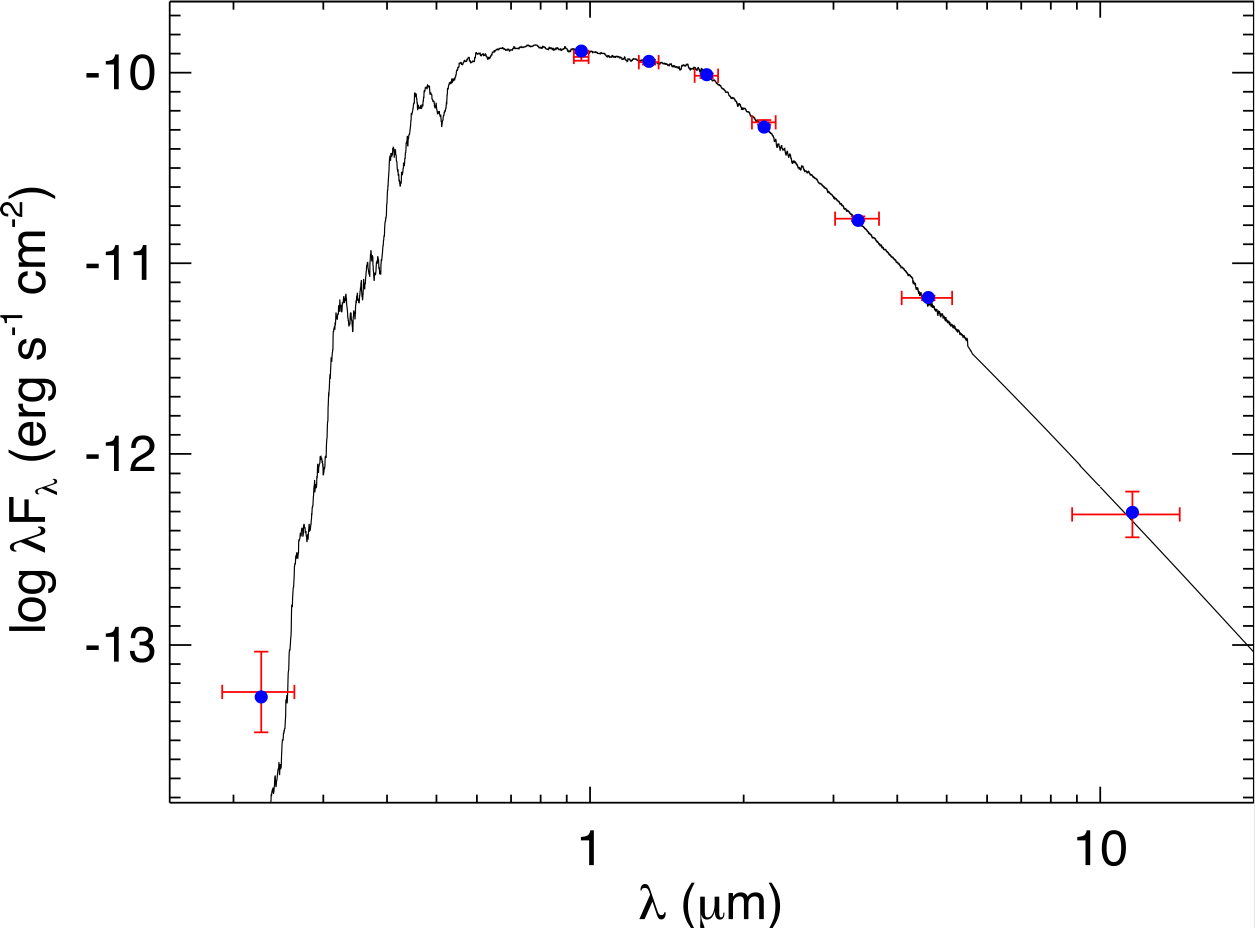}
    \caption{TOI-5300}
\end{subfigure}
\caption{The spectral energy distributions (SEDs). Red symbols represent the observed photometric measurements, and the horizontal bars represent the effective width of the passband. Blue symbols are the model fluxes from the best-fit PHOENIX atmosphere model (black). The insets show the absolute flux-calibrated \textit{Gaia} spectrophotometry as a gray swathe overlaid on the model (black). \label{fig:sed}}
\end{figure}
\FloatBarrier

\newpage
\section{\textit{Gaia} DR3 Photometry}
\label{app:GaiaPhotometry}
\begin{figure}[H]
    \centering
    \begin{subfigure}[b]{0.5\linewidth}
    \includegraphics[width=\linewidth]{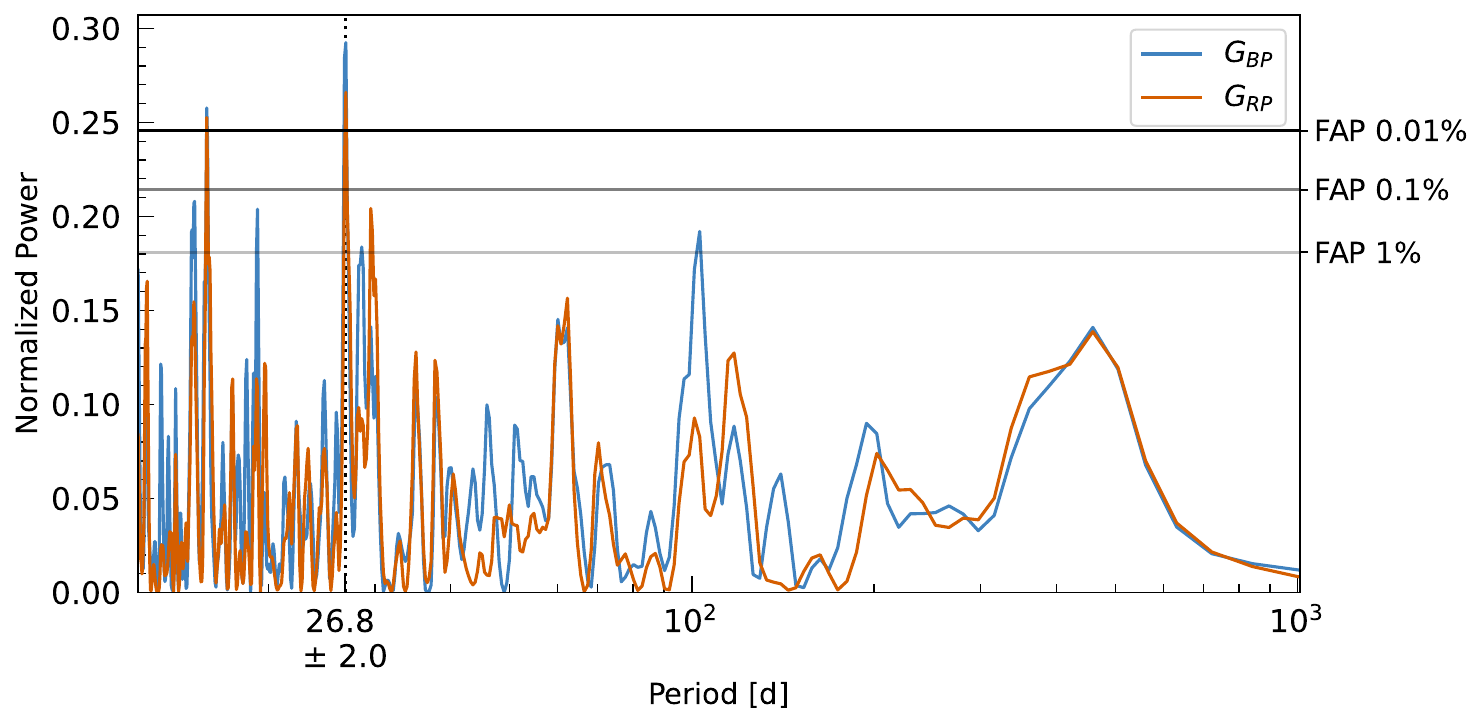}
    \caption{TOI-2969}
    \end{subfigure}
    \begin{subfigure}[b]{0.5\linewidth}
    \includegraphics[width=\linewidth]{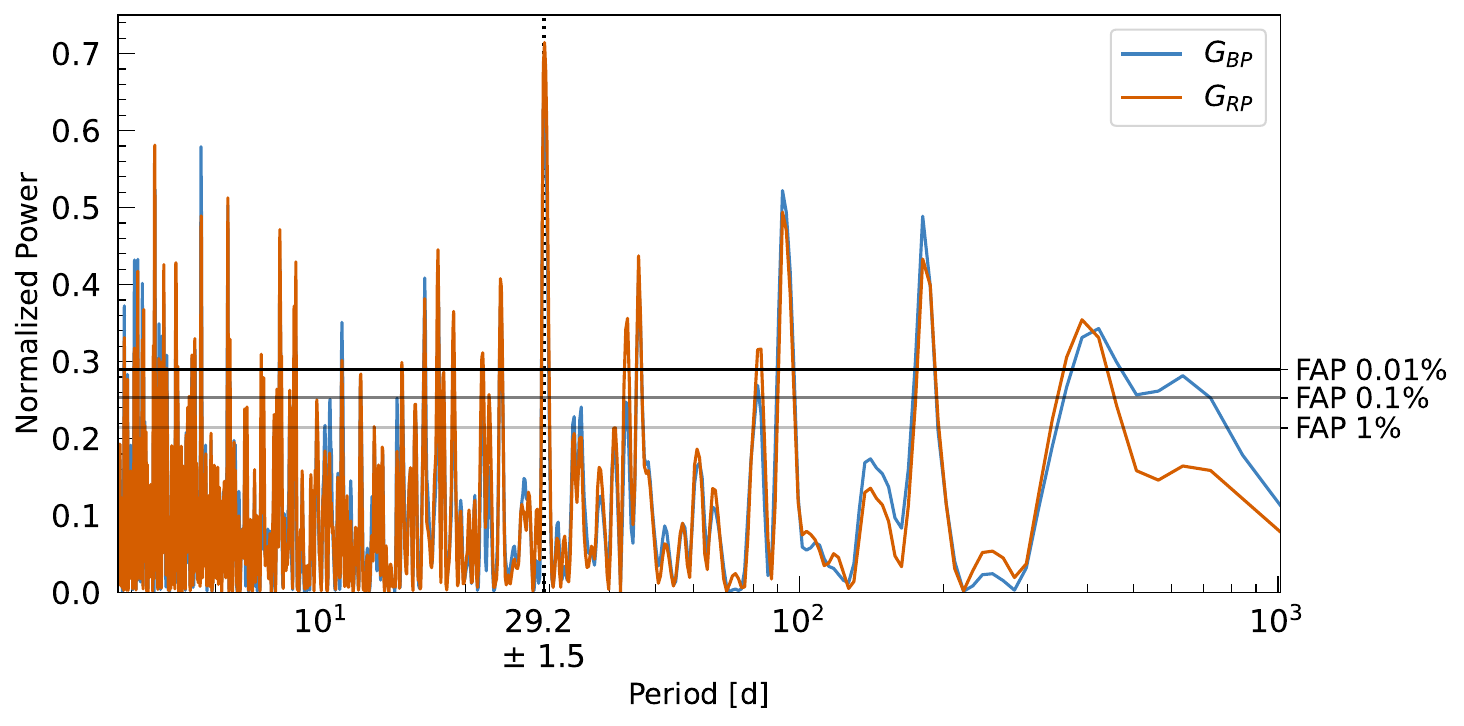}
    \caption{TOI-2989}
    \end{subfigure}
    
    \caption{The Lomb-Scargle periodogram of the \textit{Gaia} DR3 photometric observations. The observations are filtered based on flags for photometry and variability. The False Alarm Probability (FAP) levels are overplotted for the combined data sets.}
\end{figure}

\section{WASP Periodograms}
\label{app:wasp_period}
\begin{figure}[H]
	\centering
    \begin{subfigure}[b]{0.35\linewidth}
    \includegraphics[width=\linewidth]{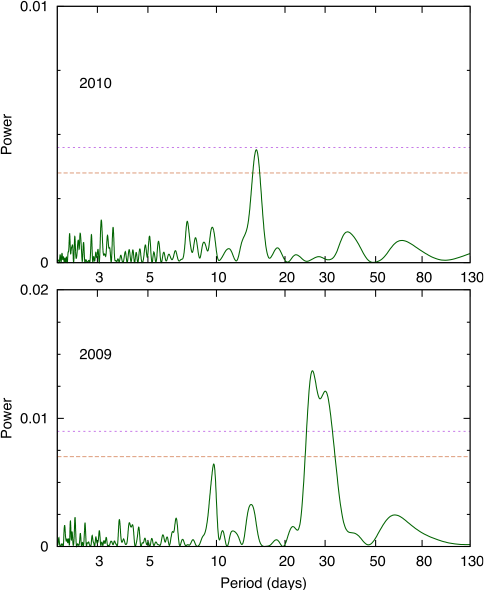}
    \caption{TOI-2989}
    \end{subfigure}
    \begin{subfigure}[b]{0.35\linewidth}
    \includegraphics[width=\linewidth]{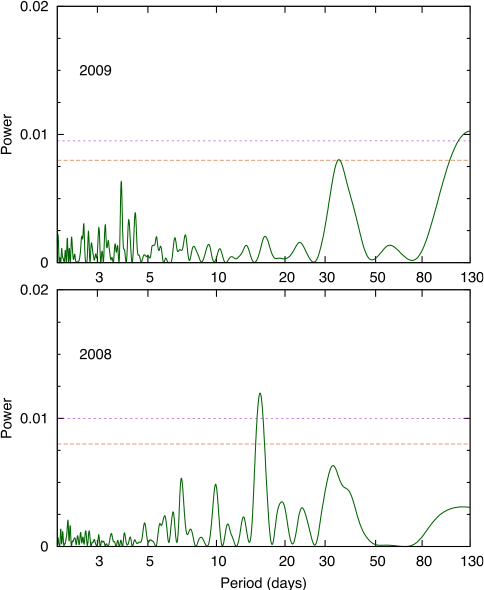}
    \caption{TOI-5300}
    \end{subfigure}
  \caption{Periodograms of the WASP lightcurves. The dashed horizontal lines show the estimated 10\%\- and 1\%-likelihood false-alarm levels.}
\label{fig:wasp}
\end{figure}
\FloatBarrier

\newpage
\section{Joint modeling priors}
\label{app:priors}
\begin{table}[H]
    \caption{Priors for the joint modeling of photometric and RV data.}
    \label{table:priors}
    \centering
    \begin{tabular}{l l l  c c c}
    \hline\hline
    \multicolumn{2}{l}{Parameter} & Distribution  & TOI-2969 b & TOI-2989 b & TOI-5300 b \\ \hline
    $P$ & [days]  & Uniform & (0, 2) & (2, 4) & (1, 3)\\
    $T_0$ & [BJD]  & Normal & (2459303, 1) & (2459302, 1) & (2459470, 1)\\
    $R_\mathrm{pl}/R_{\star}$ &  & Uniform & (0, 1) & (0, 1) & (0, 1)\\
    $b$ &  & Uniform & (0, 1) & (0, 1) & (0, 1)\\
    $\rho_{\star}$ & [kg m$^{-3}$]  & Normal & (2136, 618) & (2092, 660) & (2345, 773)\\
    $q_\mathrm{1, TESS^{(a)}}$ &  & Normal & (0.43, 0.02) & (0.45, 0.02) & (0.45, 0.02)\\
    $q_\mathrm{2, TESS^{(a)}}$ &  & Normal & (0.37, 0.03) & (0.39, 0.03) & (0.38, 0.03)\\
    $q_\mathrm{1, phot}$ & & Normal & (0,1) & (0,1) & (0,1)\\
    $q_\mathrm{2, phot}$ & & Normal & (0,1) & (0,1) & (0,1)\\
    $e$ &  & Fixed & 0 & 0 & 0\\
    $\omega$ & [$^{\circ}$]  & Fixed & 90 & 90 & 90\\
    $K$ & [km s$^{-1}$]  & Uniform & (0, 100) & (0, 100) & (0, 100)\\
    $\gamma_\mathrm{TESS}$ &  & Normal & (0.0, 0.1) & (0.0, 0.1) & (0.0, 0.1)\\
    $\sigma_\mathrm{TESS}$ & [ppm]  & Loguniform & (0.1, 1000) & (0.1, 1000) & (0.1, 1000)\\
    $\gamma_\mathrm{phot}$ & & Normal & (0.0, 0.1) & (0.0, 0.1) & (0.0, 0.1) \\
    $\sigma_\mathrm{phot}$ & [ppm] & Loguniform & (0.1, 1000) & (0.1, 1000) & (0.1, 1000) \\
    $\theta_{0,\mathrm{phot}}$ & & Uniform & (-100, 100) & (-100, 100) & (-100, 100)\\
    $\gamma_\mathrm{CORALIE}$ & [km s$^{-1}$]  & Uniform & (-100, 100) & (-100, 100) & (-100, 100)\\
    $\sigma_\mathrm{CORALIE}$ & [km s$^{-1}$]  & Loguniform & (0.001, 100) & (0.001, 100) & (0.001, 100)\\ 
    \hline
    \end{tabular}
\begin{flushleft}
\textbf{Notes:} All ground-based photometric observations have the same priors for the limb darkening coefficients ($q_\mathrm{1,phot}$ and $q_\mathrm{2,phot}$), the offset relative flux ($\gamma_\mathrm{phot}$), the jitter ($\sigma_\mathrm{phot}$), and the linear regressor ($\theta_\mathrm{0,phot}$). A dilution factor was included only for TOI-2969, using a uniform prior from 0 to 1 for the ground-based photometry and the QLP lightcurves, and fixed to 1 for the other two targets and for the \textit{TESS}-SPOC lightcurves.\\
$^{(a)}$ For TOI-2969 the $q_1$ and $q_2$ priors are shared between \textit{TESS} and \textit{QLP}, as it should be the same.
\end{flushleft}
\end{table}
\FloatBarrier

\newpage
\section{Limb darkening and photometric instrumental parameters}
\label{app:limb_instrument_table}
\begin{multicols}{2}
\begin{table}[H]
\caption{Fitted limb darkening parameters for the companions presented in this paper.}
\label{table:limb_darkening}
\centering
\begin{tabular}{l l c c c}
\hline\hline
\multicolumn{2}{l}{} &  TOI-2969 b & TOI-2989 b & TOI-5300 b \\ \hline
\multirow{2}{*}{\footnotesize{$\circ\,\,$TESS}} & & & \\ 
	$\quad q_\mathrm{1, TESS}$ &  & $0.45\pm 0.02$  & $0.46\pm 0.02$  & $0.45\pm 0.02$ \\
    	$\quad q_\mathrm{2, TESS}$ &  & $0.37_{-0.03}^{+0.02}$  & $0.40\pm 0.03$  & $0.38\pm 0.03$ \\
\multirow{2}{*}{\footnotesize{$\circ\,\,$QLP}} & & & \\
	$\quad q_\mathrm{1, QLP}$ &  & $0.42\pm 0.02$  &  & \\
    	$\quad q_\mathrm{2, QLP}$ &  & $0.37\pm 0.02$  &  & \\
\multirow{2}{*}{\footnotesize{$\circ\,\,$El Sauce (R)}} & & & \\
	$\quad q_\mathrm{1, R_1}$ &  & $0.42_{-0.09}^{+0.12}$  & $0.65_{-0.09}^{+0.13}$  & \\
   	$\quad q_\mathrm{2, R_1}$ &  & $0.6\pm 0.3$  & $0.8\pm 0.1$  & \\
\multirow{2}{*}{\footnotesize{$\circ\,\,$PEST (R)}} & & & \\
	$\quad q_\mathrm{1, R_2}$ &  & $0.7\pm 0.2$  &  & \\
    	$\quad q_\mathrm{2, R_2}$ &  & $0.4\pm 0.2$  &  & \\
\multirow{2}{*}{\footnotesize{$\circ\,\,$Brierfield (R)}} & & & \\
	$\quad q_\mathrm{1, R_3}$ &  &  &  & $0.3_{-0.2}^{+0.3}$ \\
    	$\quad q_\mathrm{2, R_3}$ &  &  &  & $0.2_{-0.2}^{+0.3}$ \\
\multirow{2}{*}{\footnotesize{$\circ\,\,$LCO-CTIO ($i'$)}} & & & \\
	$\quad q_{1, i'_1}$ &  & $0.2_{-0.1}^{+0.2}$  &  & \\
    	$\quad q_{2, i'_1}$ &  & $0.4\pm 0.3$  &  & \\
\multirow{2}{*}{\footnotesize{$\circ\,\,$LCO-CTIO ($g'$)}} & & & \\
	$\quad q_{1, g'_1}$ &  &  &  & $0.7_{-0.3}^{+0.2}$ \\
    	$\quad q_{2, g'_1}$ &  &  &  & $0.6_{-0.3}^{+0.2}$ \\
\multirow{2}{*}{\footnotesize{$\circ\,\,$LCO-HAL ($i'$)}} & & & \\
	$\quad q_{1, i'_2}$ &  &  &  & $0.6\pm 0.2$ \\
    	$\quad q_{2, i'_2}$ &  &  &  & $0.3_{-0.2}^{+0.3}$ \\
\multirow{2}{*}{\footnotesize{$\circ\,\,$LCO-HAL ($g'$)}} & & & \\
	$\quad q_{1, g'_2}$ &  &  &  & $0.6\pm 0.2$ \\
    	$\quad q_{2, g'_2}$ &  &  &  & $0.5\pm 0.2$ \\
\multirow{2}{*}{\footnotesize{$\circ\,\,$LCO-SAAO ($g'$)}} & & & \\
	$\quad q_{1, g'_3}$ &  & $0.8_{-0.2}^{+0.1}$  &  & \\
    	$\quad q_{2, g'_3}$ &  & $0.5\pm 0.3$  &  & \\
\multirow{2}{*}{\footnotesize{$\circ\,\,$TRAPPIST ($z'$)}} & & & \\
	$\quad q_{1, z'}$ &  & $0.4\pm 0.1$  &  & \\
    	$\quad q_{2, z'}$ &  & $0.3\pm 0.2$  &  & \\
\multirow{2}{*}{\footnotesize{$\circ\,\,$TRAPPIST (I+z)}} & & & \\
	$\quad q_\mathrm{1, I+z}$ &  &  & $0.5\pm 0.2$  & \\
    	$\quad q_\mathrm{2, I+z}$ &  &  & $0.2_{-0.1}^{+0.2}$  & \\
\multirow{2}{*}{\footnotesize{$\circ\,\,$TRAPPIST (V)}} & & & \\
	$\quad q_\mathrm{1, V}$ &  &  &  & $0.8\pm 0.2$ \\
    	$\quad q_\mathrm{2, V}$ &  &  &  & $0.6_{-0.1}^{+0.2}$ \\
\multirow{2}{*}{\footnotesize{$\circ\,\,$TRAPPIST (B)}} & & & \\
	$\quad q_\mathrm{1, B_1}$ &  & $0.8\pm 0.1$  & $0.8\pm 0.1$  & \\
    	$\quad q_\mathrm{2, B_1}$ &  & $0.6\pm 0.2$  & $0.8\pm 0.1$  & \\
\multirow{2}{*}{\footnotesize{$\circ\,\,$SUTO (B)}} & & & \\
	$\quad q_\mathrm{1, B_2}$ &  & $0.8\pm 0.1$  &  & \\
    	$\quad q_\mathrm{2, B_2}$ &  & $0.5\pm 0.2$  &  & \\ \hline
\end{tabular}
\end{table}

\begin{table}[H]
\caption{Fitted photometric instrumental parameters for the companions presented in this paper.}
\label{table:photometric_instrument}
\centering
\begin{tabular}{l l c c c}
\hline\hline
\multicolumn{1}{l}{} &  & TOI-2969 b & TOI-2989 b & TOI-5300 b \\ \hline
$\gamma_\mathrm{TESS}$ & [$\times10^{-6}$]  & $-33_{-27}^{+26}$  & $-26_{-31}^{+28}$  & $-30\pm 39$ \\
$\sigma_\mathrm{TESS}$ & [ppm]  & $4.5_{-4.0}^{+33.3}$  & $3.9_{-3.6}^{+33.3}$  & $3.5_{-3.1}^{+33.1}$  \\
$\gamma_\mathrm{QLP}$ & [$\times10^{-6}$]  & $-135_{-19}^{+20}$  &  & \\
$\sigma_\mathrm{QLP}$ & [ppm]  & $81_{-75}^{+167}$  &  &  \\
$\gamma_\mathrm{R_1}$ & [$\times10^{-3}$]  & $-53_{-8}^{+9}$  & $-11\pm 2$  & \\
$\sigma_\mathrm{R_1}$ & [ppm]  & $849_{-245}^{+110}$  & $172_{-171}^{+728}$  & \\
$\theta_\mathrm{0,R_1}$ & [$\times10^{-3}$]  & $-47_{-7}^{+8}$  & $4\pm 2$  & \\
$\gamma_\mathrm{R_2}$ & [$\times10^{-3}$]  & $2\pm 1$  &  & \\
$\sigma_\mathrm{R_2}$ & [ppm]  & $18_{-16}^{+183}$  &  & \\
$\theta_\mathrm{0,R_2}$ & [$\times10^{-3}$]  & $1.3_{-0.7}^{+0.8}$  &  & \\
$\gamma_\mathrm{R_3}$ & [$\times10^{-3}$]  &  &  & $2.8\pm 1.0$ \\
$\sigma_\mathrm{R_3}$ & [ppm]  &  &  & $9_{-9}^{+200}$ \\
$\theta_\mathrm{0,R_3}$ & [$\times10^{-3}$]  &  &  & $1.7\pm 0.5$ \\
$\gamma_{i'_1}$ & [$\times10^{-3}$]  & $22\pm 4$  &  & \\
$\sigma_{i'_1}$ & [ppm]  & $800_{-696}^{+160}$  &  & \\
$\theta_{0,i'_1}$ & [$\times10^{-3}$]  & $15\pm 2$  &  & \\
$\gamma_{g'_1}$ & [$\times10^{-3}$]  &  &  & $37_{-8}^{+7}$ \\
$\sigma_{g'_1}$ & [ppm]  &  &  & $16_{-15}^{+238}$ \\
$\theta_{0,g'_1}$ & [$\times10^{-3}$]  &  &  & $21_{-5}^{+4}$ \\
$\gamma_{i'_2}$ & [$\times10^{-3}$]  &  &  & $1\pm 2$ \\
$\sigma_{i'_2}$ & [ppm]  &  &  & $20_{-19}^{+580}$ \\
$\theta_{0,i'_2}$ & [$\times10^{-3}$]  &  &  & $1_{-1}^{+2}$ \\
$\gamma_{g'_2}$ & [$\times10^{-3}$]  &  &  & $24\pm 6$ \\
$\sigma_{g'_2}$ & [ppm]  &  &  & $689_{-684}^{+257}$ \\
$\theta_{0,g'_2}$ & [$\times10^{-3}$]  &  &  & $29\pm 5$ \\
$\gamma_{g'_3}$ & [$\times10^{-3}$]  & $21\pm 5$  &  & \\
$\sigma_{g'_3}$ & [ppm]  & $13_{-12}^{+147}$  &  & \\
$\theta_{0,g'_3}$ & [$\times10^{-3}$]  & $14\pm 3$  &  & \\
$\gamma_{z'}$ & [$\times10^{-3}$]  & $25\pm 1$  &  & \\
$\sigma_{z'}$ & [ppm]  & $704_{-698}^{+237}$  &  & \\
$\theta_{0,z'}$ & [$\times10^{-3}$]  & $14.6\pm 0.6$  &  & \\
$\gamma_\mathrm{I+z}$ & [$\times10^{-3}$]  &  & $-9_{-1}^{+2}$  & \\
$\sigma_\mathrm{I+z}$ & [ppm]  &  & $875_{-521}^{+97}$  & \\
$\theta_\mathrm{0,I+z}$ & [$\times10^{-3}$]  &  & $4.2_{-0.8}^{+0.9}$  & \\
$\gamma_\mathrm{V}$ & [$\times10^{-3}$]  &  &  & $-26_{-8}^{+9}$ \\
$\sigma_\mathrm{V}$ & [ppm]  &  &  & $5_{-5}^{+93}$ \\
$\theta_\mathrm{0,V}$ & [$\times10^{-3}$]  &  &  & $-23\pm 7$ \\
$\gamma_\mathrm{B_1}$ & [$\times10^{-3}$]  & $32_{-1}^{+2}$  & $3\pm 8$  & \\
$\sigma_\mathrm{B_1}$ & [ppm]  & $3_{-3}^{+77}$  & $9_{-9}^{+185}$  & \\
$\theta_\mathrm{0,B_1}$ & [$\times10^{-3}$]  & $19.5\pm 0.5$  & $13\pm 7$  & \\
$\gamma_\mathrm{B_2}$ & [$\times10^{-3}$]  & $2\pm 2$  &  & \\
$\sigma_\mathrm{B_2}$ & [ppm]  & $28_{-26}^{+207}$  &  & \\
$\theta_\mathrm{0,B_2}$ & [$\times10^{-3}$]  & $1.5_{-0.9}^{+1.0}$  &  & \\ \hline
\end{tabular}
\begin{flushleft}
\textbf{Notes:} Where $\gamma$ is the offset relative flux, $\sigma$ the jitter, and $\theta_0$ the linear regressor. The filter subscripts correspond to the instruments specified for the limb darkening coefficients in Table \ref{table:limb_darkening}. 
\end{flushleft}
\end{table}
\FloatBarrier
\end{multicols}

\newpage
\section{Corner plots}
\label{app:Corner}

\begin{figure}[H]
	\centering
	\includegraphics[width=\linewidth]{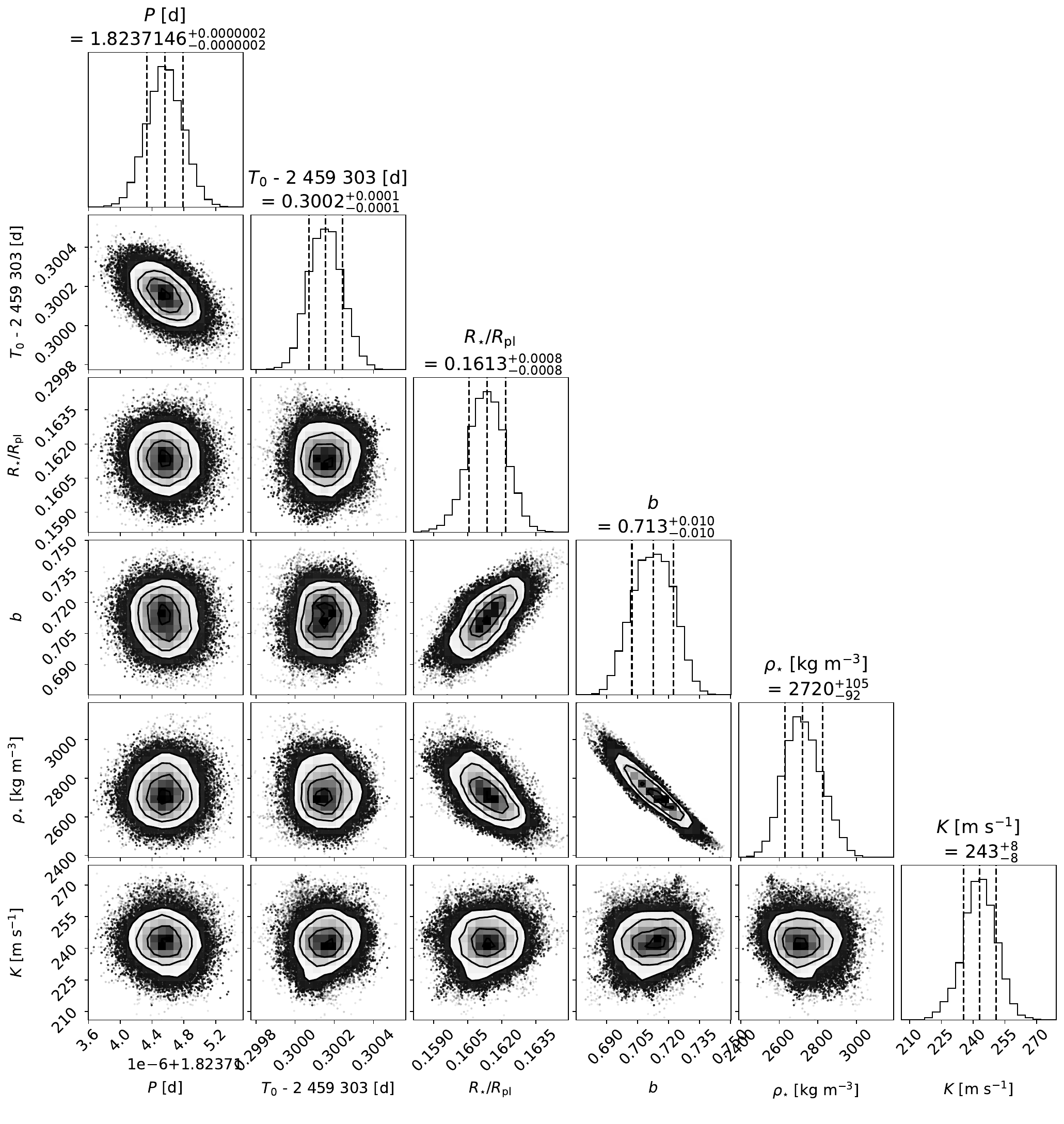}
	\caption{The corner plot for the \texttt{Juliet} results of TOI-2969.}
	\label{fig:TOI2969_corner}
\end{figure}

\begin{figure}
	\centering
	\includegraphics[width=\linewidth]{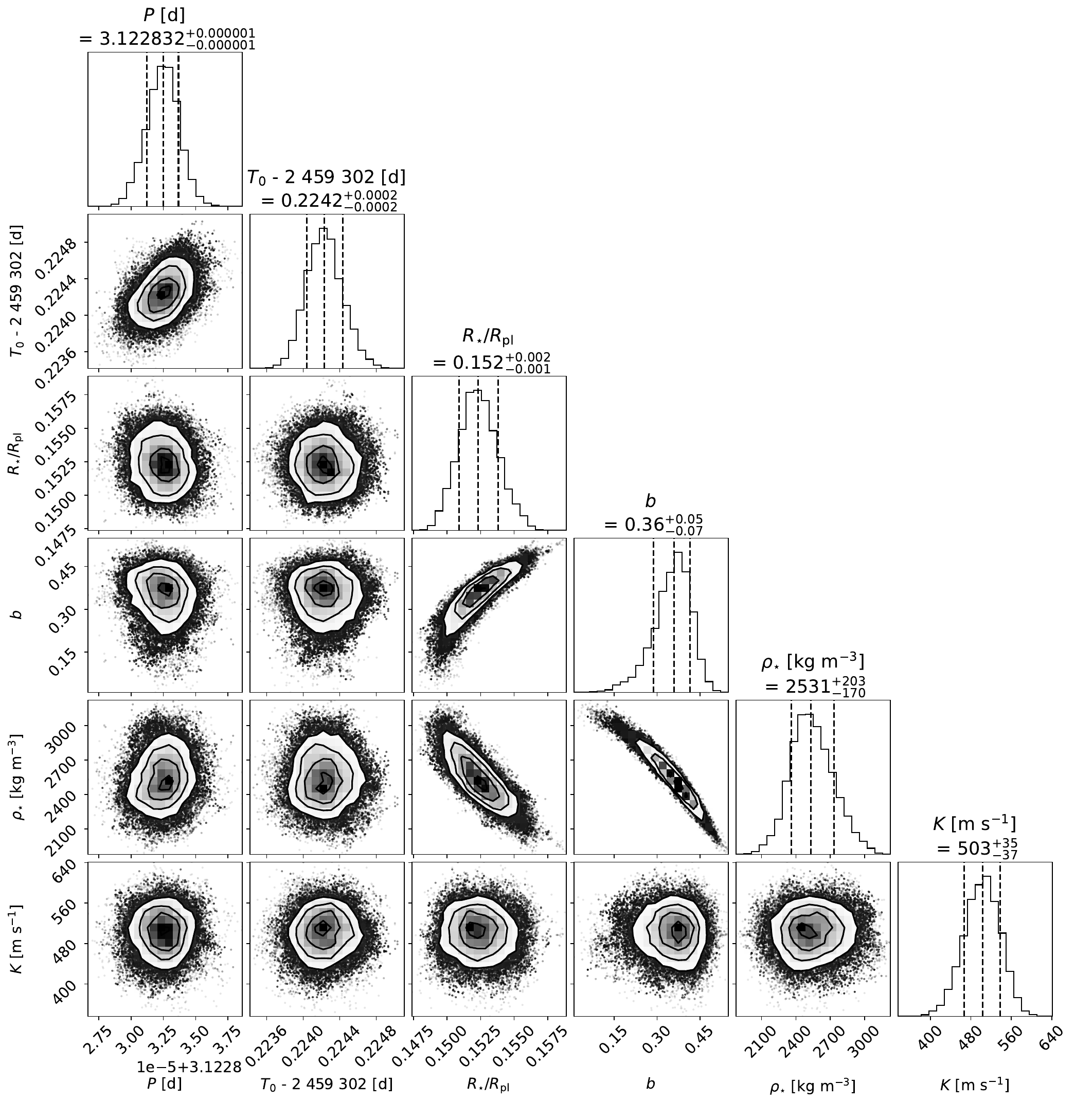}
	\caption{The corner plot for the \texttt{Juliet} results of TOI-2989.}
	\label{fig:TOI2989_corner}
\end{figure}

\begin{figure}
	\centering
	\includegraphics[width=\linewidth]{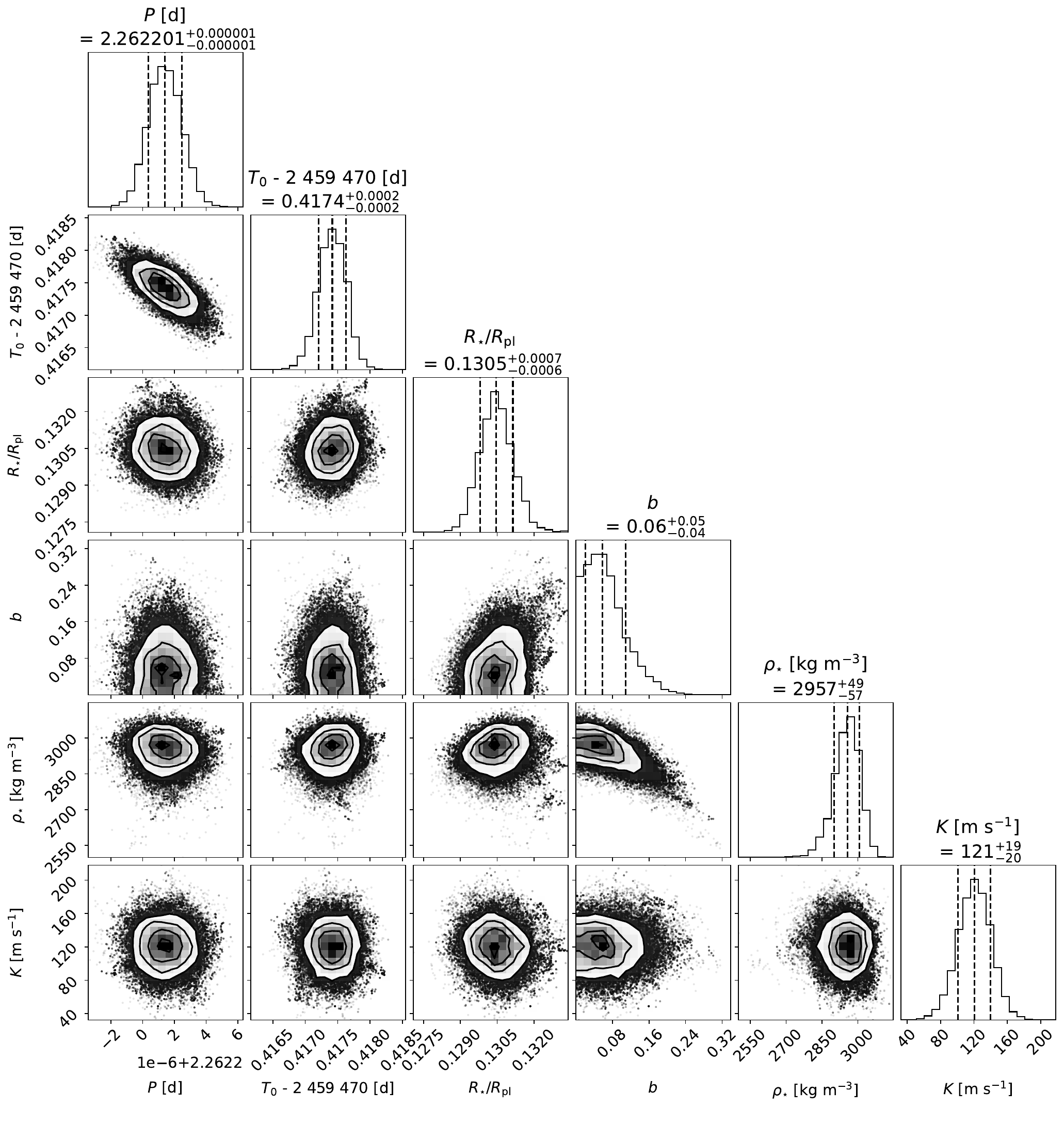}
	\caption{The corner plot for the \texttt{Juliet} results of TOI-5300.}
	\label{fig:TOI5300_corner}
\end{figure}
\FloatBarrier

\end{document}